\documentclass[manuscript]{aastex}
\usepackage{natbib}
\setcitestyle{numbers}

\usepackage{graphicx}% Include figure files
\usepackage{dcolumn}% Align table columns on decimal point
\usepackage{bm}% bold math
\usepackage{multirow}
\usepackage{amsmath}
\usepackage{amsfonts}

\shorttitle{Accretion variability in AK Sco}
\shortauthors{G\'omez de Castro et al.}

\begin{document}

\title{Accretion and inter-cycle variations in the PMS interacting binary AK Sco}
\author{ Ana I. G\'omez de Castro$^{1,2}$, Juan Carlos Vallejo$^{1,2}$, Ada Canet-Varea$^{1,2}$, Parke Loyd$^{3}$, Kevin France$^{4}$}
\author{ Ana I. G\'omez de Castro, Juan Carlos Vallejo, Ada Canet-Varea}
\affil{ [1] Joint Center for Ultraviolet Astronomy, Universidad Complutense de Madrid, Spain}
\affil{  [2] U.D. Astronomia y Geodesia, Fac. CC Matem\'aticas, Universidad Complutense de Madrid, Spain }
\author{Parke Loyd}
\affil{[3] School of Earth and Space Exploration, Arizona State University, Tempe, AZ 85287 U.S.A. }
\author{Kevin France}
\affil{[4] Laboratory for Atmospheric and Space Physics,University of Colorado, Boulder, CO 80309 U.S.A. }
\begin{abstract}
There are only a handful of known short-period pre-main sequence spectroscopic binaries with significant accretion rates (Class II sources). AK Sco stands out in this list because the
system is composed of two equal mass F5 stars in a highly eccentric orbit thus both stars get as close as 11 stellar radii at periastron passage. This configuration is optimal for accretion
studies because enhance accretion events can be precisely timed at periastron passage. In this work, we present the results from the monitoring of the AK Sco system with Hubble during
three consecutive periastron passages. These data provide a unique dataset to spectroscopically characterize accretion and evaluate the inter-cycle variability of the system. Clear
evidence of accretion rate enhancement was observed in cycle1 and 3: the blueing of the near UV continuum, the sudden flux increase of important accretion tracers, such as the N V, 
Si IV and C IV lines, and also of neutral/singly ionized species such as O I and C II. Also, variations in the Si III]/C III] ratio reveals an enhancement of the electron density by an order of
magnitude during the periastron passage. Moreover, in cycle 3, the spectral resolution of the observations obtained with the Cosmic Origins Spectrograph enabled discerning that the flow
was channelled preferentially into one of the two components.
The most remarkable feature in the cycle-to-cycle variations was the detection of a notable increase of the UV flux from cycle 1 to 2 that was not accompanied by enhanced accretion
signatures.

\end{abstract}
%--------------------------------------------------------------------------------------------------------------------

\section{Introduction}

Understanding the connection between accretion and outflow during pre-main sequence (PMS) evolution is a major endeavor in stellar astrophysics and it is required to understand planetary systems formation. The high energy spectrum (UV to X-ray) produced  through accretion plays a key role in the photoevaporation of the gas in the young planetary disks (see Clarke, 2011 and Alexander et al. 2014 for recent reviews) setting the transition between the accreting Class~II sources and the weak line T Tauri (WTTS) phase. There are plenty of studies both from the theoretical and observational point of view addressing this process (e.g. Hollenbach et al. 1994, Font et al. 2004, Alexander et al. 2004b, Gorti \& Hollenbach 2009, Ercolano et al. 2009). There is however, a need for observations of close binary systems where tidal forces are very significant in the transport of angular momentum hence, in tapping the accretion flow. An additional source of interest relates with the formation and evolution of exoplanetary systems hosting the so-called hot Jupiters; close binaries with low mass ratio\footnote{The mass ratio, $q$, is defined as $q = M_2/M_1$, being $M_1$, the mass of the primary and $M_2$, the mass of the secondary.}, such as UZ~Tau, provide important clues on giant planets formation and migration.

Most of the PMS-spectroscopic binaries (PMS-SBs) are low mass systems with primary spectral type G0 or later. In general, the known PMS-SBs are rather evolved objects containing very active young stars (WTTSs) surrounded by thin disks, often debris disks (see Melo et al. 2001 and G\'omez de Castro \& Marcos-Arenal 2015 for a recent compilation).   There are only a handful of PMS-SB that belong to Class II, namely V4046~Sgr, AK~Sco, DQ~Tau, UZ~Tau~E with orbital period smaller than 20 days, GW Ori with period 241.9 days and CS~Cha with period longer than 2482~days (Guenther et al. 2007). AK Sco stands out in this compilation because the system is composed of two equal mass F5 stars in a highly eccentric orbit (e=0.47) that get as close as 11 stellar radii at periastron passage (Alencar et al. 2003). In PMS close binary systems, accretion disks can either take up or release angular momentum and the details of the evolution depend on the mass ratio between the two stars and on the orbit eccentricity (Artymowicz \& Lubow, 1994; Bate \& Bonnell, 1997; Hanawa et al 2010, de Val-Borro et al 2011, Shi et al. 2012). In particular, AK Sco's highly eccentric orbit favors the formation of spiral waves within the inner disk;  the variable gravitational potential produced by the binary acts as a gravitational piston that drags material efficiently from the inner border of the disk at apastron to release it onto the stars preferentially at periastron (see Figure~1, from G\'omez de Castro et al. 2013, hereafter Paper I).  Observational confirmation to this prediction has been provided recently by a Hubble monitoring campaign with the Cosmic Origins Spectrograph (COS) (G\'omez de Castro et al. 2016, hereafter Paper II); coincident with periapsis, a marked 10\% decrease in the H$_2$ emission from  the disk 
was detected  caused by an infalling filament of gas that absorbs the stellar Ly$\alpha$ photons and shades some of the H$_2$ molecules on the disk surface. The light curves also showed  an associated enhancement of the main accretion tracers, namely,  C IV, Si IV, N V, C III, Si III resonance UV multiplets (Paper II). 
AK Sco was also monitored for two more cycles with the Space Telescope Imaging Spectrograph (STIS). All together, the data acquired during these three consecutive cycles provide a unique dataset  to  spectroscopically characterize accretion and evaluate the inter-cycle variability of the system. The aim of this work is to present the results of this analysis.

The article is organized as follows. A comprehensive summary of the Hubble observing campaign is provided in Section 2. AK Sco was monitored during the 
first two cycles with the Space Telescope Imaging Spectrograph (STIS) providing low dispersion, high sensitivity spectra in the full 1140~\AA\ - 3184~\AA\ spectral range; the results from this STIS-based monitoring are described in Section 3. During the last cycle, AK Sco was monitored with COS in the 1159~\AA\ - 1762~\AA\ range with much better spectral dispersion (19,000) to study the kinematics of the line emitting plasma and resolve properly the H$_2$ emission features. The main results from this cycle are gathered in Paper II however, the detailed description of the kinematics of the radiating plasma is included
in Section 4. In Section 5, the data from all the three cycles are analyzed together and compared with the predictions from numerical simulations of the 
dynamical evolution of the binary.  STIS data also show evidence of an extended diffuse envelope around the system radiating in Ly$\alpha$ as will be
shown in Section 6. All observations were obtained in photon counting mode, this enabled studying in detail the light curves especially, during the
long COS observations. The analysis is included in Section 7 and shows no evidence for short time scale fluctuations ($\tau \leq 800$~s)  
neither in the C IV light curve nor in the overall far UV spectrum. To conclude, a short summary with the main results is in Section 8.

\section{Observing program and spectral overview}

The observations were obtained in July-August 2014 with the instruments Space Telescope Imaging Spectrograph (STIS) and Cosmic Origins Spectrograph (COS), the log of observations is in Table~2. As shown in Fig.~2, AK~Sco was tracked during periastron passage in three consecutive periods. During the first two periods, it was observed in low dispersion with STIS and gratings G140L and G230L, to get full coverage in the range: 1140~\AA\ -- 3184~\AA\ range with high sensitivity and spatial 
information. In the last period only the 1159~\AA\ -- 1762.490~\AA\ range was observed but with higher dispersion (19,000) using the gratings G130M and G160M in COS. Observations were carried out in photon counting mode to preserve as much temporal information as possible and enable the time series analysis of the results. 

The average spectrum of AK~Sco during the observing run is displayed in Fig.~3. Emission lines formed at a broad range of plasma temperatures are detected as otherwise, usually observed in the ultraviolet spectrum of the T Tauri stars (TTSs) (G\'omez de Castro, 2009): emission from $H_2$ molecular bands,  from the resonance transitions of highly ionized species (N~V, C~IV, Si~IV),  intermediate ionization species (C~III, Si~III, O~III) and singly ionized or neutral plasma (C~I, C~II, O~I, O~II, Mg~II, Si~II, Fe~II). Contributions from the stellar atmosphere, the outflow and the accretion flow are expected in all of them. The unique feature about Fig.~3 is the unprecedented SNR of the UV continuum obtained after co-adding all exposures. In Fig.~4,  the reddening corrected  average spectrum ($A_V = 0.5, R=4.3$, see Table 1) is compared with that of the nearby F-type main sequence stars,  HD~139664 and HD~22879 (see Appendix A for details on the UV spectrum of F stars). HD~139664 (F5) is known to have a debris disk but thin enough to have a tiny contribution to the infrared IRAS flux (see Fig.~8 in Scheneider et al. 2014). Some of the main photospheric features of an F5 star are readily recognized in AK~Sco UV spectrum, as well as the significant excess below 2000~\AA . The overall UV spectral energy distribution is closer to a later spectral type (F9) star that to an F5.

\section{ Results from the STIS monitoring}

\subsection{UV Continuum variability}

The UV continuum of F stars in the $1640-3100$~\AA\ is very sensitive to the spectral type, as shown in detail in Appendix A.  This trend is quantified in Fig.~5 where the ratio between the integrated fluxes in the the bands: F1 ($1640-2400$~\AA ), F2 ($2400-2775$~\AA ) and F3 ($2830-3100$~\AA ) has been computed for the F stars in Table~A1 and used to build the rates $R1 = F1/F2$ and $R2 = F2/F3$ represented in the Figure. Spectral types of main sequence stars are indicated, they define a clear trend parallel to the extinction arrow; extinction is negligible for most of them (see Table~A1). Each band traces a main component: F1 is the most significantly affected by extinction (also includes the UV bump), F2 is dominated by the Fe~II multiplets and F3 by the Balmer continuum and the stellar photosphere. The 55~\AA\ separation between F2 and F3 bands has been set to avoid the very strong Mg~II feature.  The extinction arrow has been drawn  for the average ISM extinction law (Fiztpatrick \& Massa, 2007), as well for a modified extinction law with $R=4.3$ that according to Manset et al. (2005) is representative of the circumstellar environment in AK~Sco. Notice that the extinction arrow runs roughly parallel to the spectral types, i.e., there is degeneracy between spectral type and extinction.

We have computed $R1$ and $R2$ for all STIS observations of AK~Sco (cycles 1 and 2) and over-ploted them on the figure. AK Sco is located closer to late F stars than to early F in the diagram though it is classified as an F5 star from its optical spectrum. AK Sco extinction is small, A$_V = 0.5$ (see Table 1) or E(B-V)$\simeq 0.12$, and thus extinction alone cannot
account for the observed SED. Henceforth, according to its UV spectrum, AK Sco should be classified as an F8-F9 star instead of the F5 type assigned on the base of its optical spectrum. Moreover, the excess contribution from the accretion flow pushes AK~Sco away from the  main sequence band. The  variations found during our observing campaign go basically perpendicular to the extinction arrow and are most likely caused by variations in the accretion rate. The integrated UV flux in the $1640 - 3100$~\AA\ spectral range increased by a 17\% from cycle~1 to cycle~2.

During cycle~1, the excursion of AK~Sco in the diagram has a non-negligible component parallel to the extinction arrow of  $E(B-V) \simeq 0.01$, corresponding to a variation in $A_V$ of -0.043 ($A_V =  R*E(B-V)$); this variation is towards a decreasing extinction or increasing blueing of the spectrum in the middle of the cycle's observations. 

\subsection{Spectral lines variability}

In figure 6, the spectral variability during cycle~1 and 2 is represented. The variability has been computed as the standard deviation of the mean of the spectra for each cycle\footnote{Let us denote as $S_i$ each spectrum and N the number of spectra obtained in the cycle. The mean spectrum is computed as 
$<S> = (\sum _{i=1}^{i=N} S_i)/N$ and the standard deviation of the mean as, $ \sigma = ((\sum _{i=1}^{i=N} (S_i^2-<S>^2))/(N(N-1)))^{1/2}$}; as displayed it is concentrated in the cores of the emission lines but there are differences between both cycles, being the main,

\begin{itemize}
\item The variability of the most prominent UV lines is a factor of 2 higher during cycle 1 than during cycle 2.
\item During cycle 1, the UV continuum (2630 \AA\ - 3150  \AA\ ) varies by less than 0.8\%; during cycle 2 this variation rises to 2.2\%. 
\item Not all spectral lines follow the same behaviour. For instance, N~V variability during cycle 1 is significantly higher than during cycle 2. 
The same occurs for O III] and Mg~II.
\end{itemize}

The light curves of the main tracers are plotted in Figure~7. Cycle~1 flux variations are reminiscent of  those observed in the high dispersion spectra of cycle 3 (see Paper II); at phase $\sim 1$, at the periapsis, there is a sudden increase in the flux that it is observed in N V, Si IV, C IV but  also in singly ionized species such as C II or O I. During cycle 2, variations are softer and within the error bars for most tracers. Clearly,  the observed trends change from one cycle to the next. It is noteworthy, that the semiforbidden transitions do not follow by the same trends than the permitted lines and their variability is not correlated.

\subsection{Flux-flux relations and accretion shock diagnostics}

Radiation from an accretion shock has a markedly different spectral energy distribution than radiation from a cool star atmosphere. 
In cool stars, magnetic energy is transported from the stellar surface onto the atmosphere where it is dissipated into three
main regions: the warm (T$\sim 10^4$ K) chromosphere, the hot (T$\geq 10^6$ K) corona and the transition region (TR) between them.
The TR is a very thin layer (some $10^4$ km thickness in the Sun) where the temperature increases by two orders of magnitude and it can only 
be observed at UV wavelengths. There are well characterized correlations between the flux radiated in the various spectral tracers of these regions; chromospheric (neutral and  singly ionized species), TR (C III, Si III, SI IV, C IV, N V, He II...) and coronal spectral tracers (highly ionized species, O VI, X-ray flux...). These so-called flux-flux relations are used to model energy transport in cool stars and call for a universal mechanism operating in them (Ayres et al. 1995, Mihalas 1978). 

Flux-flux relations have also been studied in TTSs (Hu\'elamo et al. 1998, Johns-Krull et al. 2000, Yang et al. 2012, G\'omez de Castro \& Marcos-Arenal 2012, hereafter GdCMA). When compared with their main sequence analogues, it becomes evident that there is excess radiation from low 
ionization species (C II, Mg II, OI) with respect to the highly ionized ones (C IV, Si IV, He II). This indicates that radiation is released by a
different mechanism than in cool stars. The  most successful models propose that the excess gravitational energy of the accreting
matter is released into heating at  accretion shocks where the temperature reaches 0.3-1 MK, i.e., coronal-like temperatures, driving
a photoionization cascade that results in the observed scalings (Calvet \& Gullbring 1998, G\'omez de Castro \& Lamzin 1999). 

AK Sco monitoring provides a unique chance to evaluate how these scalings behave during an accretion event and hence test the theoretical models. Of the many possible flux-flux relations we should focus in four: C IV versus O I, C IV versus C II, C IV versus He II and Si III] versus C III].  To study AK Sco behaviour in the context of the TTSs properties, all observations of TTSs obtained with HST and STIS G140L and G230L gratings have been downloaded from the Hubble archive and the  fluxes 
of the O I, C II, C IV, He II, Si III]$_{1892}$ and C III]$_{1908}$ lines have been determined. 

The C IV versus OI flux-flux relation is optimal to evaluate the relative abundance between highly ionized and neutral species in the TTSs environment; it shows the largest deviation from  the trend observed in main sequence stars (see GdCMA). Figure 8 shows that AK Sco falls in the TTSs trend and that from cycle 1 to cycle 2 moves along the trend with the largest fluxes being observed in cycle 2. Both cycles are neatly separated in the plot. Moreover, the intra-cycle variations are significantly smaller than the inter-cycle ones. 
Note that AK Sco fluxes are the sum of the contribution of the two components of the system. To evaluate the contribution from
each component is not trivial since according to numerical simulations the accretion flow may be preferentially channeled in any of them (Paper I); 
splitting the flux in equal parts between the two components is equivalent to shifting by 0.3 dex the location of AK Sco in the diagram. 

Cycles 1 and 2 are markedly different. During cycle 2, there are not significant variations in the line fluxes however, this is not the case in cycle 1; the OI flux
increases by a 20\% from the beginning to the end of the cycle while the C IV flux increases by a $\pm 7$\% in the same time lapse. The overall flux increase could be
accounted by a decrease in the extinction but the markedly decrease of the C IV flux at the beginning of the cycle cannot be explained by an extinction effect.

The  trends observed in the C IV-O I flux-flux diagram are reproduced in the C IV-C II flux-flux  and the C IV-He II flux-flux diagrams (see, Figure 9 and 10, respectively). AK Sco observations are located on the regression line of the TTSs in the C IV-C II diagram and they are slightly up in the C IV-He II diagram, similarly to what observed in the C IV-O I diagram.

The ratio Si III]/C III] is density sensitive and  it is often used to measure the electron density of warm plasma in  the atmosphere of late-type stars and TTSs (Brown et al. 1984, G\'omez de Castro \& Verdugo 2001, 2003). In AK Sco,  Si III]/C III]$\simeq 2$, similar to what observed in the TTSs (see Figure 11), 
and indicates  that the electron density of the emitting plasma is in the range  $10^9$~cm$^{-3}$ to $10^{11}$~cm$^{-3}$ depending on the 
precise plasma temperature (see Figure 4, in G\'omez de Castro \& Verdugo, 2001). 
The Si III]/C III] ratio does not vary significantly from cycle to cycle ($2.2 \pm 0.6$ in cycle 1 and $2.0 \pm 0.3$ in cycle 2) even though the line fluxes do vary.
The first exposure in each cycle has very good SNR and it can be used as a pivot to measure variations during cycle. Again, a markedly different behavior is observed between cycle 1 and cycle 2. During cycle 2, no significant variations are observed (1-$\sigma$ and 2-$\sigma$ confidence ellipse are displayed in the plot). However, in cycle 1 there is a significant variation between the first exposure (phase 0.9939) and that at phase 0.9993, when the lines ratio increases by a 37\% and thus, the electron density of the plasma.

\section{Results from COS monitoring}

The mean profiles of the main transitions are displayed in Figure~12; the profiles are very broad and centered at rest
(the radial velocity of the system is 1.3 km~s$^{-1}$). The two spectroscopic components are not resolved in spite of their high 
relative velocity at periastron passage; from 170 and 190 km~s$^{-1}$ during the monitored phase interval  (Alencar et al. 2003).
The highest asymmetry is observed in the He II transition; as shown in Figure~13, after periastron passage, the line becomes significantly
redshifted (95 km~s$^{-1}$).  This redshift  cannot be caused by mass infall; though He II is a very sensitive tracer of accretion,
He II observations of TTSs show that it is only slightly redshifted, if at all (G\'omez de Castro 2013).  However, 95 km~s$^{-1}$ is the 
expected radial velocity of one of the two components of the system at periapsis thus, the observed shift  
of the He II flux enhancement indicates that at periastron accretion is preferentially driven into one of the two components. 

A similar trend is observed in the rest of the accretion tracers though blurred by the large broadening of the profiles. 
To visualize it better, we have computed the {\it excess} profiles obtained by subtracting the first observation (phase=0.9920 for G130M and phase=0.9930 for G160M) from the rest. In Figure 14, the excess profile is represented as a hyper-surface in phase and wavelength for the main lines. 
The increase of the red-wards shifted component flux from phase 1.0034 on is clearly noticeable in the Si~IV,  C~V, N~V and Si~III lines. 
At the same time, the blue-wards shifted component becomes dimmer. This behavior is observed in all hot gas lines, regardless of the COS grating setting or detector segment from which the data were extracted.  It is also in marked contrast to the COS observations of H$_{2}$. The uncertainty of the COS wavelength solution is $\approx$~15 km s$^{-1}$ (Holland et al. 2014, {\it COS Instrument Handbook}), therefore we consider the systematic redshifts of the emission lines to be a real effect.  
The profiles of the {\it CIV excess} are also plotted in Figure 15.  They show an increasing absorption in the blue edge of the line that it is caused by a progressive decrease of the flux in the blue-wards shifted part of the profile compared with the beginning of the periastron passage.  One may naively think that this effect is caused by a decrease in the C~IV flux (hence the accretion rate) from the companion ($V_{rad} \in [-95,-85]$~km~s$^{-1}$) however, the sharp edge at 286 km~s$^{-1}$ precludes this interpretation.
There is also an H$_2$ emission  line on the blue-wing of C IV 1548 (see Herczeg et al. 2002, France et al. 2014) however, variations in the 
H$_2$ could not caused the observed edge since the line is very narrow and H$_2$  variability is decoupled from the atomic species (see Paper II).

The light curve of the excess is displayed  in Figure 16 for the main spectral lines. The fastest rise is observed in He II; the excess rises by a factor of 4 in one hour. The C IV light curve is significantly softer ({\it excess} growth rate of 2.3 per hour). The maximum {\it excess} is observed from phase 1.01 on (3.26 hours after periapsis).

\section{Accretion and cycle-to-cycle variations}

The AK Sco binary system has a highly eccentric orbit, as a result when the system approaches periastron, the outer boundaries of the circumstellar disks (and the accretion streams passing by) get close enough one to each other to effectively lose the angular momentum, leading to an increase of the accretion rate. The predictions from numerical simulations 
for some sample cycles are shown in Figure 17; most often, the accretion flow is not evenly distributed between the two components of the system. In fact, there is 
a pronounced asymmetry in some cycles, see {\it i.e.} cycle 12 in Figure 17. Inter-cycle variations do not occur only in the total amount of the accretion rate but also on the 
details of the temporal distribution of the infall that shows in the light curves of the relevant spectral tracers.

The Hubble observations confirm these predictions:

\begin{itemize}
\item During cycle 1, there is a blueing of the near UV continuum and an increase of the line flux at phase 0.9982-0.9993 by $\sim 10$\%. The increase is pronounced 
in OI, C II, and Si IV and it is less prominent in saturated (Mg II, C IV) or weak (N V, He II) transitions.  This evolution cannot be interpreted as an increase of the 
fraction of the stellar surface affected by the accretion shock; rather, the variations observed in the flux-flux diagrams call for a variation in the accretion rate
and the electron density at the shock front. It is noteworthy that the  Si III]/C III] ratio  rises by a 37\% from phase 0.992  to phase 0.9993 ([1] and [3] in Figure 11) corresponding to an increase of the electron density by a factor of $\sim 10$ for a fiducial temperature of 50,000~K.  These observations are consistent with the theoretical prediction of enhanced mass-infall (accretion rate) at periastron.
\item During cycle 2, the flux is higher but there are not significant variations during the monitored time lapse. 
\item During cycle 3, the behavior observed in cycle 1 seems to be reproduced. Moreover, the He II and CIV profiles show evidence of the accretion rate enhancement being channeled preferentially onto one of the two components of the system. 
\end{itemize}

The UV radiation studied in this work is mainly produced at accretion shocks or very close to the stellar surface. 
Numerical calculations of the structure of accretion shocks in TTSs indicate that the C~III], O~III] and Si~III] lines should have comparable intensities and that their ratios can be reliably used to derive the density, accretion infall velocity and hence, accretion rate on these stars (G\'omez de Castro \& Lamzin 1999). Though C~III] and Si~III] transitions could also be excited at the base of the jets (G\'omez de Castro \& Verdugo 2001),  the profiles of these lines in AK~Sco clearly indicate that any contamination by a possible jet is negligible (G\'omez de Castro 2009), see also Sect.~6.  The location of AK Sco in the Si III]/O III] versus Si III]/C III] diagram
is displayed in Figure 18. There are not significant variations during the HST monitoring and, in all cases, the observations indicate that the emission is produced by low temperature (mild shock) and low density (low accretion rate) plasma. The electron density inferred is $3-4 \times 10^{10}$~cm~s$^{-3}$ in good agreement with the predictions
from generic collisional plasma diagnosis.

The Si III]/O III] ratio varies from $3.2 \pm 0.4$ in the first observation of cycle 1 (the one with the highest SNR) to $2.6\pm 0.3$ 
in the first observation of cycle 2; though this variation is marginal, it is suggestive of a change in the electron temperature from one cycle to another. The Si III]/O III] ratio is temperature sensitive and in the accretion shock scenario, this variation is associated with a small change in the shock velocity;  from $200~km~s^{-1}$ to 215~km~s$^{-1}$. 
Material in the shock front is heated by the release of the gravitational energy of the infalling material at the impact point; roughly $T \simeq  \mu v^2_{shock}/3k_B $. The higher the shock velocity, the higher the electron temperature in the Si III], C III] lines formation region. Small variations in the end shock speed are to be expected since the angular momentum
of the material in the innermost orbit might suffer slight variations due to the dynamics of the system (see Paper I).

\section{Extended Ly~$\alpha$ emission}

Ly~$\alpha$, Mg~II and semiforbidden line radiation (C~II], C~III, Si~III]) has been detected from jets and Herbig-Haro objects in TTSs (Coffey et al. 2004, L\'opez-Mart\'inez \& G\'omez
de Castro, 2015). We used the STIS 52\arcsec~$\times$~0.2\arcsec\ slit to enable the possibility of a serendipitous detection of extended outflows from the AK Sco system.  The STIS long-slit was oriented at position angles of 28.5\arcdeg\ and  36.6\arcdeg\ for visits 01 and 02, respectively.  The objective here was to build up signal to look for the spectral signature of extended emission, and consequently all of the STIS G140L and G230L observations were coadded to maximize the chance of detection.  The two dimensional spectra were aligned by fitting a Gaussian profile to the cross-dispersion profile of bright emission regions, \ion{C}{4} and the NUV continuum for G140L and G230L, respectively.  The centroids of all of the individual exposures were then shifted to the spatial centroid of the initial exposure and the data were stacked.  Centroid shifts were $\leq$ 2 pixels in all cases.  Spatial profiles of several spectral regions of interest were then extracted from the coadded G140L and G230L spectral images.   

Figure~19 shows the flux-normalized spatial profiles from the G140L ($top$) and G230L ($bottom$) observations.  We created individual extractions of \ion{H}{1} Ly$\alpha$, \ion{C}{2}, \ion{Si}{4}, the 1427~--~1520~\AA\ region comprising H$_{2}$ emission and FUV continuum, \ion{C}{4}, and \ion{O}{3}] from the G140L spectra.  \ion{Mg}{2}, an adjacent continuum region to \ion{Mg}{2}, and two continuum regions dominated by Balmer continuum and stellar photosphere (2000~--~2620~\AA\ and 2900~--~3040~\AA) were extracted from the combined G230L observations.  The \ion{Mg}{2} profile is indistinguishable from the adjacent continua.  The FUV emission lines (except possibly Ly$\alpha$) are also unresolved, with lines at shorter wavelengths displaying progressively broader profiles at flux levels $\lesssim$~2~\% of the peak.  These are likely the result of the extended short-wavelength point spread function of the $HST$ optical telescope assembly, owing primarily to mid-frequency errors related to the final quality of the telescope polish.  The Ly$\alpha$ profile displays both a 1-pixel offset from the main FUV peak and evidence for spatial extension.  We caution the reader that these profiles are extracted across the bright geocoronal Ly$\alpha$ emission line~--~the dotted line profile in Figure 5, $top$, shows the Ly$\alpha$ profile prior to airglow subtraction and the solid circles are the profile following the subtraction of a constant geocoronal airglow component, the average of 100 pixels centered approximately 4.3\arcsec\ in the ``+$y$'' direction on the STIS FUV MAMA.   Taking the nominal 103 pc distance (van Leeuwen 2007) and 68\arcdeg inclination angle (Alencar et al. 2003) for AK Sco, $>$ 90~\% of the FUV line flux is located within $\pm$~9.3 AU of the center of mass of the system, with $>$~90~\% of the NUV flux originating within $\pm$~7.4 AU from the center of mass.  Our conclusion is that there is tentative evidence that AK Sco displays extended Ly$\alpha$ emission (subject to large uncertainties in the airglow subtraction), but that all of the other emissions originate within $\approx$~$\pm$~10 AU of the center of mass.

\section{Light curve analysis: search for low frequency modes}

In a previous work (Paper I), we reported the detection of an Ultra Low Frequency (ULF)
oscillation in the UV light-curve from AK Sco using the Optical Monitor (OM)
instrument on-board the XMM-Newton. The ULF oscillation was excited close to the periastron passage,
lasting only for few oscillations at the rise of the UV light curve. It had a period of 
790$^{+200}_{-150}$~s  and was detected with filter UVM2 (spectral range 2000 \AA\ - 2700 \AA). 
This bulk radiation may be produced in several distinct regions  of the TTSs environment
(chromosphere, accretion shock, photoionized gas in the inner disk...) making difficult the identification
of the source. The spectral lines detected with COS cover a broad range of densities and temperatures 
hence the analysis of their light curves has the potential of enabling the identification of the source.

For this analysis, the COS event lists datasets have been used instead of the 1-D extracted spectra (.x1d files).
The light-curve generation reads the segments A and B from FUV datafiles
following the steps taken by the light-curve COS package from Ely et al. (2015).
These light-curves were later on processed for computing the Lomb-Scargle Periodograms 
(Lomb 1976, Scargle 1982), using the standard astropy package.
This method performs well with unevenly spaced data, as is our case.
The Lomb-Scargle method essentially fits a sinusoidal model to the data at each frequency, with a 
larger power assigned to a given frequency reflecting a better fit.
The selected normalization uses the residuals of the data around the constant reference model,
leading to the resulting power $P$ to be a dimensionless quantity between $0$ and $1$.

When using a Lomb-Scargle Periodogram and deciding whether a signal contains a periodic 
component, an important consideration is the significance of the peaks. 
In a Bayesian view, the Lomb-Scargle periodogram is the optimal statistic for 
detecting a stationary sinusoidal signal in the presence of Gaussian noise.
Hence, this significance is usually expressed in terms of a false alarm probability (fap), 
which encodes the probability of measuring a peak of a given height (or higher) 
conditioned on the assumption that the data consists of Gaussian noise with no periodic component.
A fap level of, let us say $1\%$  means that, 
under the assumption that there is no periodic signal in the data, 
we will observe a peak reaching such level high or higher only $1\%$ of the time.

The COS monitoring of the periastron passage lasted about $9$hrs.
The Figure~\ref{fig_jcv_periodogram} shows the periodograms corresponding 
to the light-curves (see Paper II). 
The upper panel corresponds to the G130M data, and the bottom panel to
the G160M data.
The periodograms are built up to a maximum period of $2.0$hrs.
The peaks around the rightmost vertical line correspond to the detection
of the HST orbital period of around $1.6$hrs. The grey area shows 
the ULF interval as reported in (Gomez de Castro 2013).
The periodogram panel also plots the horizontal lines corresponding to 
the required peak height to attain a false alarm probability (fap) of $10\%$, $5\%$ and $1\%$.

In addition to the peaks linked to the HST orbital period, we observe other high peaks.
The Lomb-Scargle periodograms detect all periodicities found in the data, including those periods
corresponding to the observational sampling period and its $1/n$ multiples. That is, they will also reflect
combinations of the duration of exposures and separations between them.

In order to distinguish any real ULF periodicity from those aliased multiples, we have 
overlaid in the figure the periodograms resulting from fixing the real signal to an arbitrary 
value of $1$. In this way, the pattern of the resulting peaks will exclusively correspond 
to the observational sampling. 

The region where the ULF is suspected to be present is between $0.15$ and $0.4$hrs.
Here, all the peaks are grouped and mimic those found
at larger periods, which indicate they are alias from main frequencies. 
Unfortunately, the typical G160M exposure duration is around $0.22$, 
and these peaks likely come from this sampling duration close to the searched ULF period. 
The results with G130M data are very similar. Fixing again the counts to an arbitrary unit value, 
we get the same pattern of peaks resulting from the observational sampling. 
Finally, when computing the periodograms corresponding to the light-curves 
selecting certain wavelengths, such as C IV, Si IV (Hot species) and C II
(warm species), the results are again comparable. Any periodicity within the 
ULF region may still exist, but with a really high fap, not distinguishable from noise. Outside the
ULF region, one may search for other periodicities. But, in a similar manner, 
all the peaks with low fap are linked to the observing windows, and a clean up process of the sampling frequencies
does not improve the results.

As an alternative, we can compute the periodograms just using individual exposures. The signal will be weaker 
and the fap levels may decrease, but the main sampling multiples will not be present.
This analysis is seen in Figure~\ref{fig_jcv_periodogram_individual}. The upper panel shows the results coming from the 
G130M single exposures. We do not detect any periodicity in the ULF area.
The G130M exposure labeled as `isq' presents a peak around a period of $0.17$hrs, but the corresponding 
fap is very high, around $83\%$, indicating that this could be produced by noise. 
The G160M analysis also presents peaks with a very high fap. However, as the G160M exposures 
were taken in consecutive pairs, this analysis uses longer duration light-curves.
The results can be seen in the bottom panel of Figure~\ref{fig_jcv_periodogram_individual}.
The pair labeled `jeq-jgq' presents a peak around $0.25$hrs, with a fap of $\%24$. One may think this peak could be
an alias of the double exposure duration. But, notably, such a peak is not present in any of the remaining
curves. Hence, we may have a (faint) indication that the ULF was present at least during one of the exposures.

\section{Summary and Conclusions}

The observations provided by the dedicated Hubble's monitoring of AK Sco have enabled for the first time to track the variability of a pre-main sequence
binary with a degree of detail similar to that of the  UV observations of interacting binaries. Much of the observed behaviour was already predicted by numerical simulations (Paper I)
but this campaign has provided the highly needed experimental evidence of the erratic cycle-to-cycle variations. Also the radiative output from accretion has been
accurately measured rendering fundamental data for accretion shocks  calculations.

Two other PMS interacting binaries had previously been monitored with Hubble namely, UZ Tau~E and DQ Tau (Ardila et al 2015) but the phase coverage had significantly poorer 
temporal resolution and this is fundamental issue since, as shown in this work, the dynamics of these systems is very complex. Moreover,  the gravitational piston effect of the passage by the periapsis is less significant in them making more difficult the precise timing of the accretion events. DQ~Tau, the system most similar to AK Sco, is composed of two 
equal mass stars with mass 0.5~M$_{\odot}$, in a orbit less eccentric than AK Sco's one and with larger semi-major axis (Mathieu et al. 1997). The mass ratio of the components in UZ~Tau~E is 0.289  (Prato et al. 2003) thus mass infall is preferentially channelled in the primary and the effect is less significant.

AK Sco was monitored for two consecutive cycles in low dispersion (high sensitivity) with STIS and a third one with COS  providing good time and kinematical 
resolution at the cost of a lower SNR in the flux. 
STIS data analysis revealed the enhancement of the accretion rate during the first periastron passage, supported by:
\begin{itemize}
    \item 	Blueing of the 1640-3100 NUV continuum between 0.9982 and 0.9993 phases, in the middle of the cycle.
    \item Sudden increase in the flux of important accretion tracers, such as the  NV, Si IV and C IV lines, an also in neutral singly ionized species such as O I and C II. 
	\item Variations in the  Si III]/C III] flux-flux diagrams, revealing variations in the electron density by an order of magnitude during the periastron passage.
\end{itemize}
This behavior is reproduced as well in the third cycle, in agreement with the previous analysis showed in G\'omez de Castro et al. (2016). Moreover, the high resolution of COS make possible to recognize an increase in the red-wards shifted component in important accretion tracers (He II and C IV), while the blue-wards shifted component of these lines decreases, pointing out that accretion is preferentially driven into one of the two components of the system. 
These intra-cycle variations are in concordance with the results given in Paper I through the XMM-Newton monitoring of the AK Sco system, where the enhanced of the UV and X-ray flux in the binary was suggested to be produced by an accretion outburst.\\
 
Cycle-to-cycle variations have been measured as well, where the most remarkable feature is the notably increase in the total UV radiation of the system from cycle 1 to cycle 2. Moreover, between the two first cycles, the Si III]/O III] temperature-dependent ratio reveals a marginal variation of the plasma temperature, that could be translated into a small change in the shock velocity from 200 km/s to 215 km/s.\\
Despite the measured enhancement of the UV radiation in cycle 2, the absent of significant variations in the flux of spectral lines and in the flux-flux relations during this cycle reveals no hints about the accretion rate enhancement, in discordance with the other two cycles.

Beside accretion processes in the AK Sco system, the presence of extended emission due the presence of jets has been addressed in this study. Spatial cross-dispersion profiles of STIS data allowed us to identify hints of a diffused envelope around the AK Sco binary radiating in Ly$\alpha$. \\

Finally, inspired by the detection of ultra-low frequency oscillations in AK Sco (Paper I), we analyzed the presence of these low-frequency modes in the light-curve of the binary, however, the analysis of the UV light curve from COS data only shows a minor indication that the ULF may be present in the region where the C~IV line is produced. This result requires further confirmation with a dedicated campaign monitoring AK Sco only with COS/G160M.

\acknowledgments
This work has been partly funded by the Ministerio de Economia y Competitividad of Spain through grant 
ESP-2017 -87813-R.
The data presented here were obtained through $HST$ Guest Observing program 13372.

{\it Facilities:} \facility{HST (COS)}, \facility{HST (STIS)}.

\appendix
\section{The ultraviolet spectrum of F stars}

For the processing and modelling of the UV continuum of AK~Sco, and its variability, it is important to compare the UV spectrum of AK~Sco with that of F-type main sequence stars. For this purpose, we have searched the archive of the International Ultraviolet Explorer (IUE) mission for observations of F stars (IUE object class 41).  924 observations were obtained of stars either in the 2000-3200~\AA\ range (cameras
LWP or LWR) or in the 1200-2000~\AA\ range (camera SWP).  We submitted the target list output from the IUE archive to the Centre of Donn\'ees Stellaires de Strassbourg (program SIMBAD) for the sources cross-identification. We found that about 30\% of the observations corresponded to RS~Cannis~Venaticorum systems and spectroscopic binaries. Only single stars with well known spectral types were selected and from those, only stars with high SNR and non-saturated spectra in the 1200-3200~\AA\ (see Table~A1).

The spectra are shown in Fig.~A1. Three windows can be readily identified: W1(1640-2400~\AA ), W2(2400-2775~\AA  ) and
W3(2775-3100~\AA ). W1 contains the UV bump and the high energy tail of the spectrum; it is most prominent in early types.
W2 is dominated by the Fe~II multiplets (2,3,32-36,62,63,363). Flux in W3 increases for the late spectral types as the W1 flux decreases.
This permits using the rates between the integrated fluxes in these windows, $F(W1)/F(W2)$ and $F(W2)/F(W3)$, as spectral subtype (or effective temperature) indicators.

%%%%FIGURES

\newpage

\begin{figure}[h]
\centering

\begin{tabular}{cc}
\includegraphics[width=6cm,angle=270]{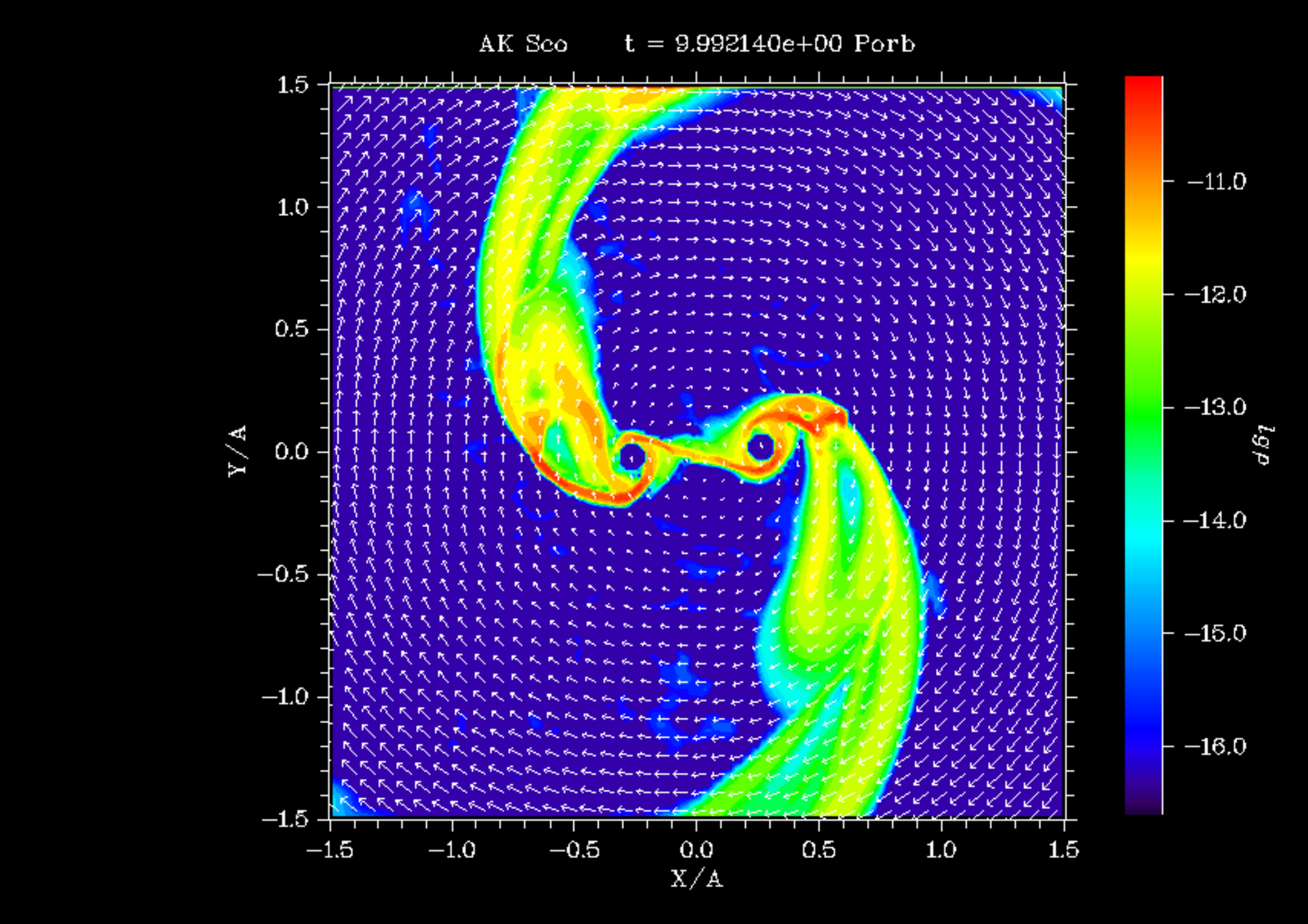} & \includegraphics[width=6cm,angle=270]{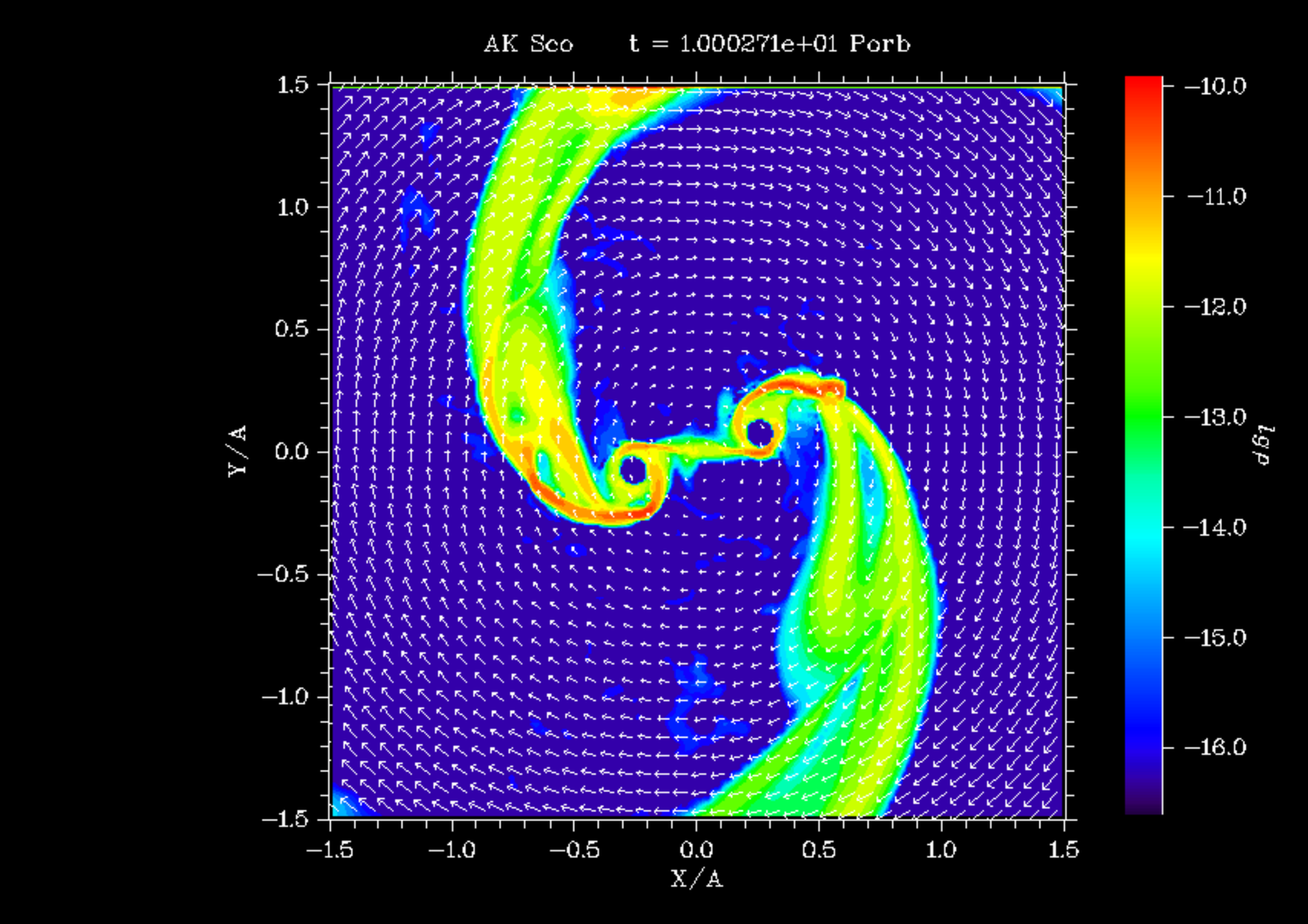} \\
\includegraphics[width=6cm,angle=270]{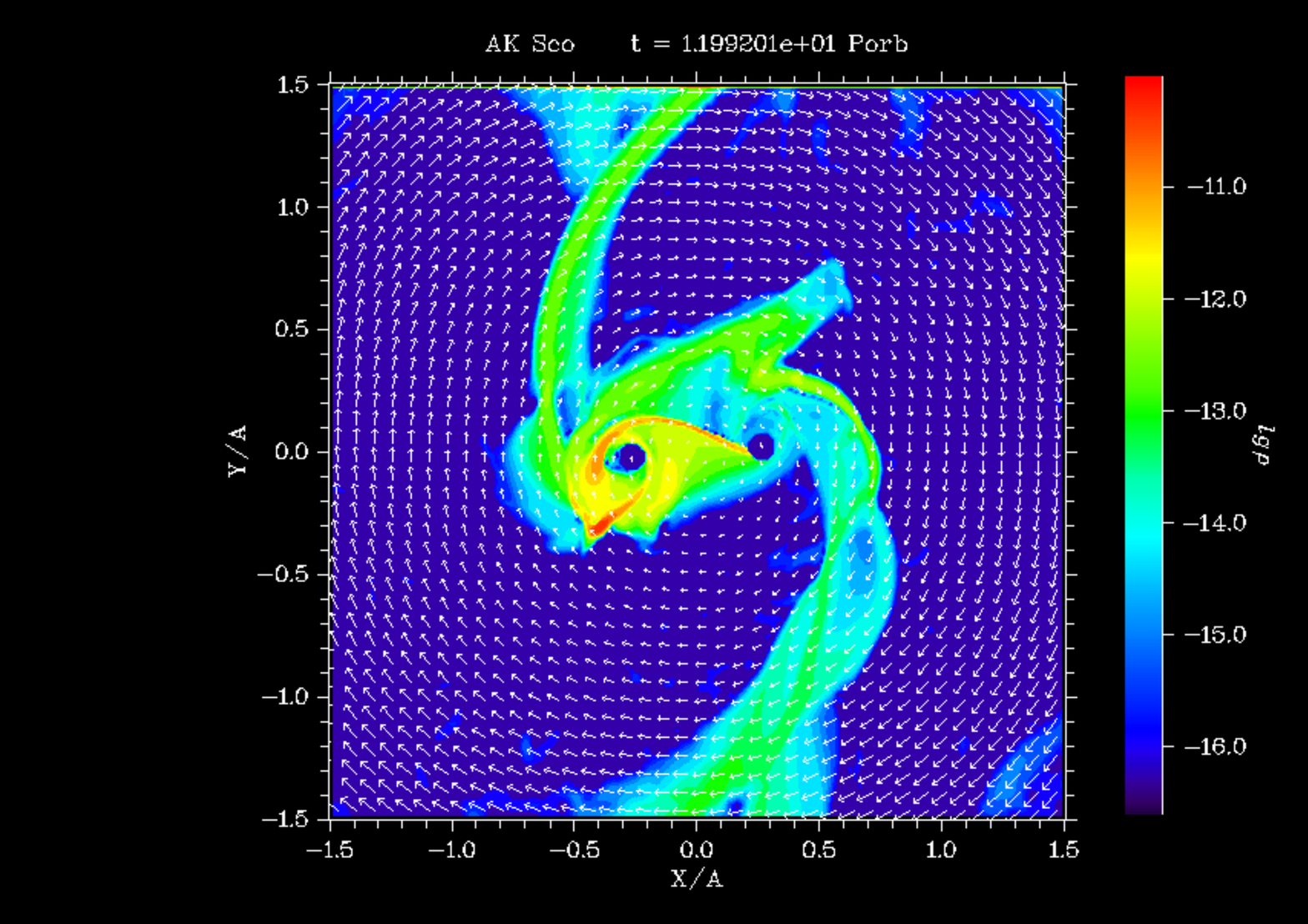} & \includegraphics[width=6cm,angle=270]{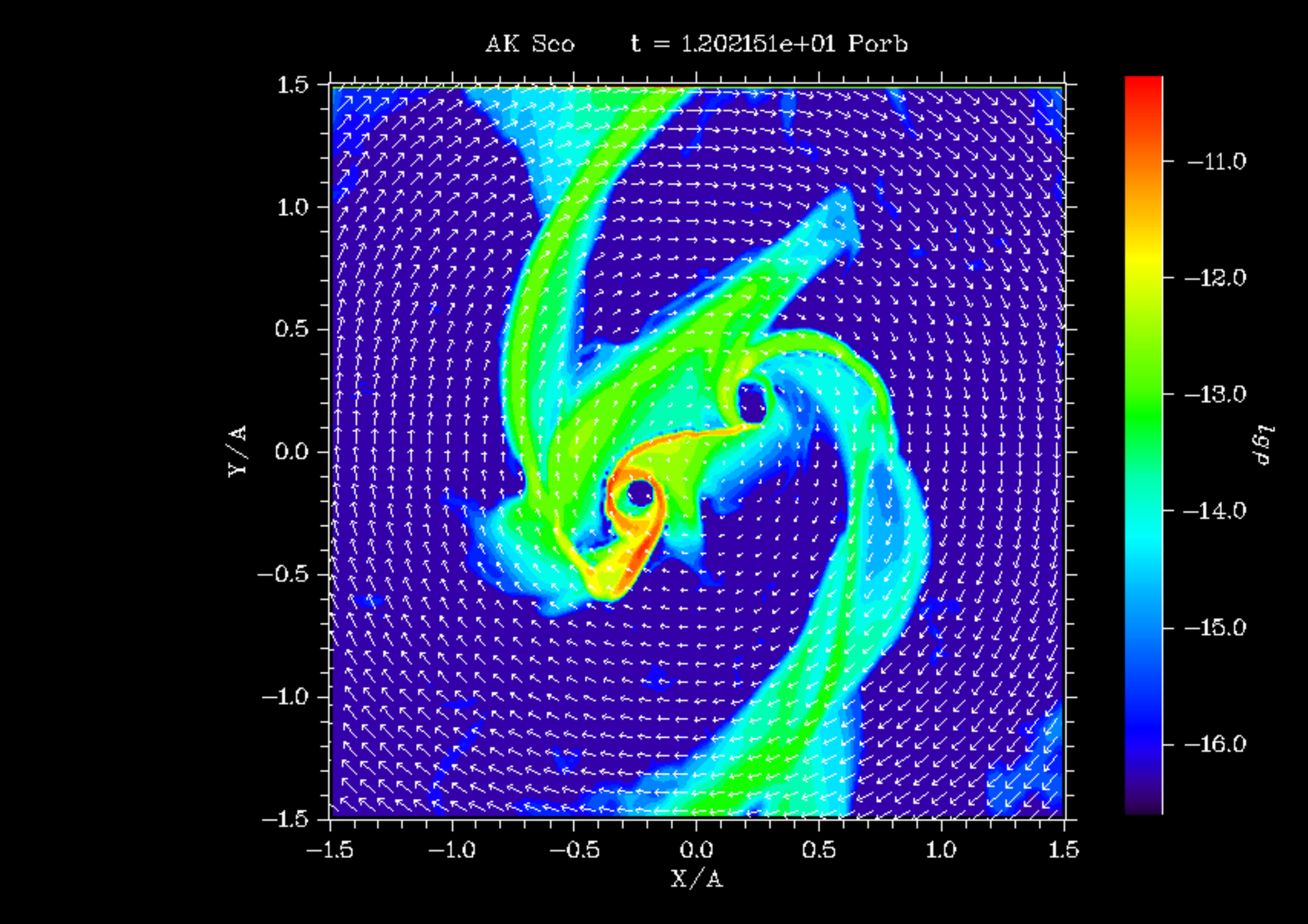} \\
\end{tabular}
\caption{Theoretical predictions of the evolution of the density distribution in the inner gap during the periastron passage
for any two cycle; cycle 1 top panels, cycle 2 bottom panels. Note that the mass infall is channelled differently from 
cycle to cycle making the physical configuration not exactly repeatable. Left panels correspond to phase
0.992 and right panels to phase 1.02 (approximately the phase coverage of the HST monitoring).
Each frame represents a physical size of $3\times 3$ times the semimajor axis of the orbit ($0.23 {\rm AU} 
\times 0.23 {\rm AU}$).   In the figure, AK Sco system is displayed from a pole-on view (i.e. $i=0^o$) for better visualization
but the inclination of the system is $65^o-70^o$ (see Table 1).}
\label{fig:simu}
\end{figure}

\newpage

\begin{figure}[h]
\centering
\includegraphics[width=14cm]{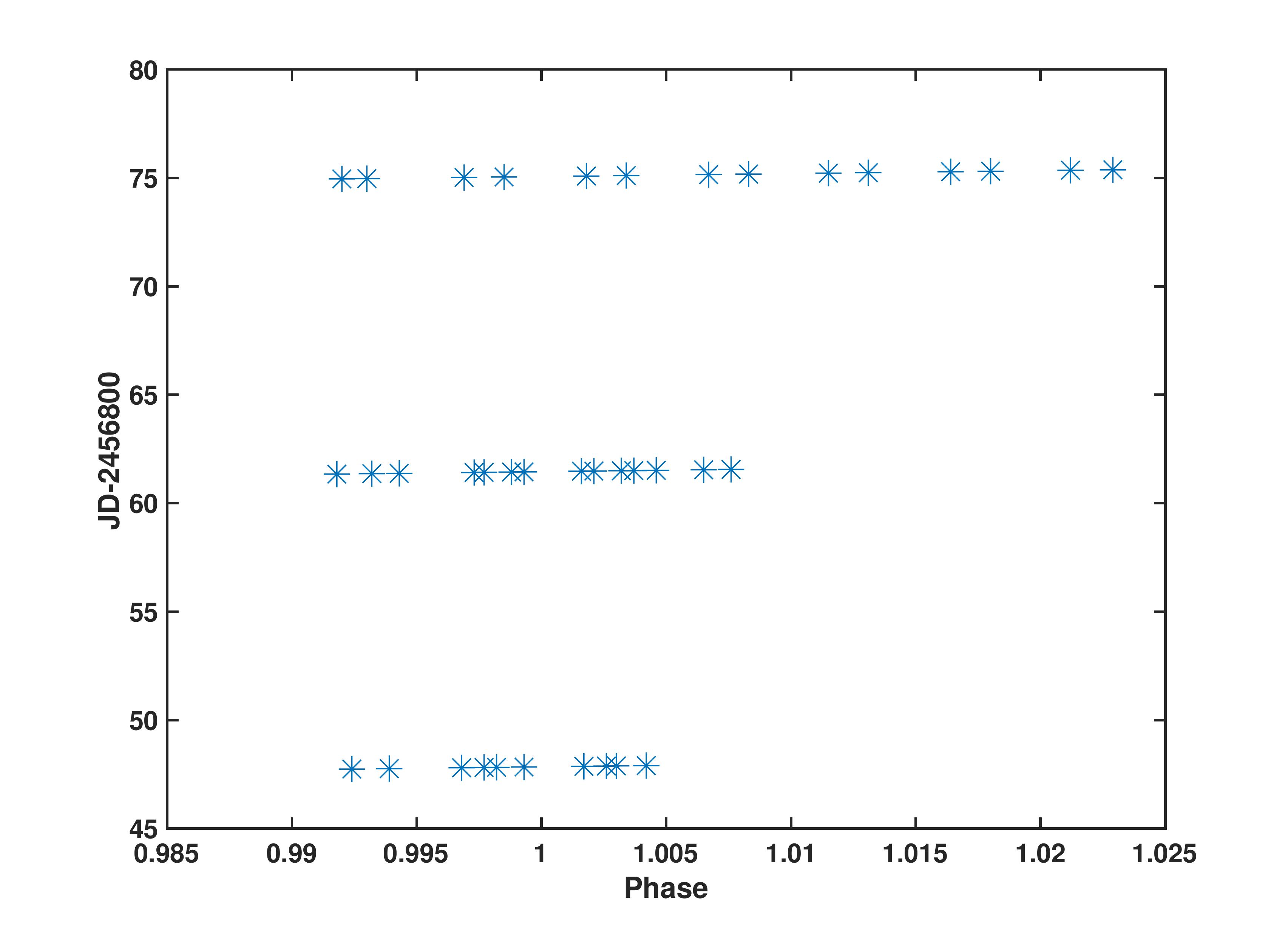}
\caption{The main characteristics of the monitoring program are outlined. AK Sco was tracked during three 
consecutive periastron passages. During the first two passages it was observed with STIS, using grating G140L
(1140~\AA - 1730~\AA ) and G240L (1568~\AA - 3184~\AA ). COS was used for the last passage covering the
1160~\AA - 1453~\AA range and then, the 1402~\AA -1762~\AA range. All observations were carried in photon 
counting mode.  }
\label{fig:obs_program}
\end{figure}

\newpage

\begin{figure}[h]
\centering
\includegraphics[width=15cm]{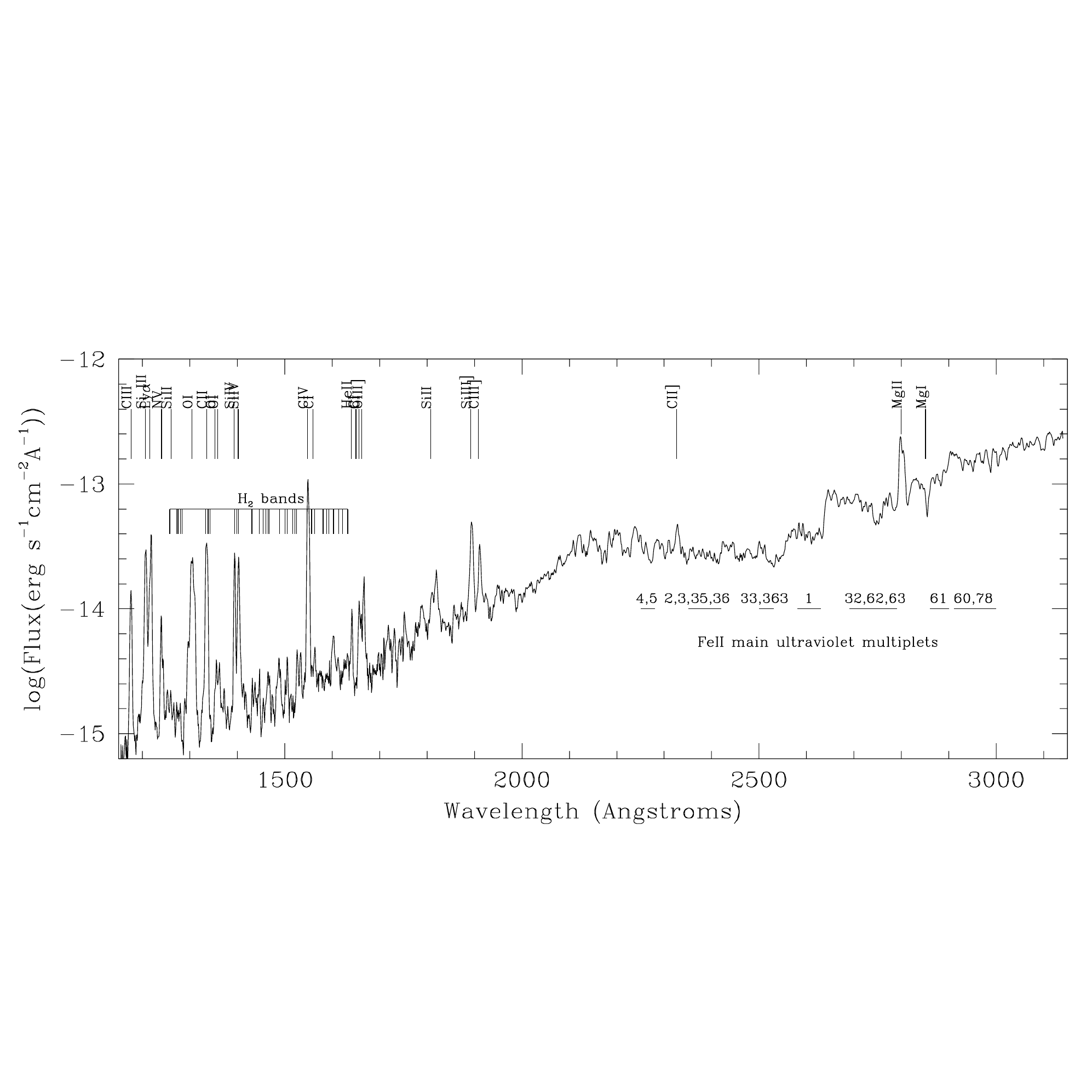}
\caption{Average spectrum of AK~Sco from STIS observations. The main spectral features are marked.}
\label{fig:aksco_SED}
\end{figure}

\newpage

\begin{figure}[h]
\centering
\includegraphics[width=15cm]{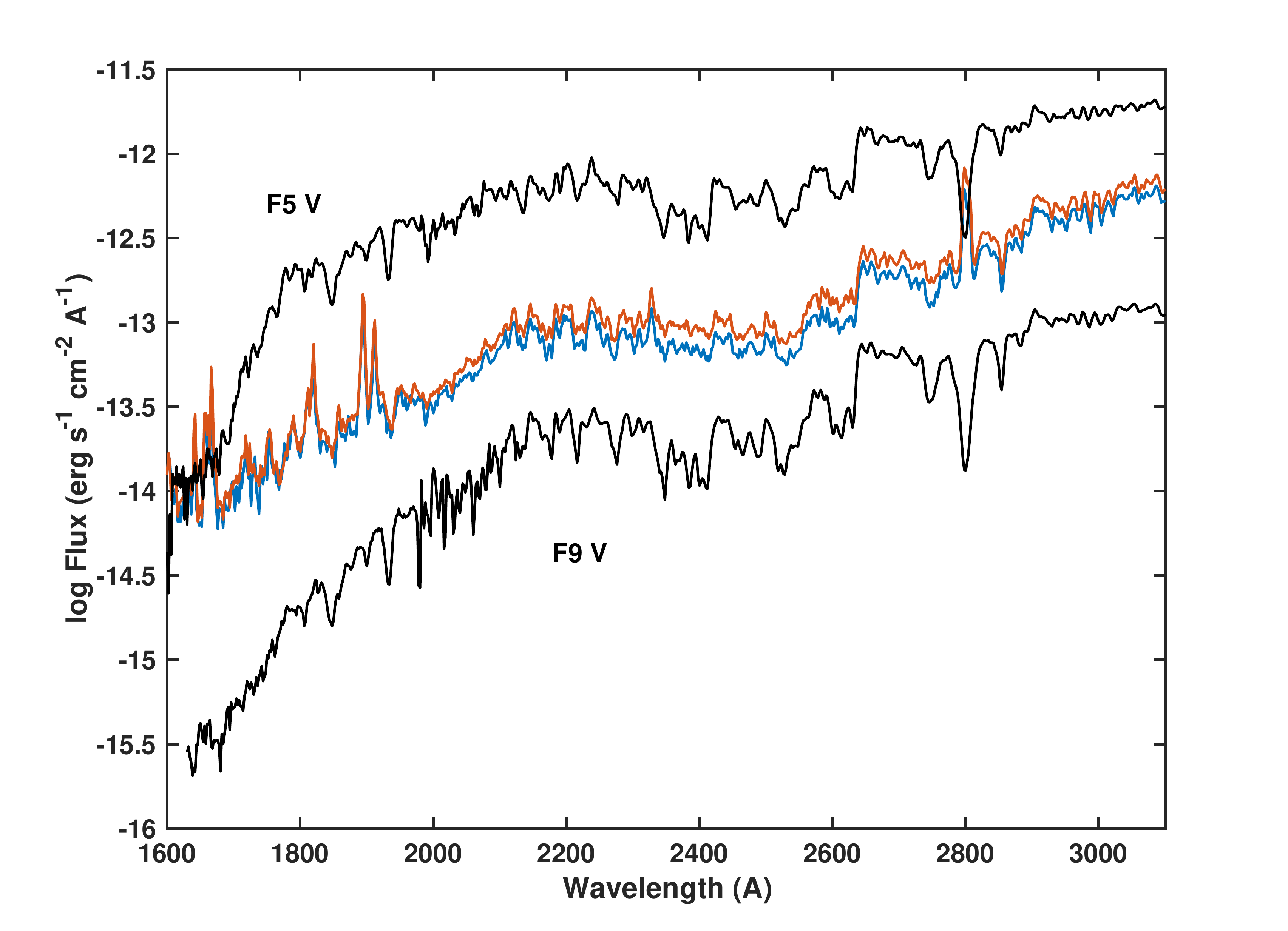}
\caption{Average spectrum of AK~Sco during cycle 1 (blue) and cycle 2 (red), STIS observations. The UV spectrum of the F5~V star, HD~139664
and the F9~V star, HD~22879 are plotted (grey) for comparison; they have been offset by -1.2 dex and -2.5 dex respectively, for the plot. }
\label{fig:avespec}
\end{figure}

\newpage

\begin{figure}[h]
\centering
\includegraphics[width=15cm]{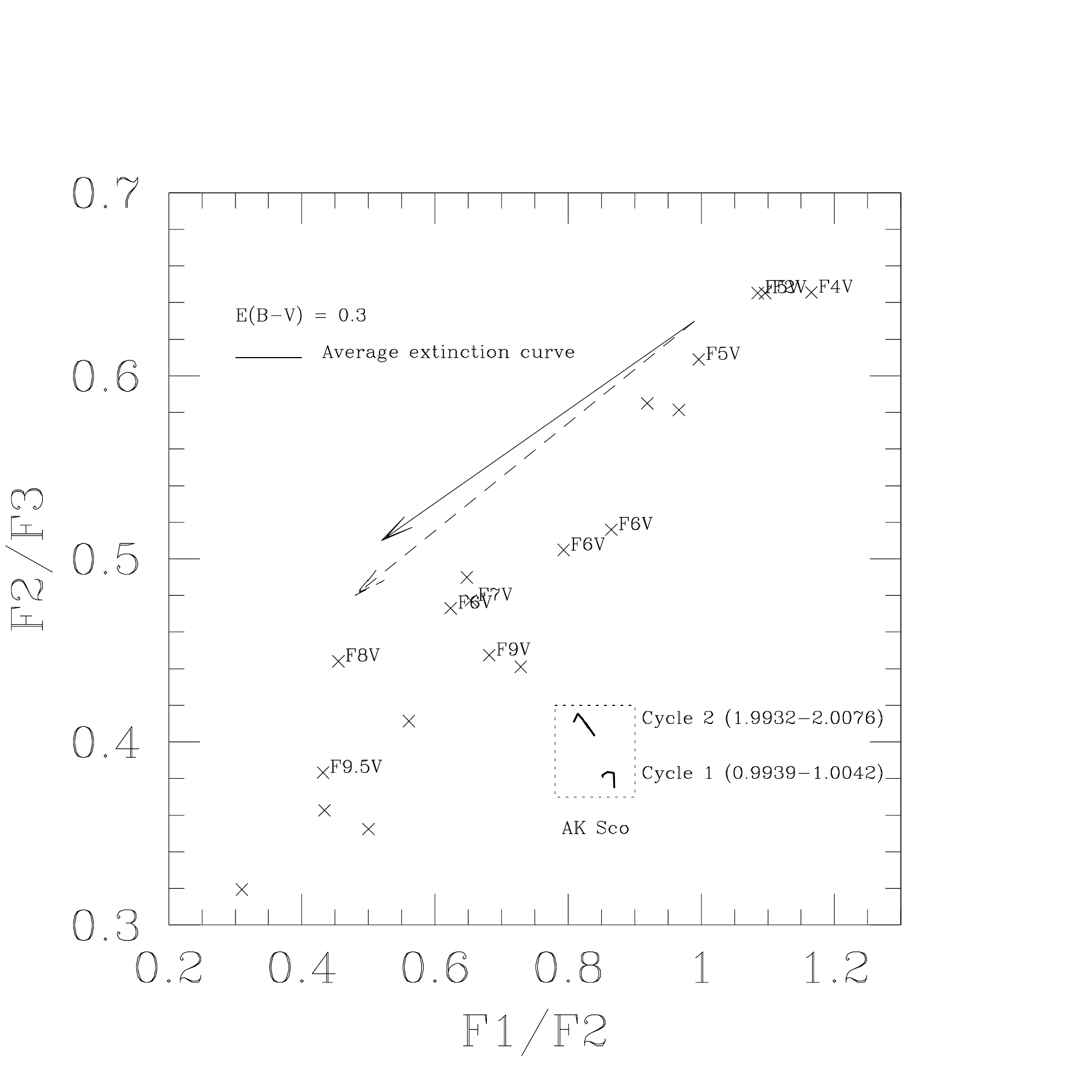}
\caption{Variation of the location of AK~Sco in the $R1 = F1/F2$ and $R2 = F2/F3$ space during the monitorings run with HST/STIS. The location of main sequence F stars is  indicated as
well as the extinction arrow (solid)  for  the average ISM extinction law (Fiztpatrick \& Massa, 2007) and the modified extinction law ($R=4.3$)  found by Manset et al. (2004) to be representative of the circumstellar environment in AK~Sco (dashed arrow). Notice that extinction  runs roughly parallel to the spectral types. AK Sco spectrum has not been dereddened
for the plot but its location close to late spectral types (F7V-F9V) instead of F5 cannot be fully ascribed to extinction; AK Sco extinction $A_V = 0.5$ or $E(B-V) = 0.12$
(Manset et al. 2005).}
\label{fig:aksco_NUVrates}
\end{figure}

\newpage

\begin{figure}[h]
\centering
\begin{tabular}{cc}
\includegraphics[width=8cm]{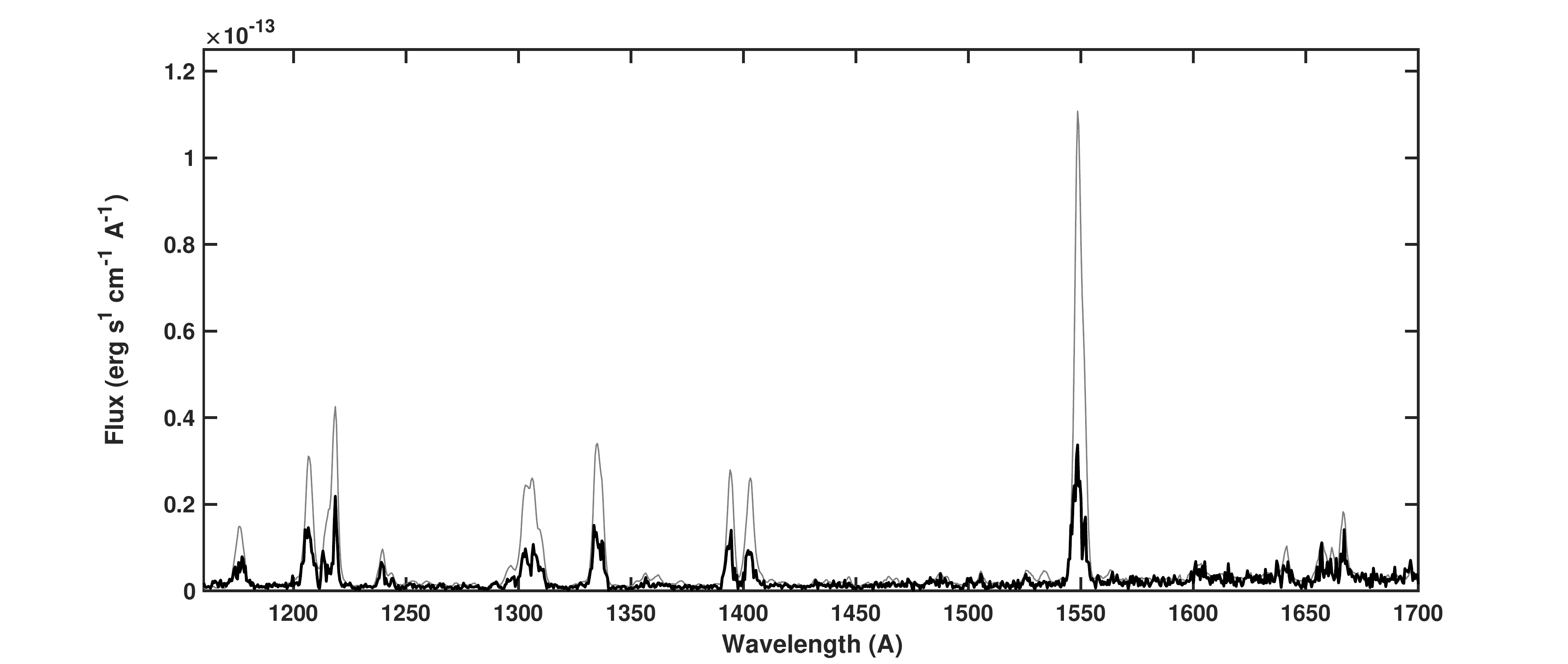} & \includegraphics[width=8cm]{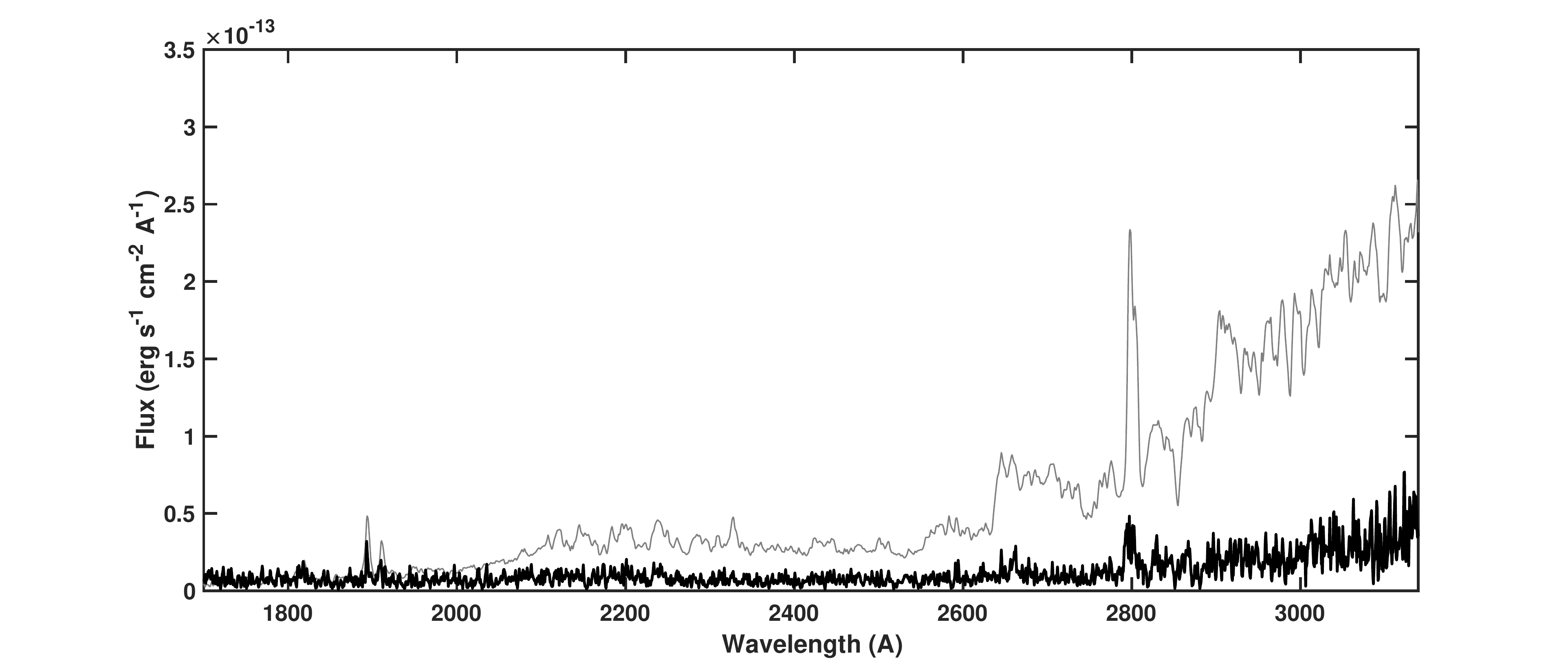} \\
\includegraphics[width=8cm]{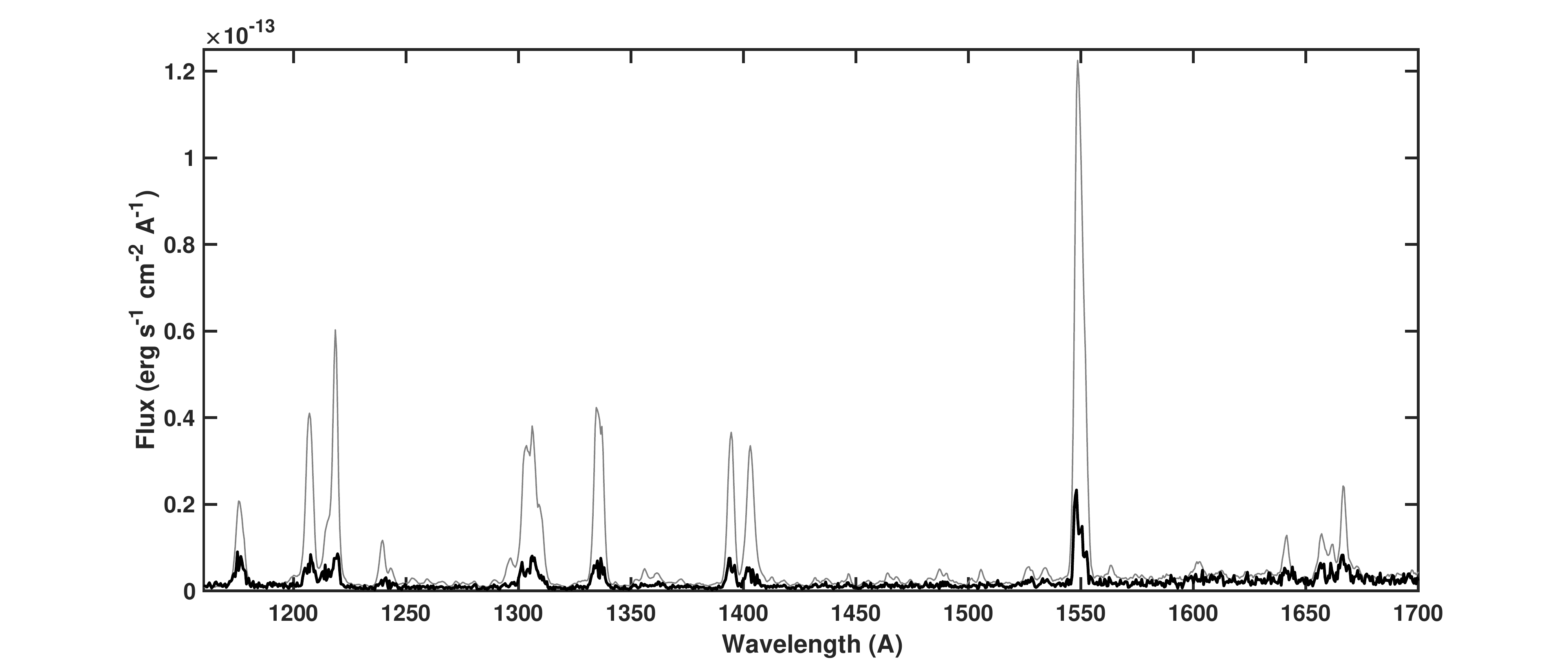} & \includegraphics[width=8cm]{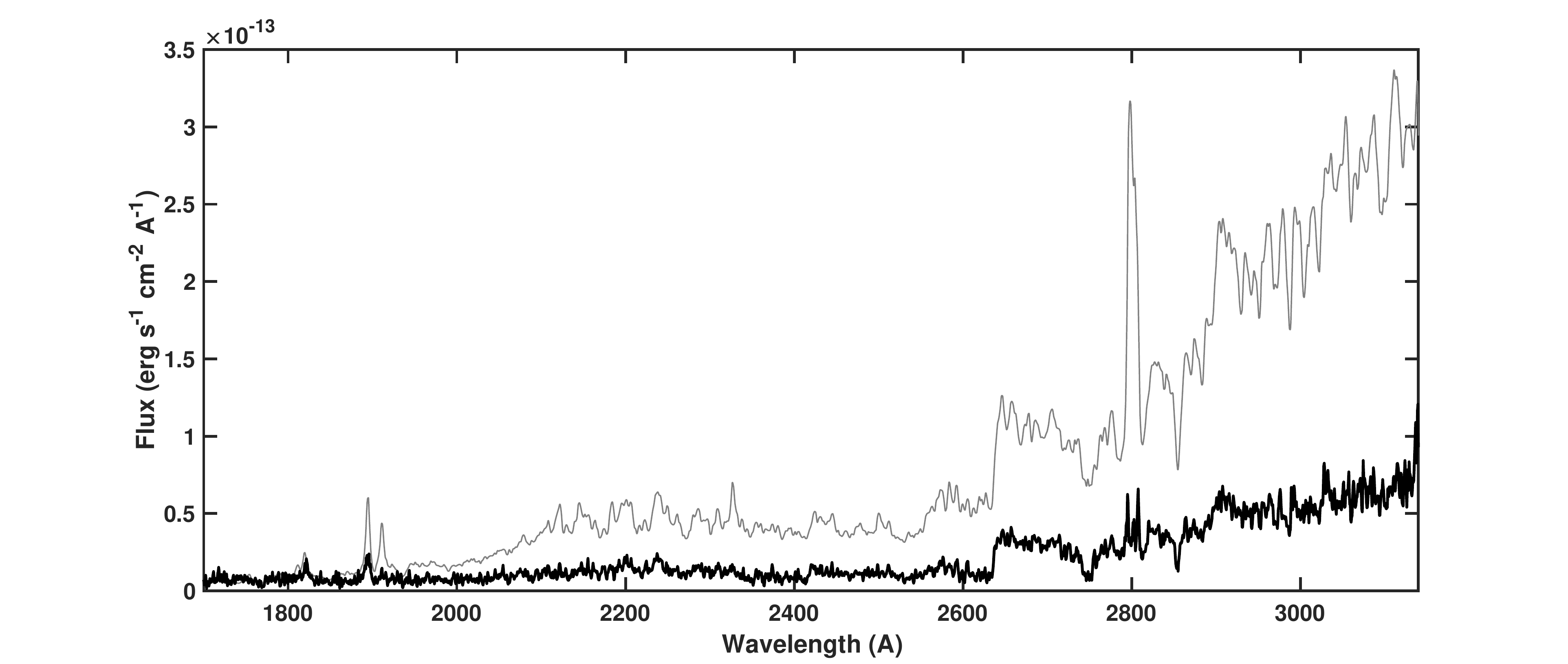} \\
\end{tabular}
\caption{Spectral variability during cycles 1 (top) and 2 (bottom). The average spectra is displayed in grey and the dispersion around the mean in thick black;
the dispersion has been multiplied by 10 for the plot (in all the panels). }
\label{fig:Ldisp_variability}
\end{figure}

\newpage

\begin{figure}[h]
\centering
\begin{tabular}{cc}
\includegraphics[width=10cm]{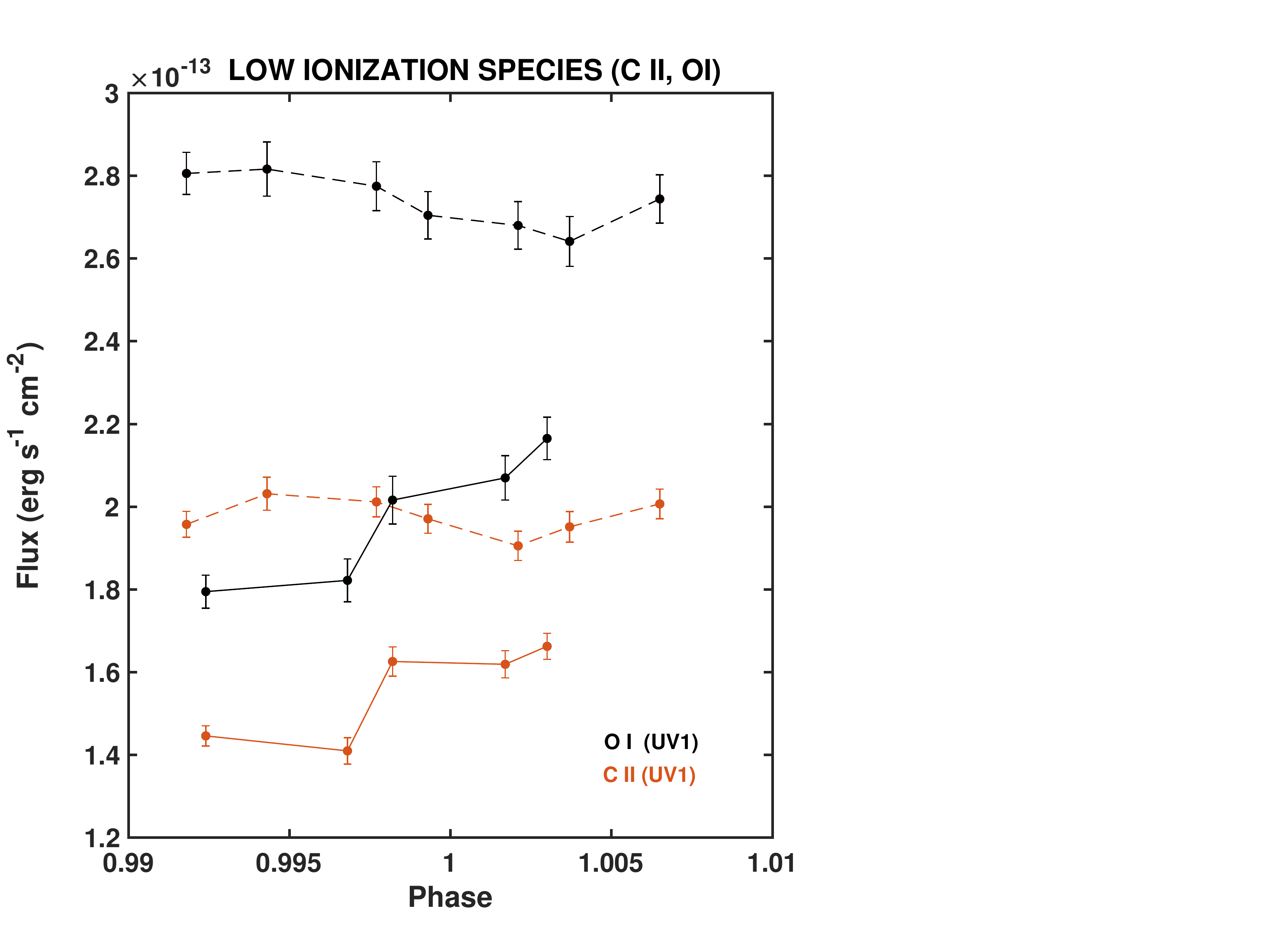} & \includegraphics[width=10cm]{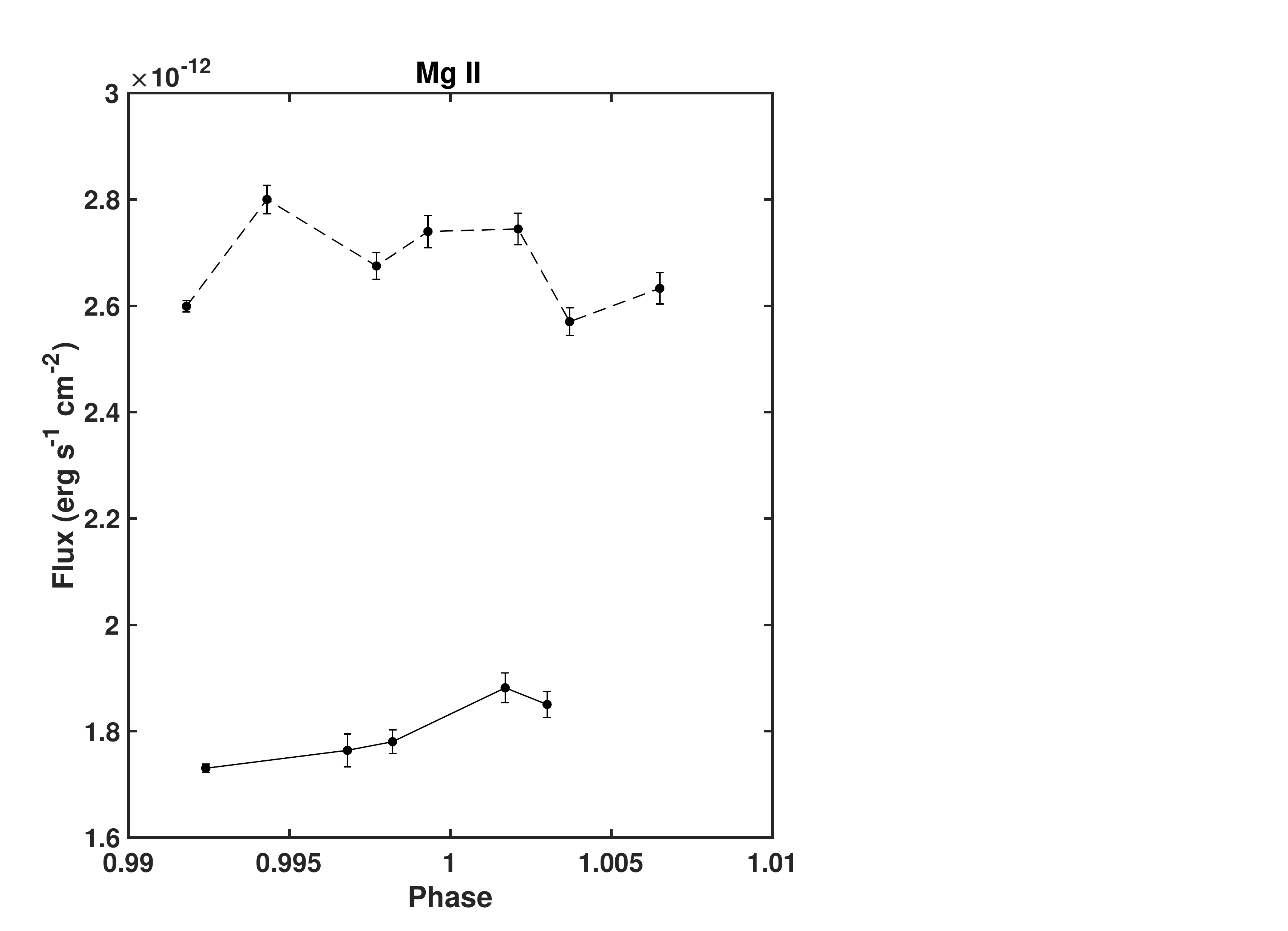} \\
\includegraphics[width=10cm]{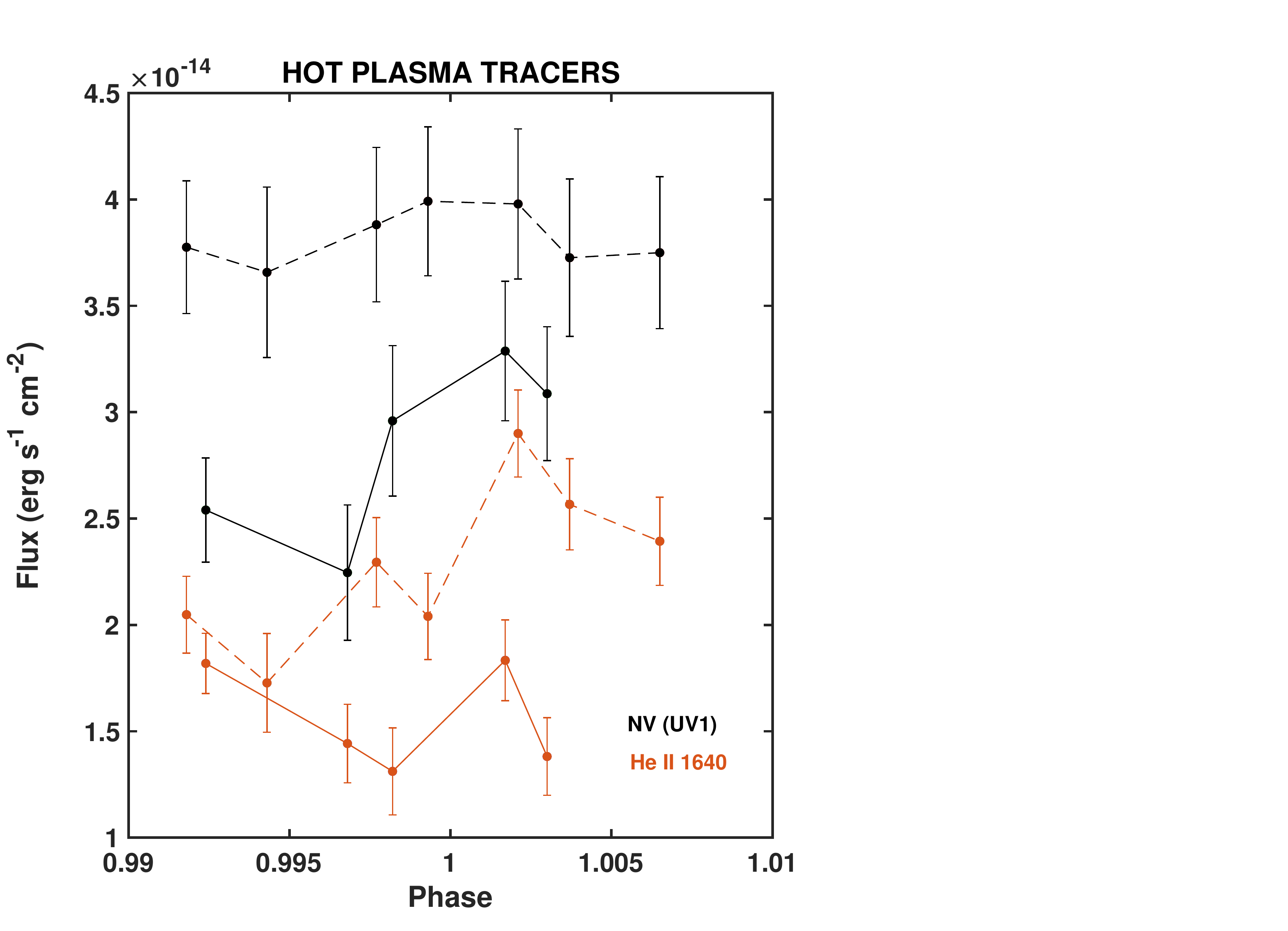} & \includegraphics[width=10cm]{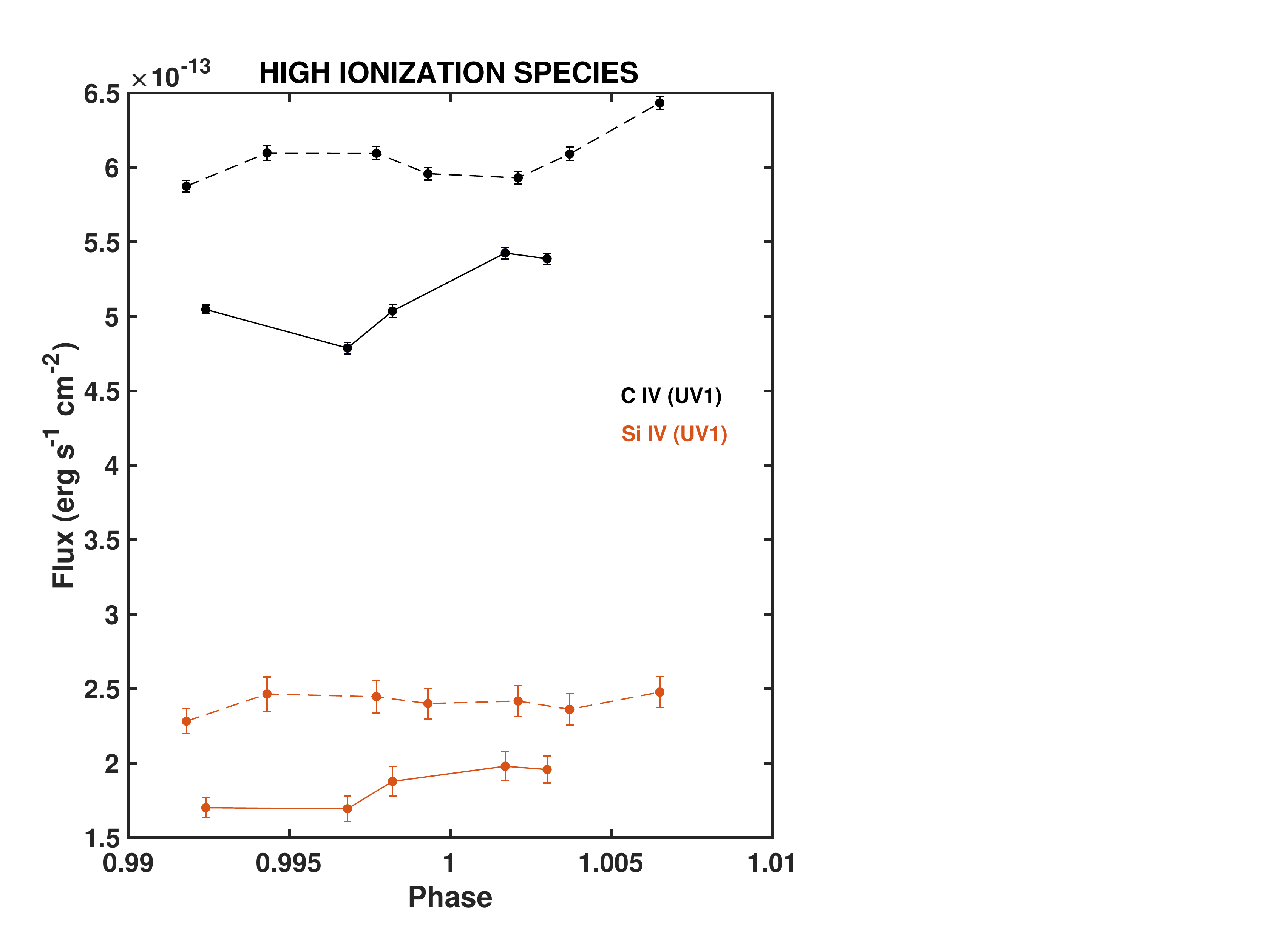} \\
\includegraphics[width=10cm]{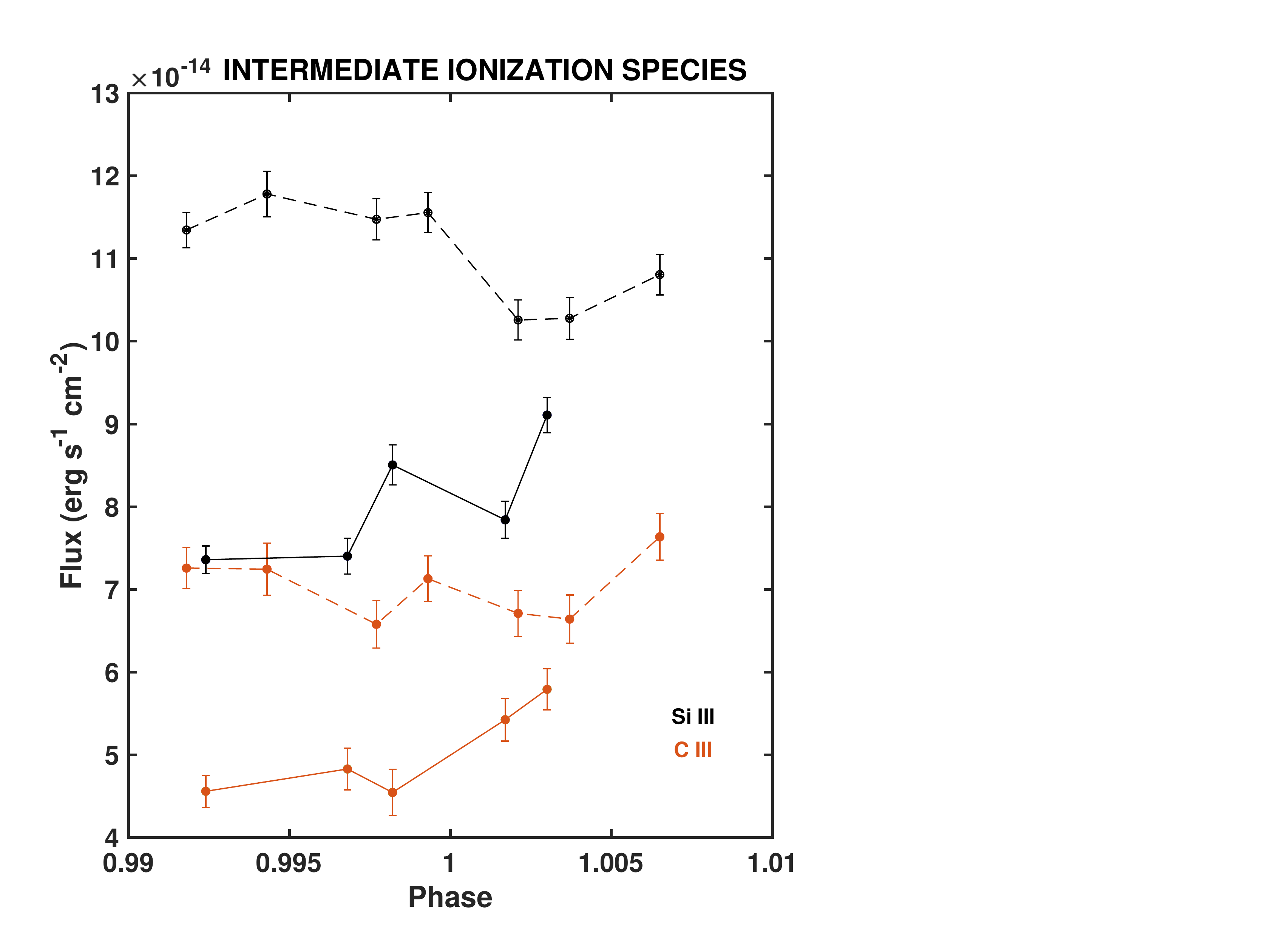} & \includegraphics[width=10cm]{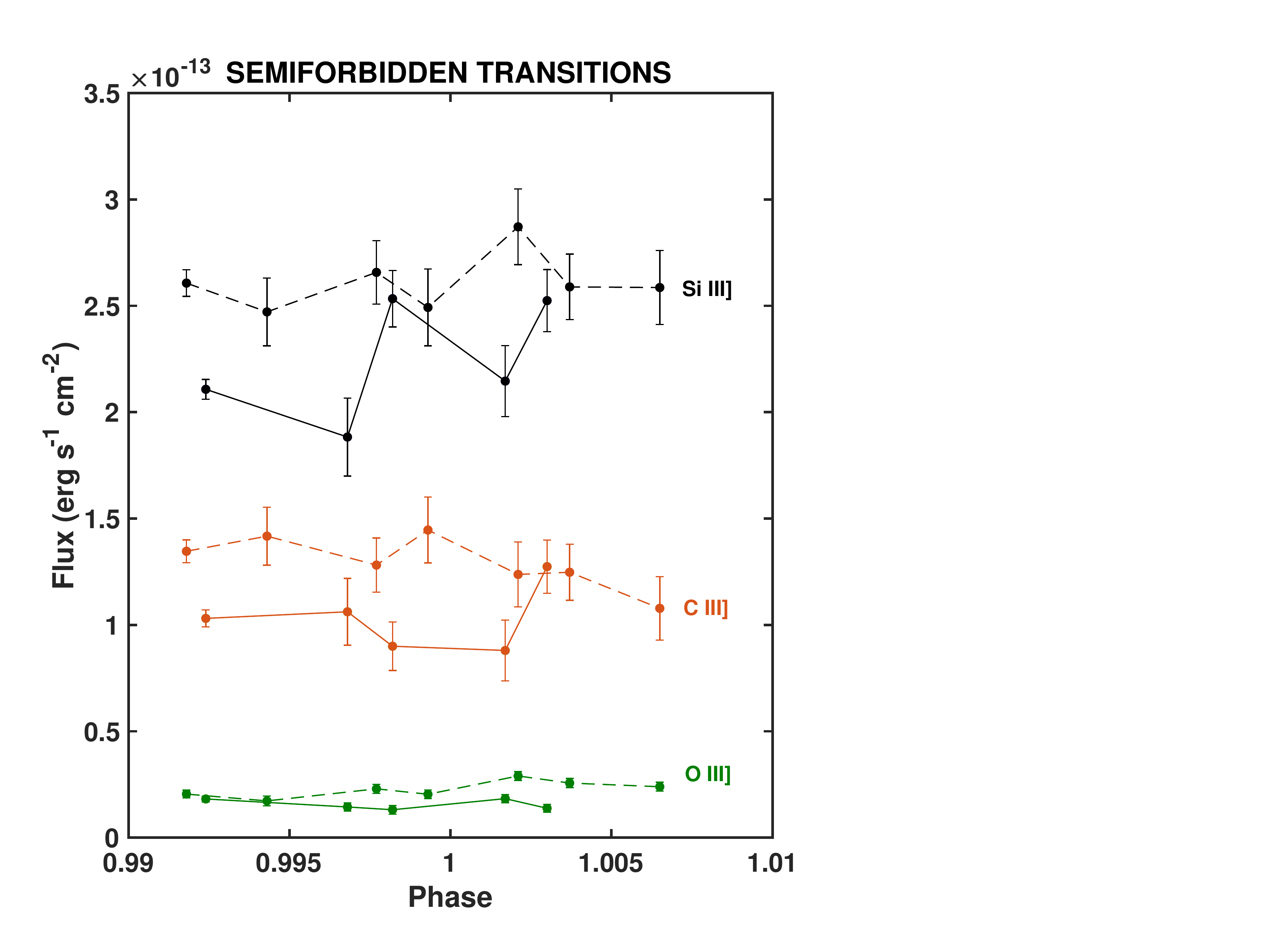} \\
\end{tabular}
\caption{Light curves of the main spectral lines during cycles 1 (solid) and 2 (dashed); 1-$\sigma$ error bars are indicated.  }
\label{fig:Ldisp_variability}
\end{figure}

\newpage

\begin{figure}[h]
\centering
\begin{tabular}{c}
\includegraphics[width=8cm]{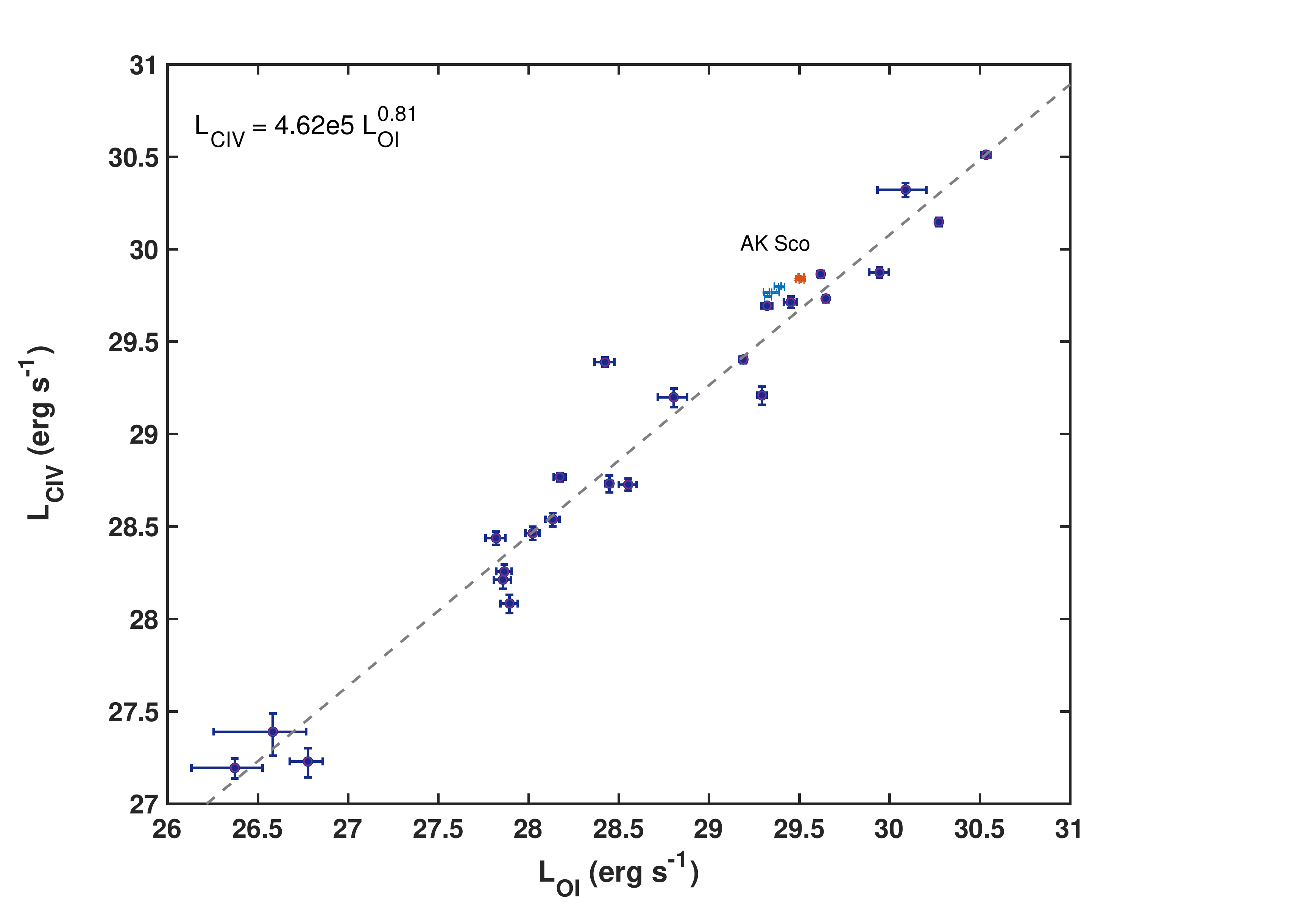} \\
\includegraphics[width=8cm]{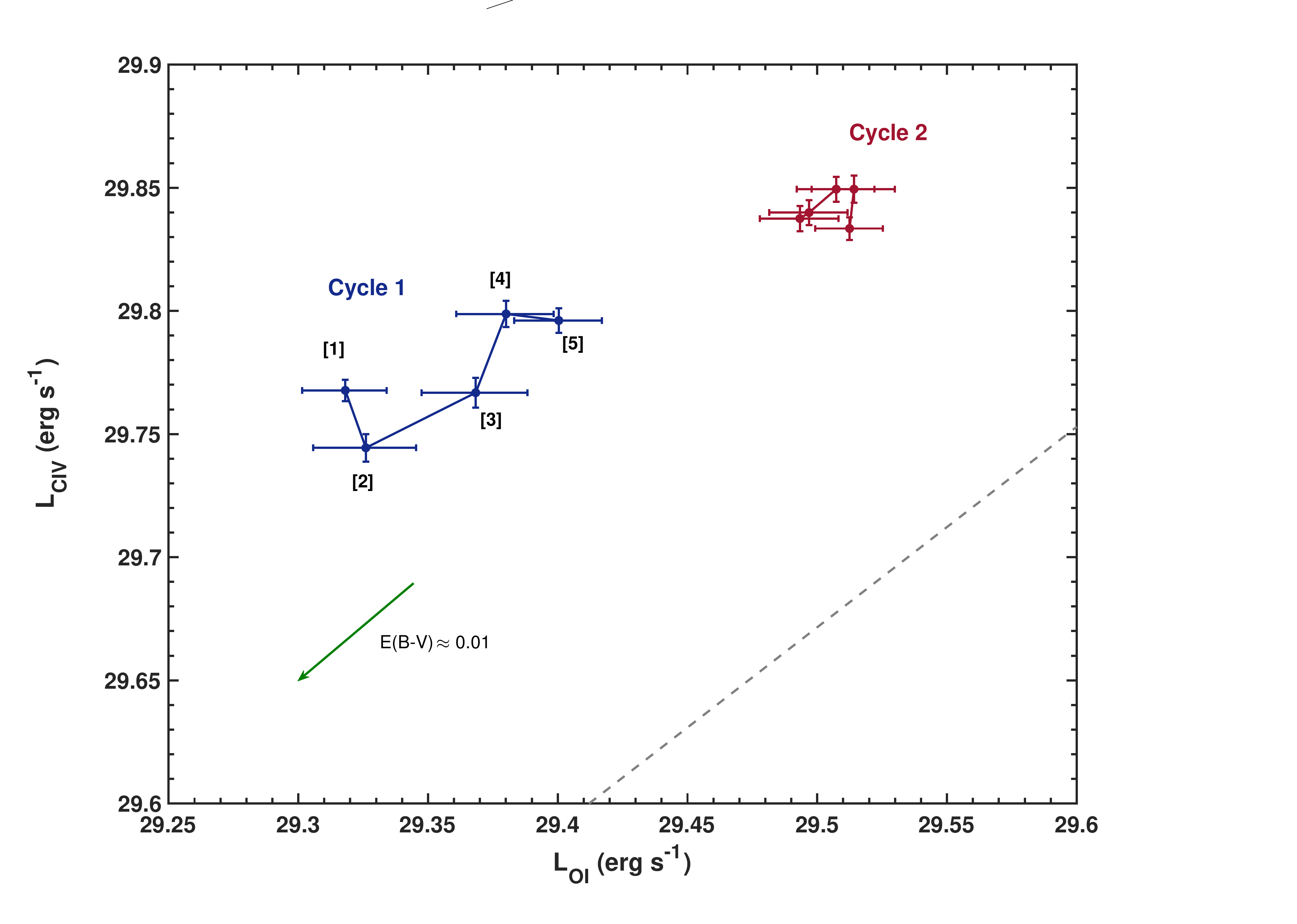} \\
\end{tabular}
\caption{C IV - OI flux-flux diagram. Top, diagram for all the T Tauri stars observed with Hubble with the 
same configuration (STIS/G140L). AK Sco observations are plotted in blue (cycle 1) and red (cycle 2).
The regression line is indicated. Bottom, zoom on the location of AK Sco observations. The order of the
observations in cycle 1 is marked as [1], [2].... . 1-$\sigma$ error bars are plotted.}
\label{fig:Ldisp_variability}
\end{figure}

\newpage

\begin{figure}[h]
\centering
\begin{tabular}{c}
\includegraphics[width=8cm]{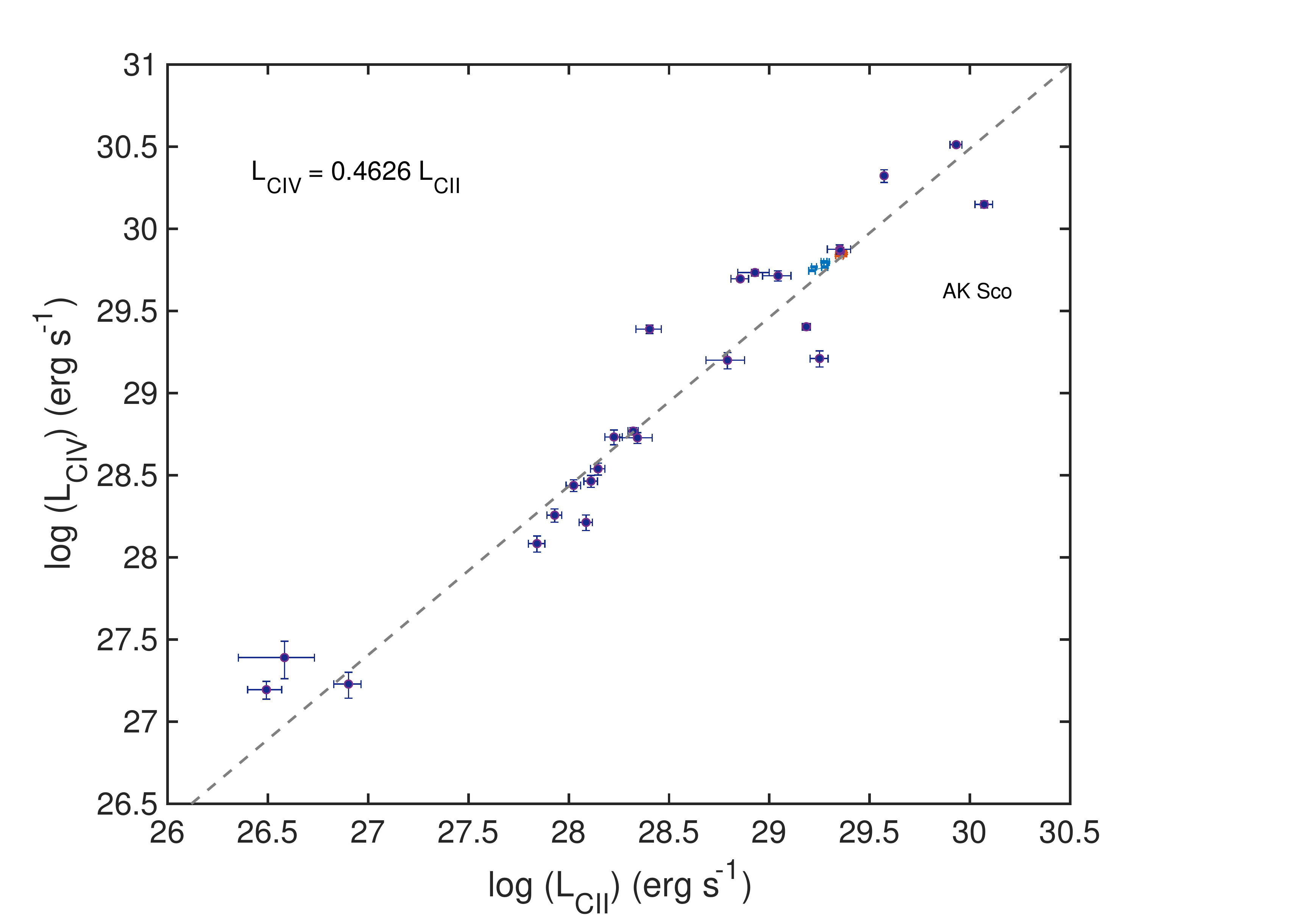} \\
\includegraphics[width=8cm]{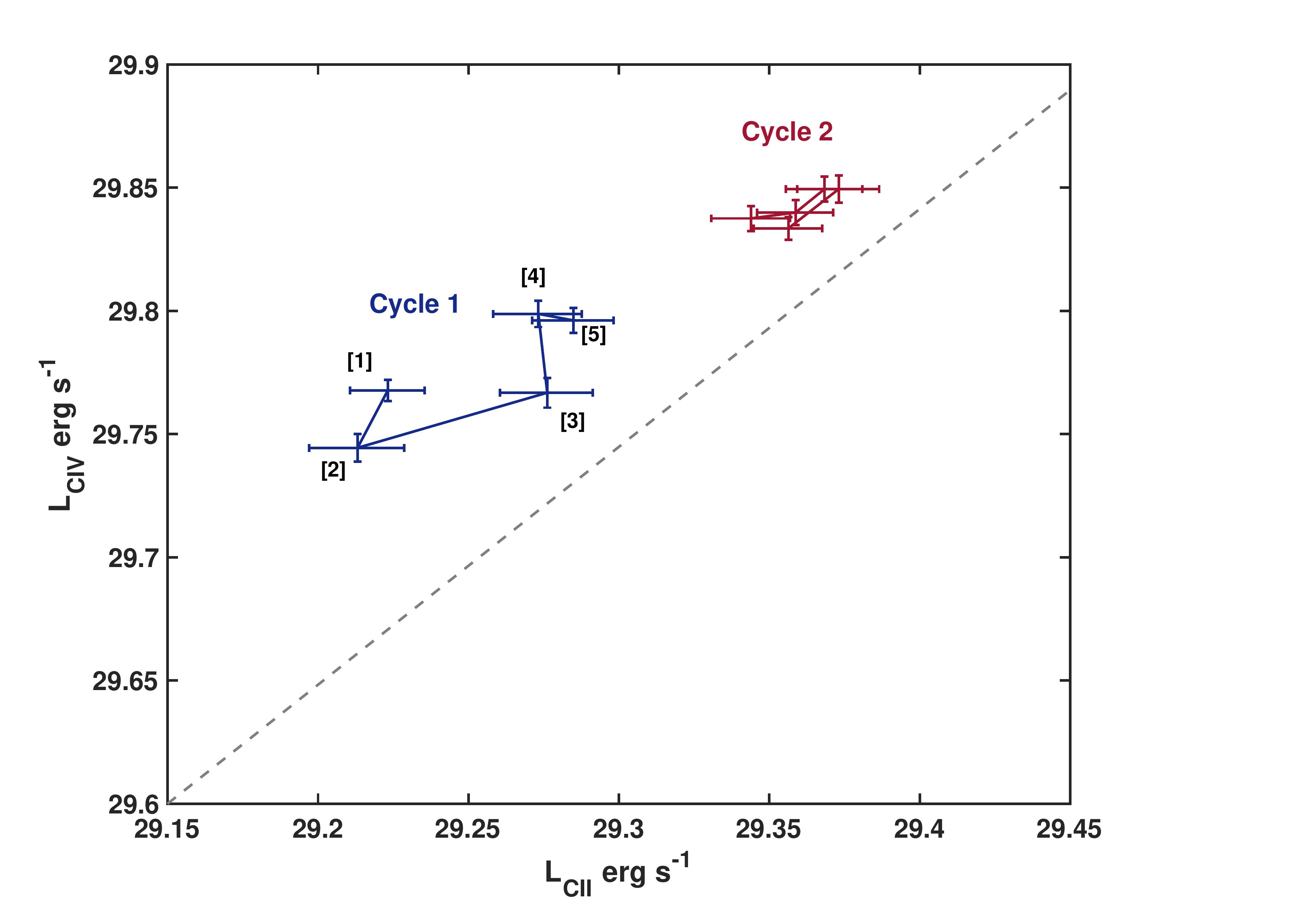} \\
\end{tabular}
\caption{C IV - C II flux-flux diagram. Top, diagram for all the T Tauri stars observed with Hubble with the 
same configuration (STIS/G140L). AK Sco observations are plotted in blue (cycle 1) and red (cycle 2).
The regression line is indicated. Bottom, zoom on the location of AK Sco observations. The order of the
observations in cycle 1 is marked as [1], [2].... 1-$\sigma$ error bars are plotted.}
\label{fig:Ldisp_variability}
\end{figure}

\newpage

\begin{figure}[h]
\centering
\begin{tabular}{c}
\includegraphics[width=8cm]{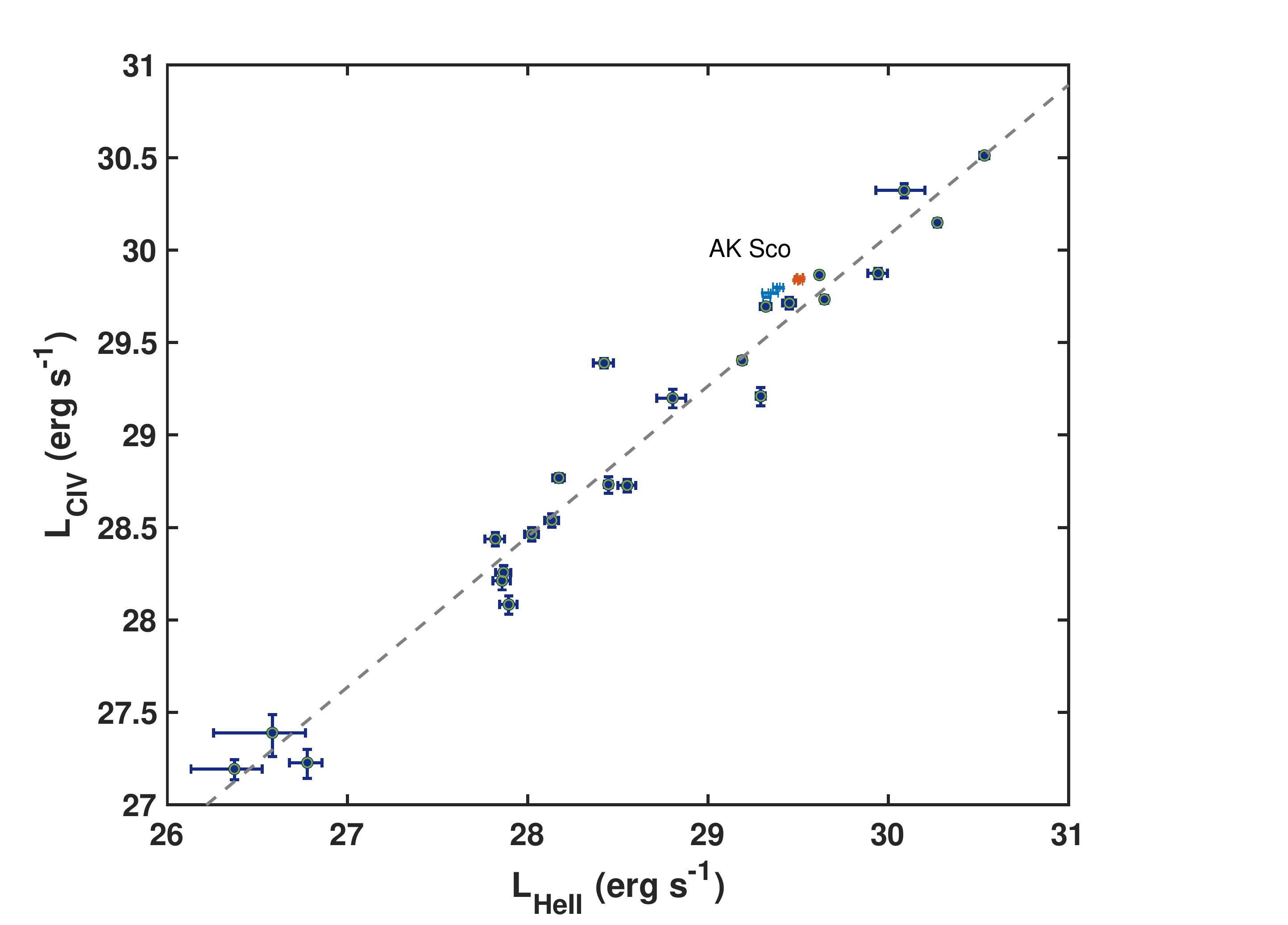} \\
\includegraphics[width=8cm]{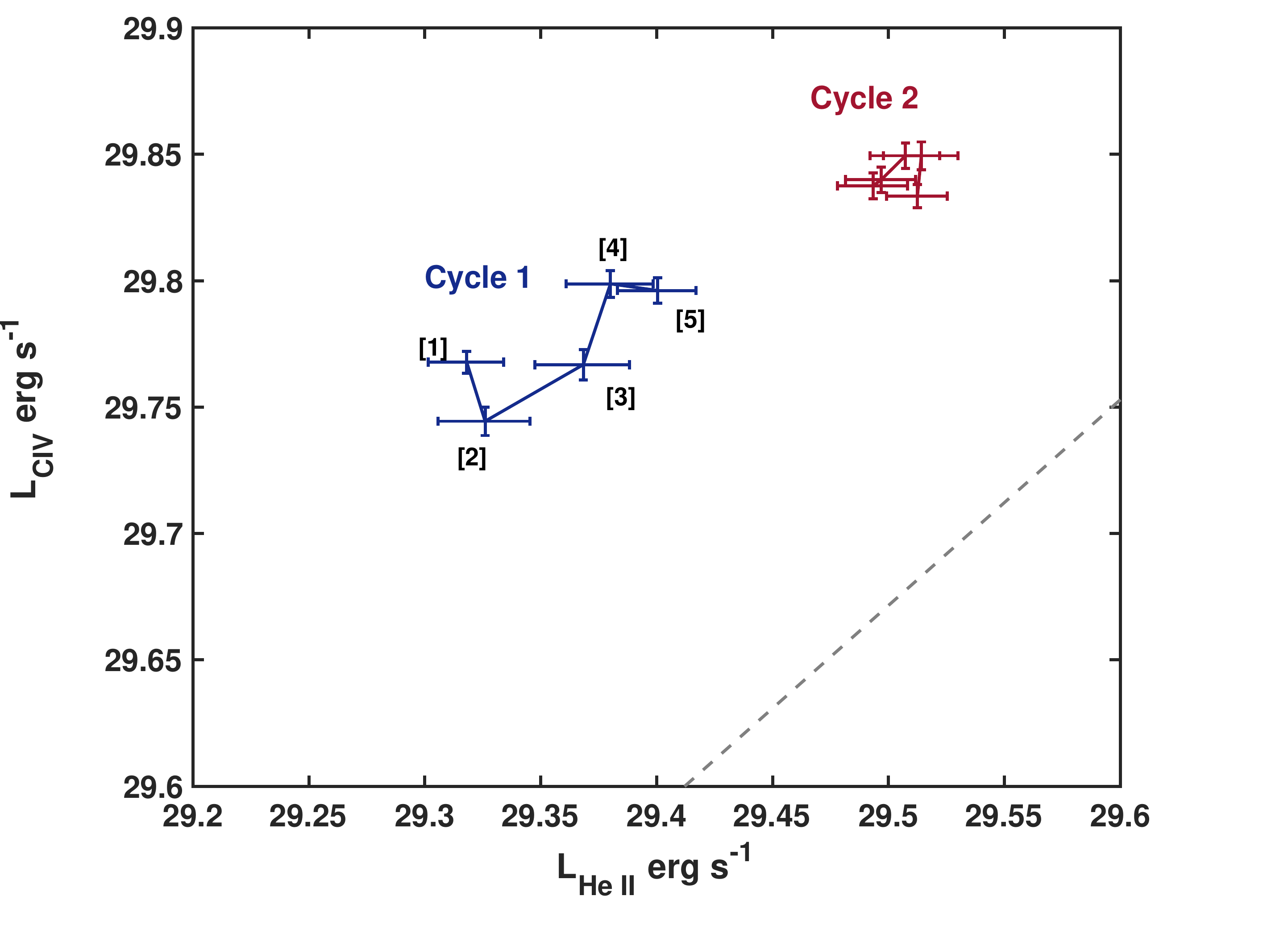} \\
\end{tabular}
\caption{C IV - He II flux-flux diagram. Top, diagram for all the T Tauri stars observed with Hubble with the 
same configuration (STIS/G140L). AK Sco observations are plotted in blue (cycle 1) and red (cycle 2).
The regression line is indicated. Bottom, zoom on the location of AK Sco observations. The order of the
observations in cycle 1 is marked as [1], [2].... 1-$\sigma$ error bars are plotted.}
\label{fig:Ldisp_variability}
\end{figure}

\newpage

\begin{figure}[h]
\centering
\begin{tabular}{c}
\includegraphics[width=8cm]{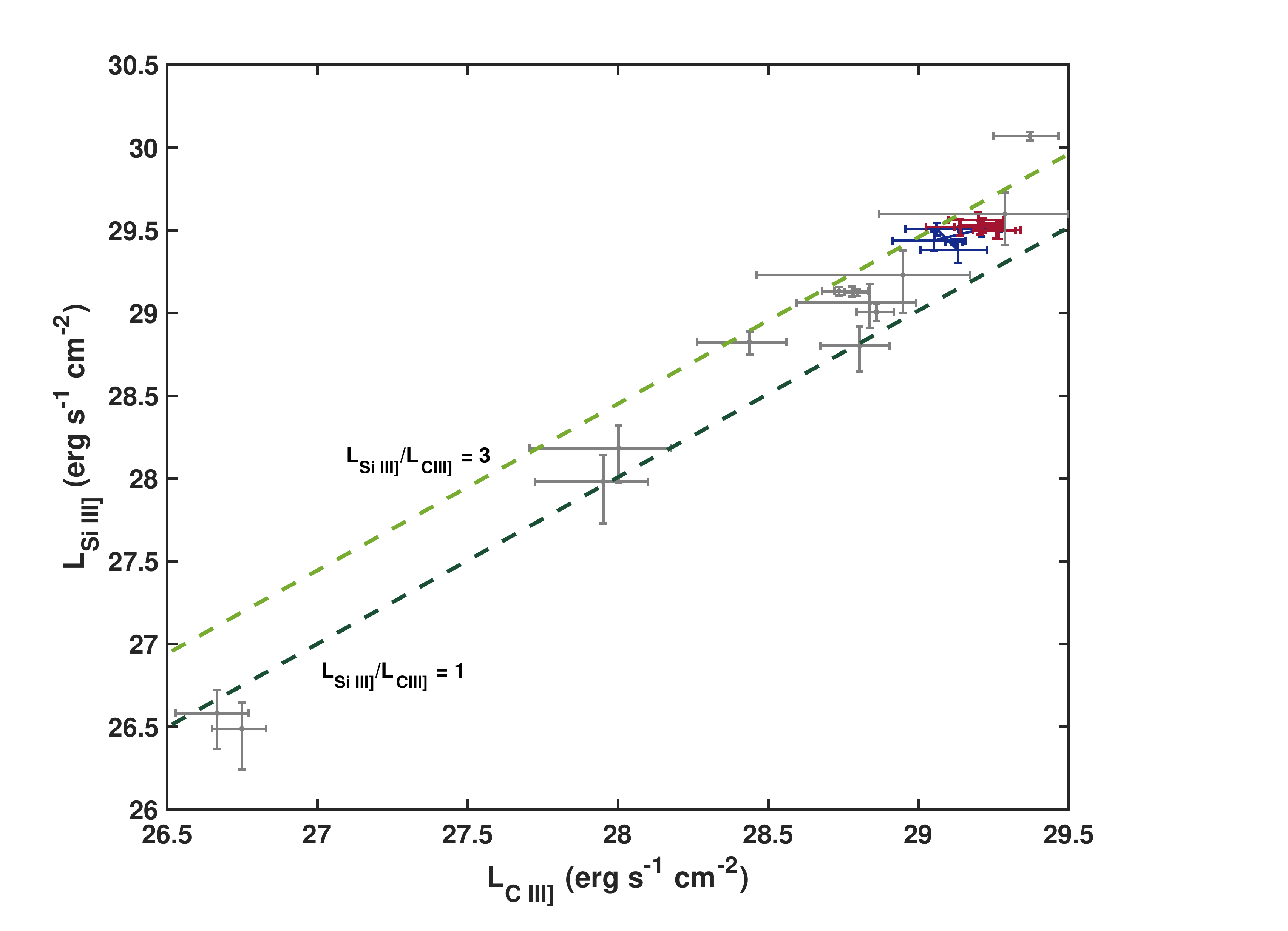} \\
\includegraphics[width=8cm]{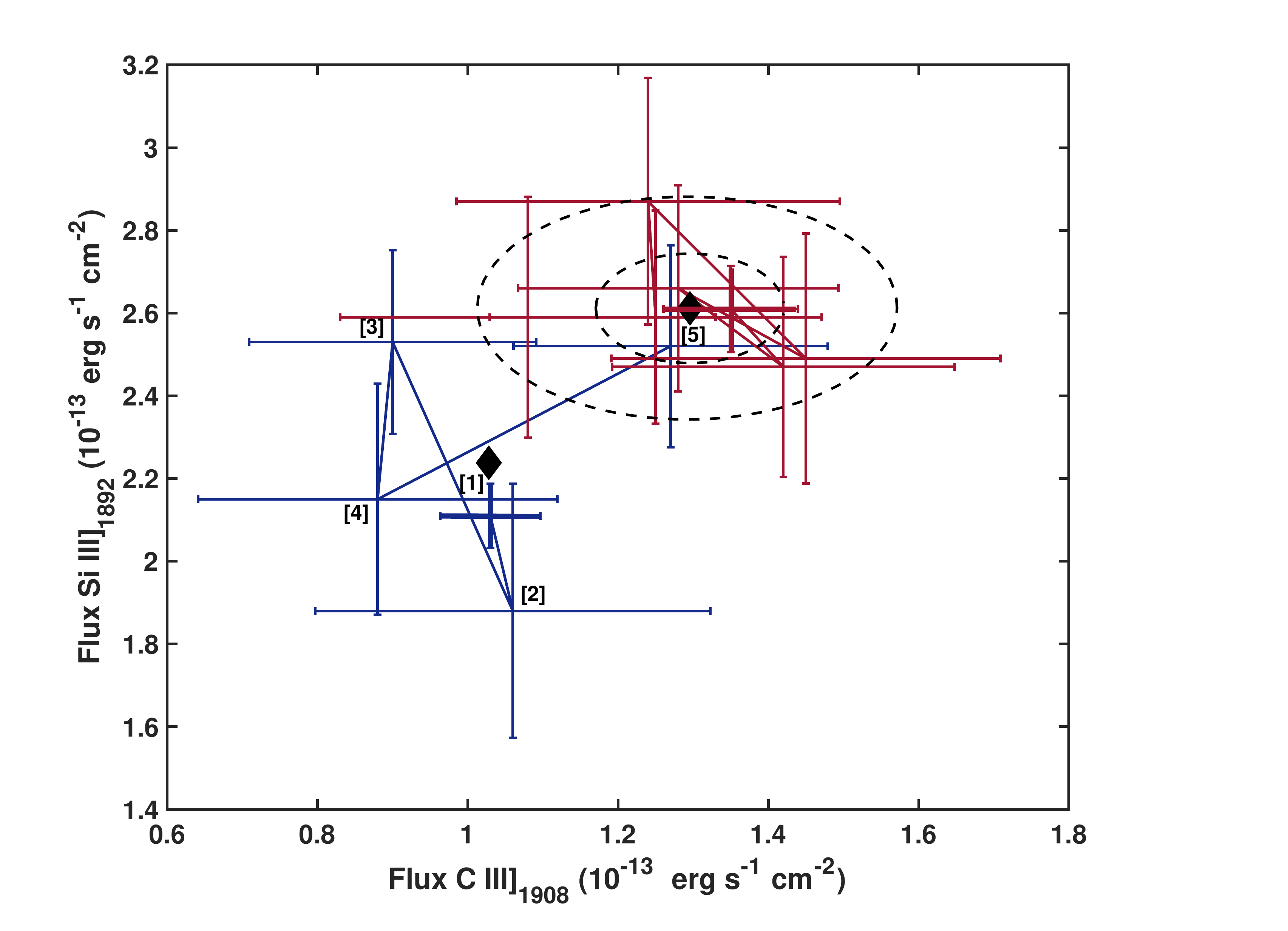} \\
\end{tabular}
\caption{Si III] - C III] flux-flux diagram. Top, diagram for all the T Tauri stars observed with Hubble with the 
same configuration (STIS/G140L). AK Sco observations are plotted in blue (cycle 1) and red (cycle 2).
The lines corresponding to Si III]/C~III]=1 and 3 are marked with dashed lines. This ratio is sensitive to the electron
temperature; for $T_e \simeq 5\times 10^4$~K this range corresponds to a variation in electron density from
10$^{10}$~cm$^{-3}$ to  10$^{11}$~cm$^{-3}$. Bottom, zoom on the location of AK Sco observations. The black
diamonds mark the location of the average values during cycle 1 and 2. The 1-$\sigma$ and 2-$\sigma$ confidence 
ellipse for cycle 2 observations is marked with dashed lines. The order of the
observations in cycle 1 is marked as [1], [2].... 1-$\sigma$ error bars are indicated for all observations.
Note the significantly better SNR of the first observation in each cycle
(bold error bars). }
\label{fig:Ldisp_variability}
\end{figure}

\newpage

\begin{figure*}[h]
\centering
\begin{tabular}{cc}
\includegraphics[width=8cm]{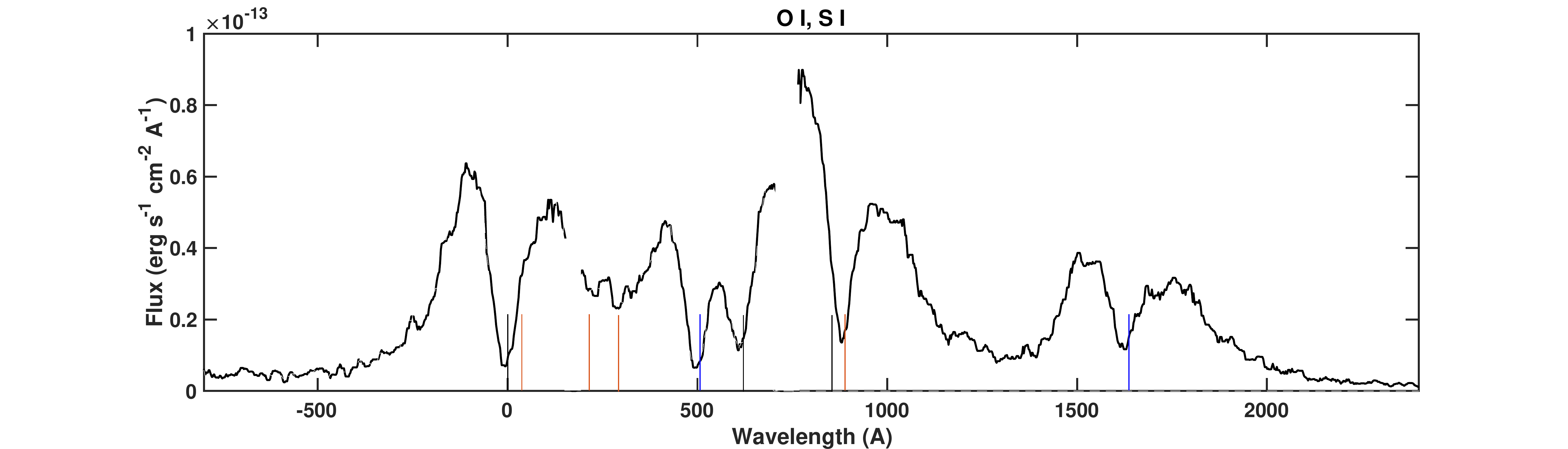} & \includegraphics[width=8cm]{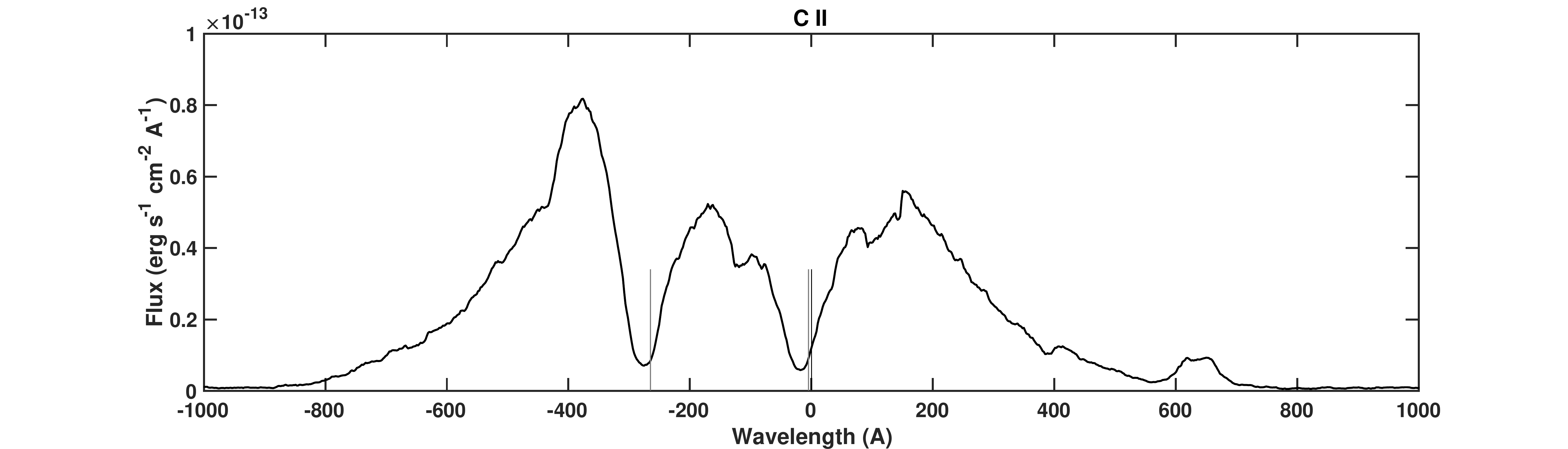} \\
\includegraphics[width=8cm]{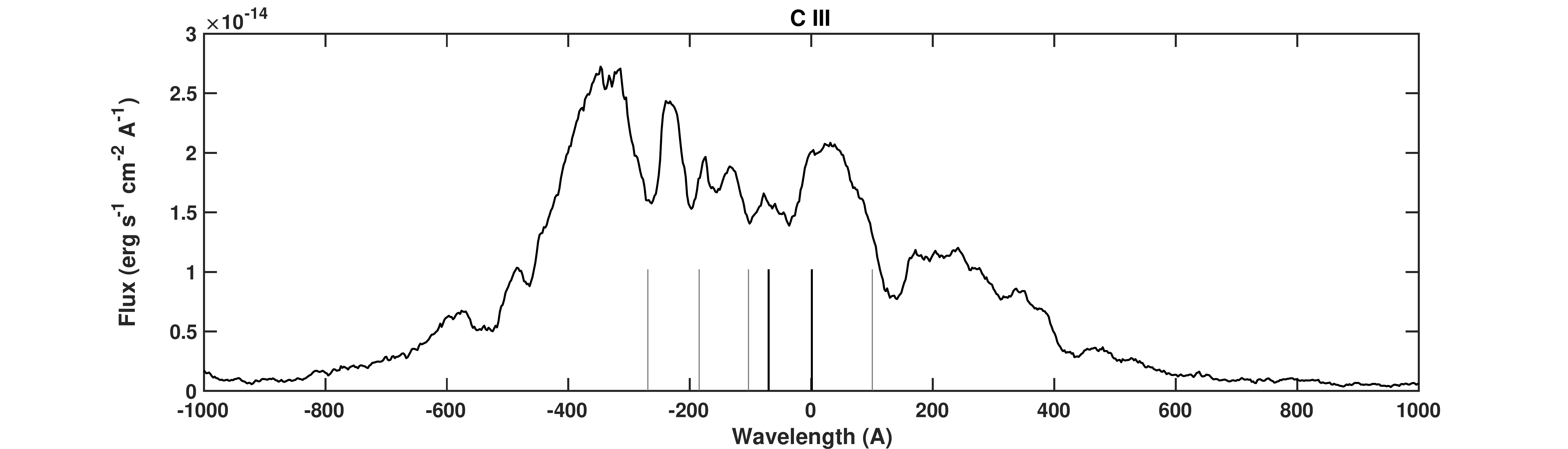} & \includegraphics[width=8cm]{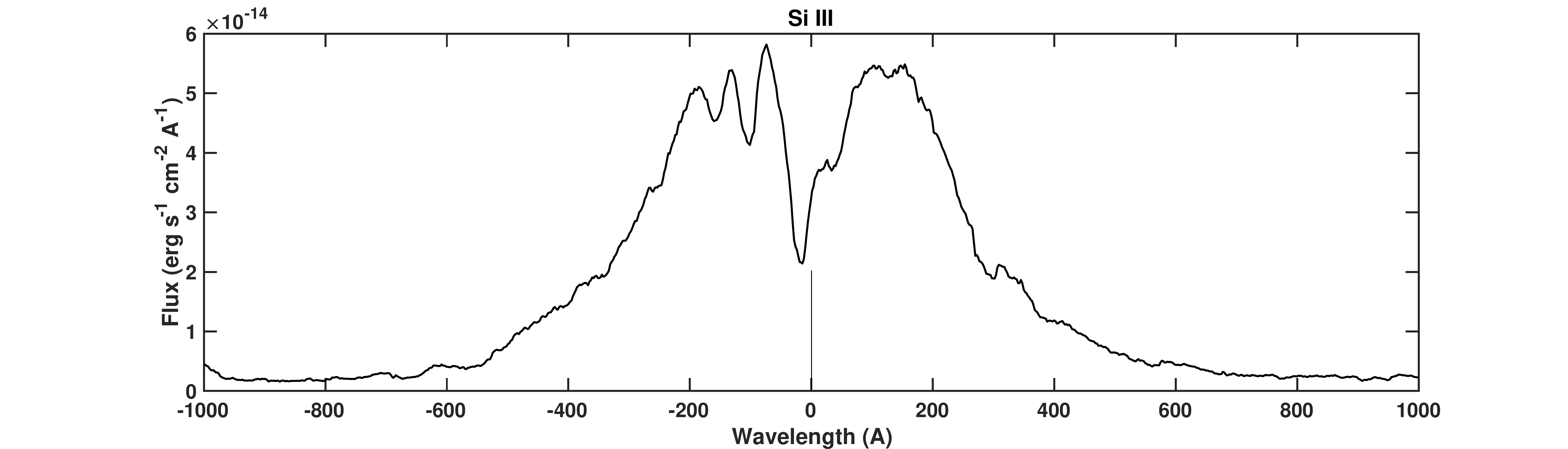} \\
\includegraphics[width=8cm]{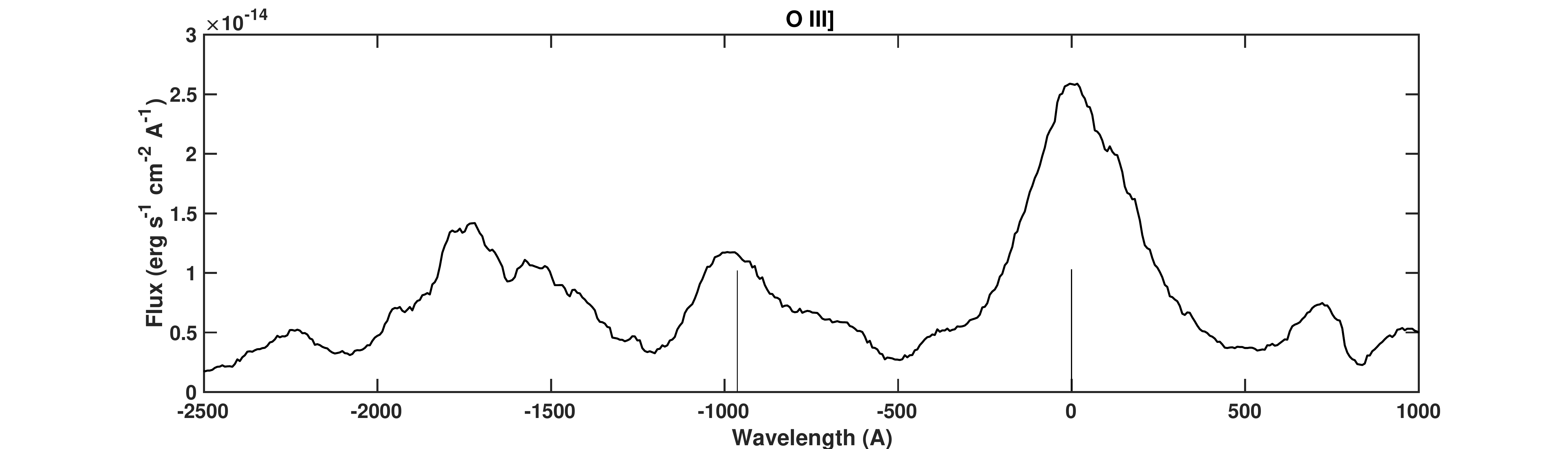} & \includegraphics[width=8cm]{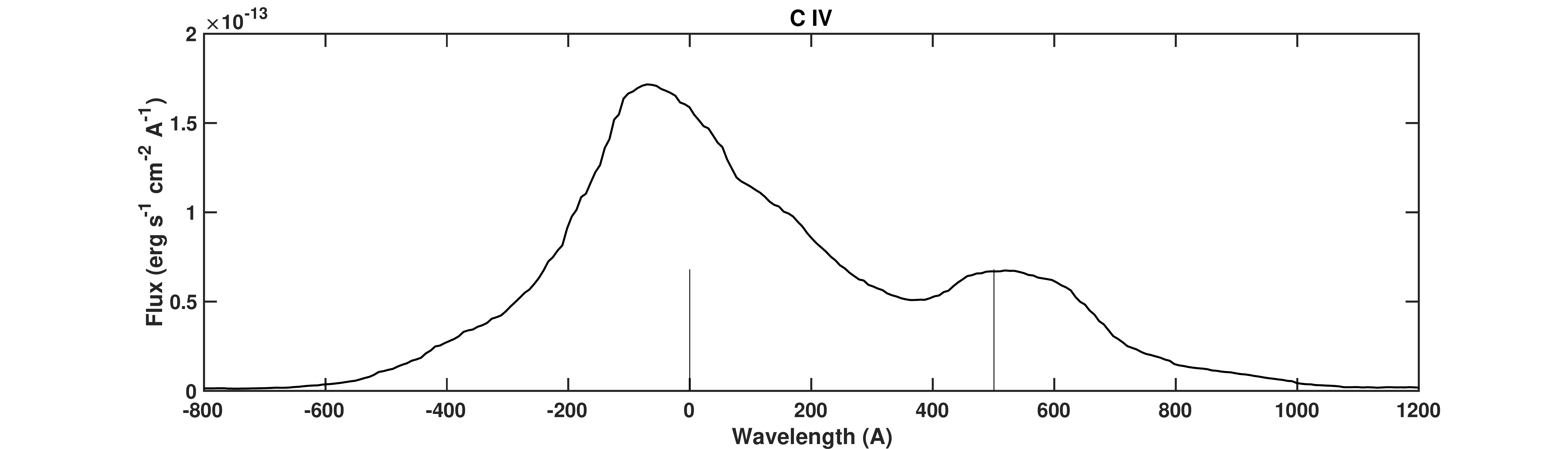} \\
\includegraphics[width=8cm]{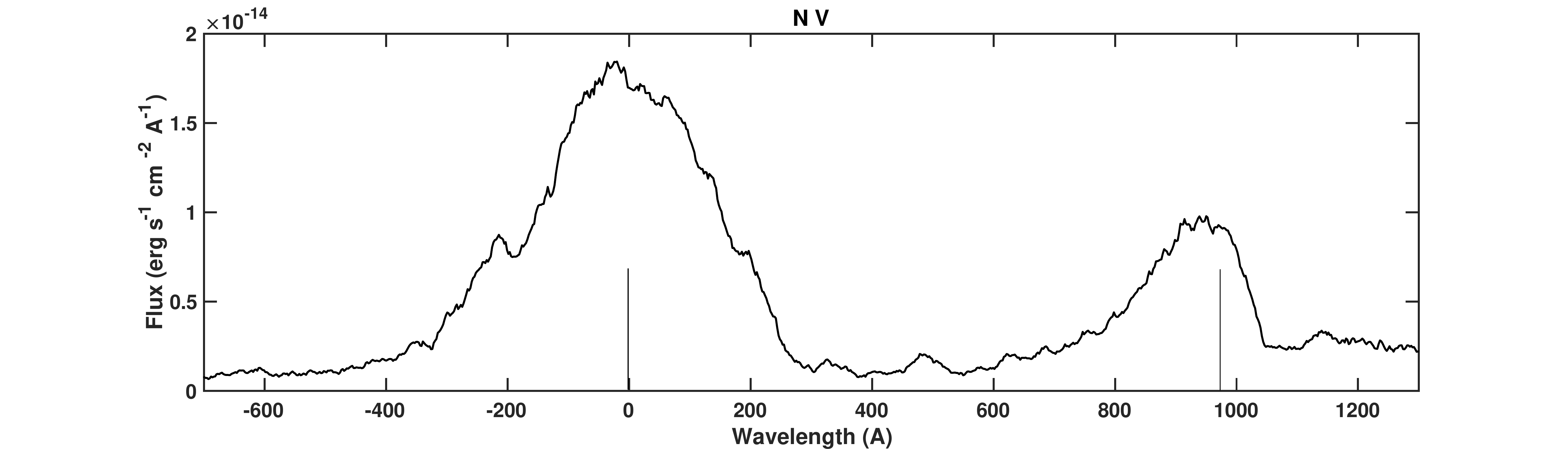} & \includegraphics[width=8cm]{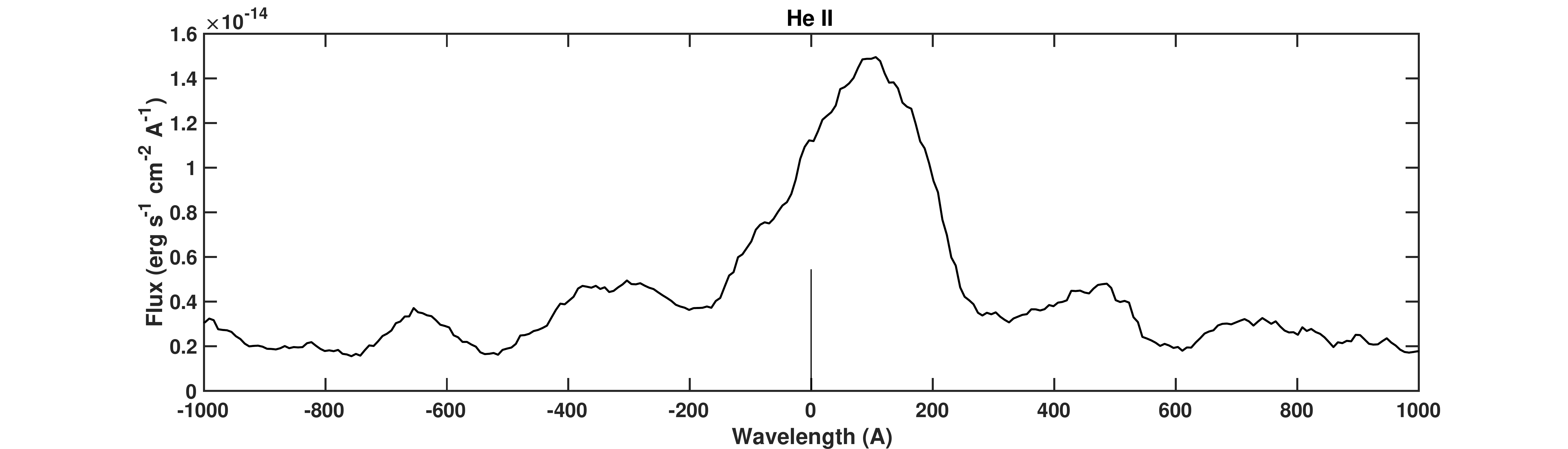} \\
\end{tabular}
\includegraphics[width=16cm]{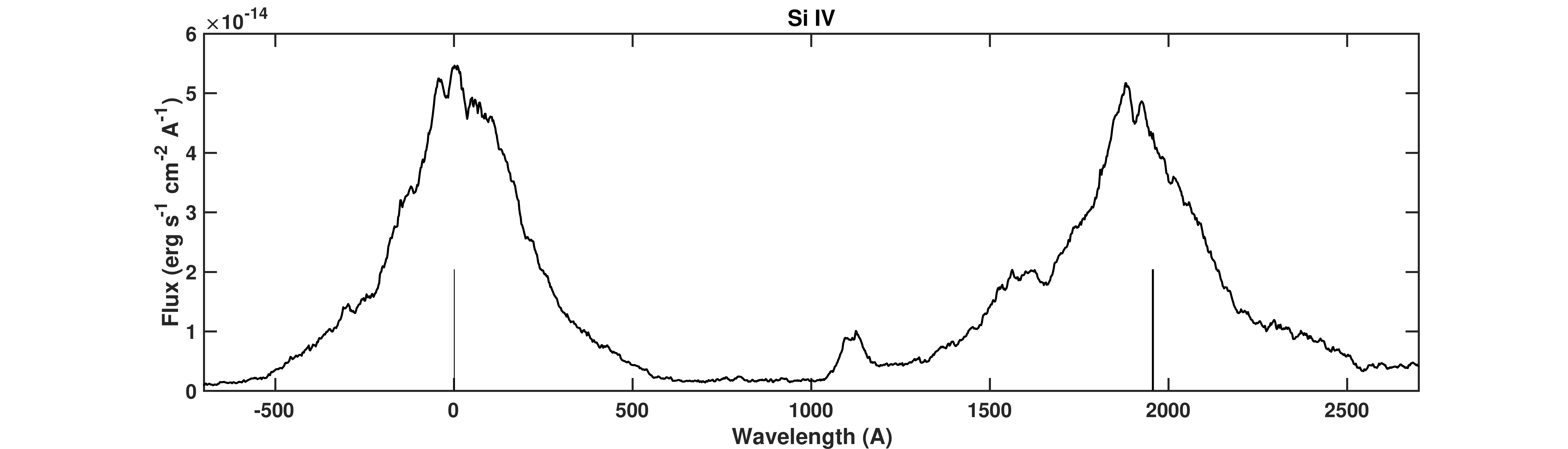} 
\caption{Mean profiles of the main atomic transitions during the COS monitoring. The rest wavelengths of the atomic transitions are marked for reference; in the OI, SI panel these references are colour coded (black OI, orange SI and blue Si II).}
\end{figure*}

\newpage

\begin{figure}[h]
\centering
\includegraphics[width=8cm]{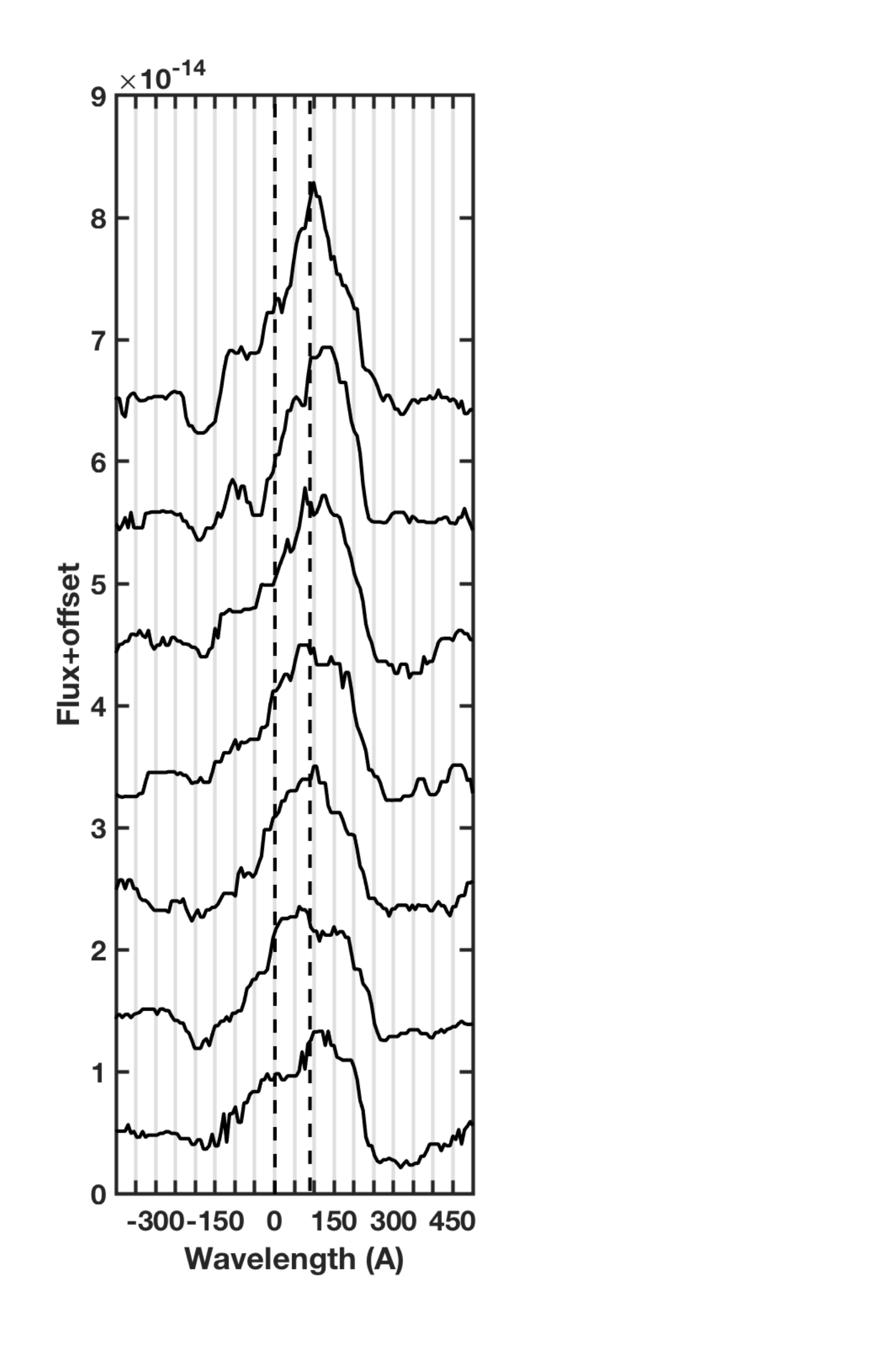} \\
\caption{Variation of the He II profile with the phase; phase increases from bottom to top. The dashed lines mark the 
radial velocity of AK~Sco (-1.3 km/s) and radial velocity 95 km/scorresponding to the orbital solution
at phase 0 for one of the two components of the binary (Alencar et al. 2003). }
\end{figure}

\newpage 

\begin{figure}
\begin{center}
\begin{tabular}{cc}
\includegraphics[width=9cm]{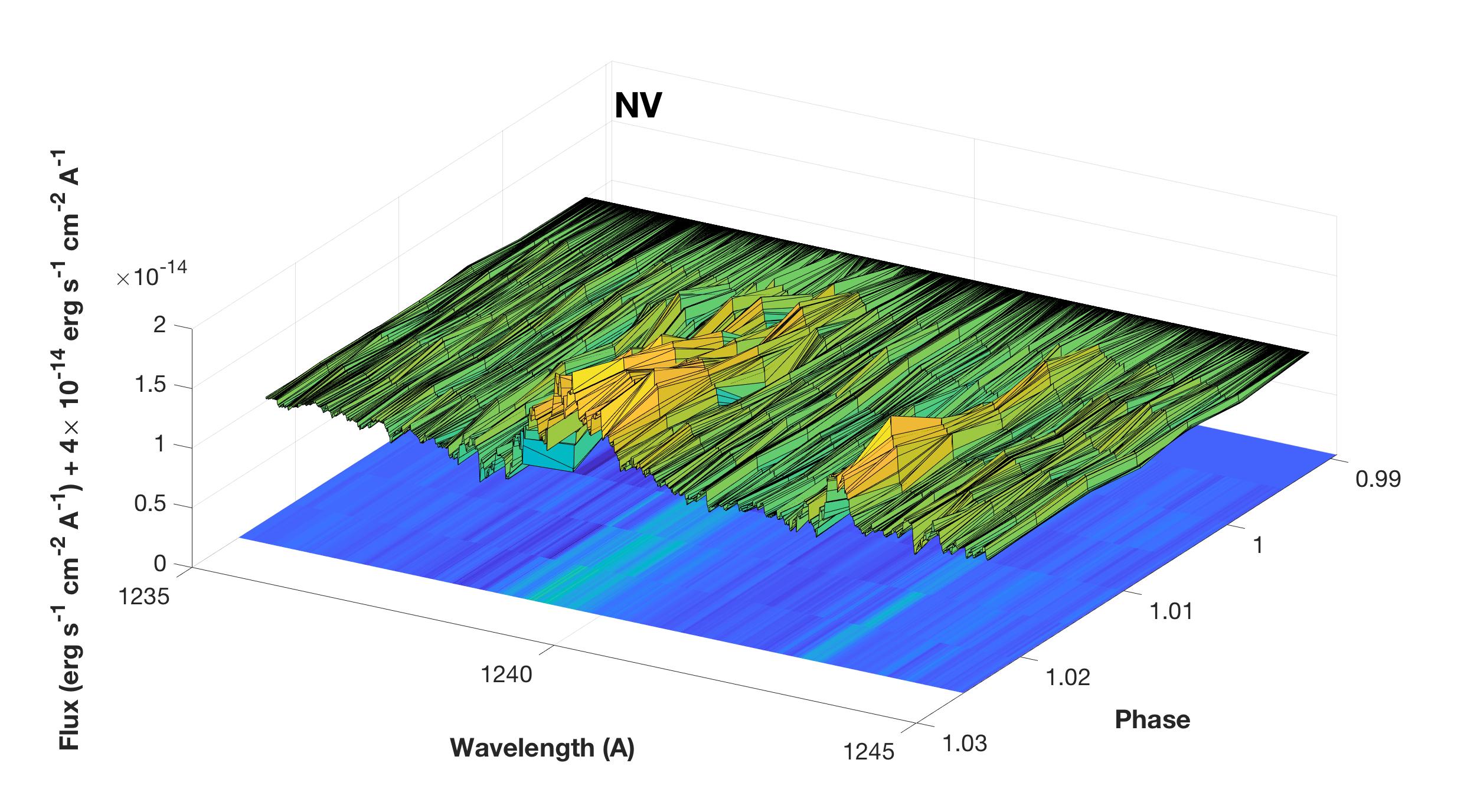}&\includegraphics[width=9cm]{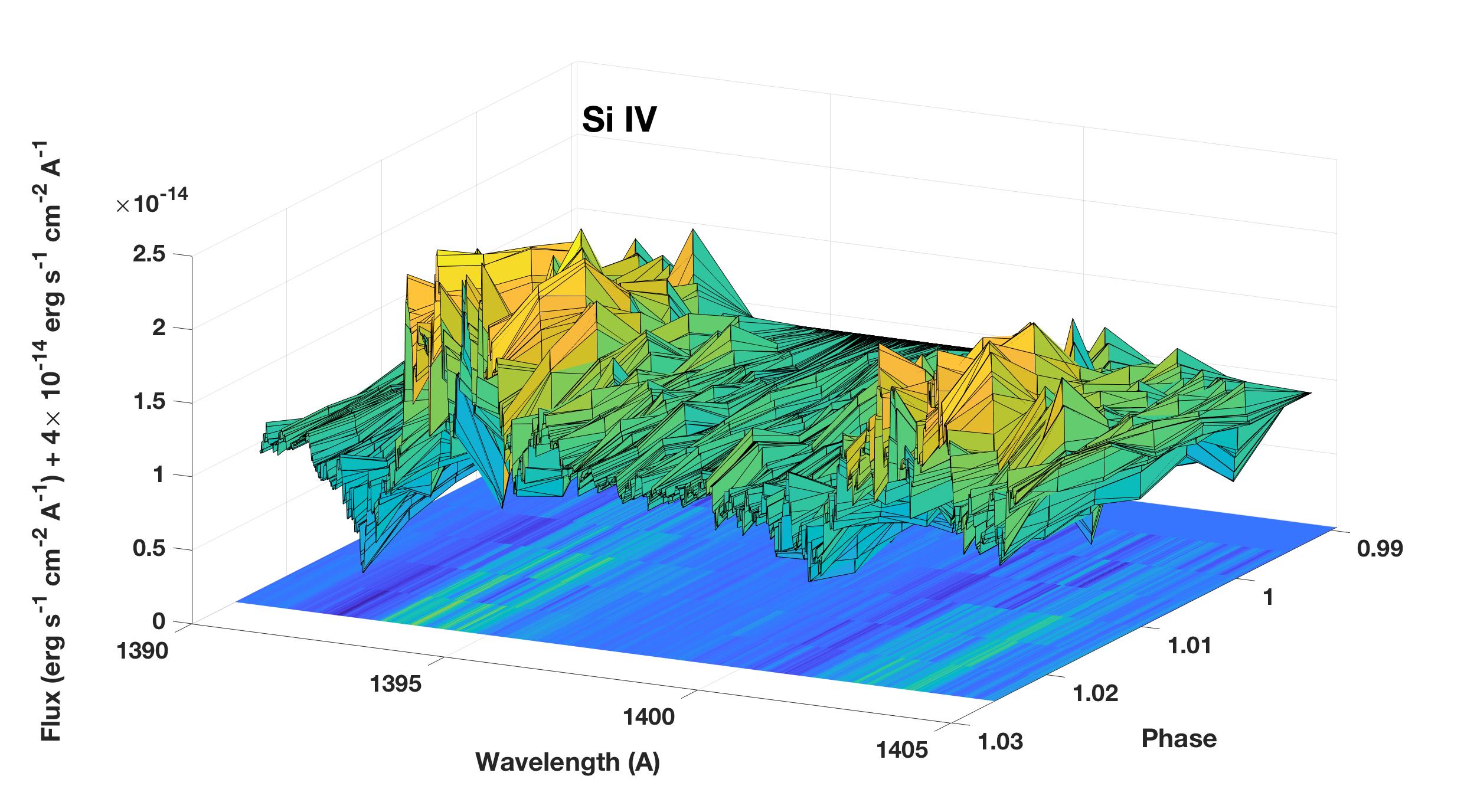}\\
\includegraphics[width=9cm]{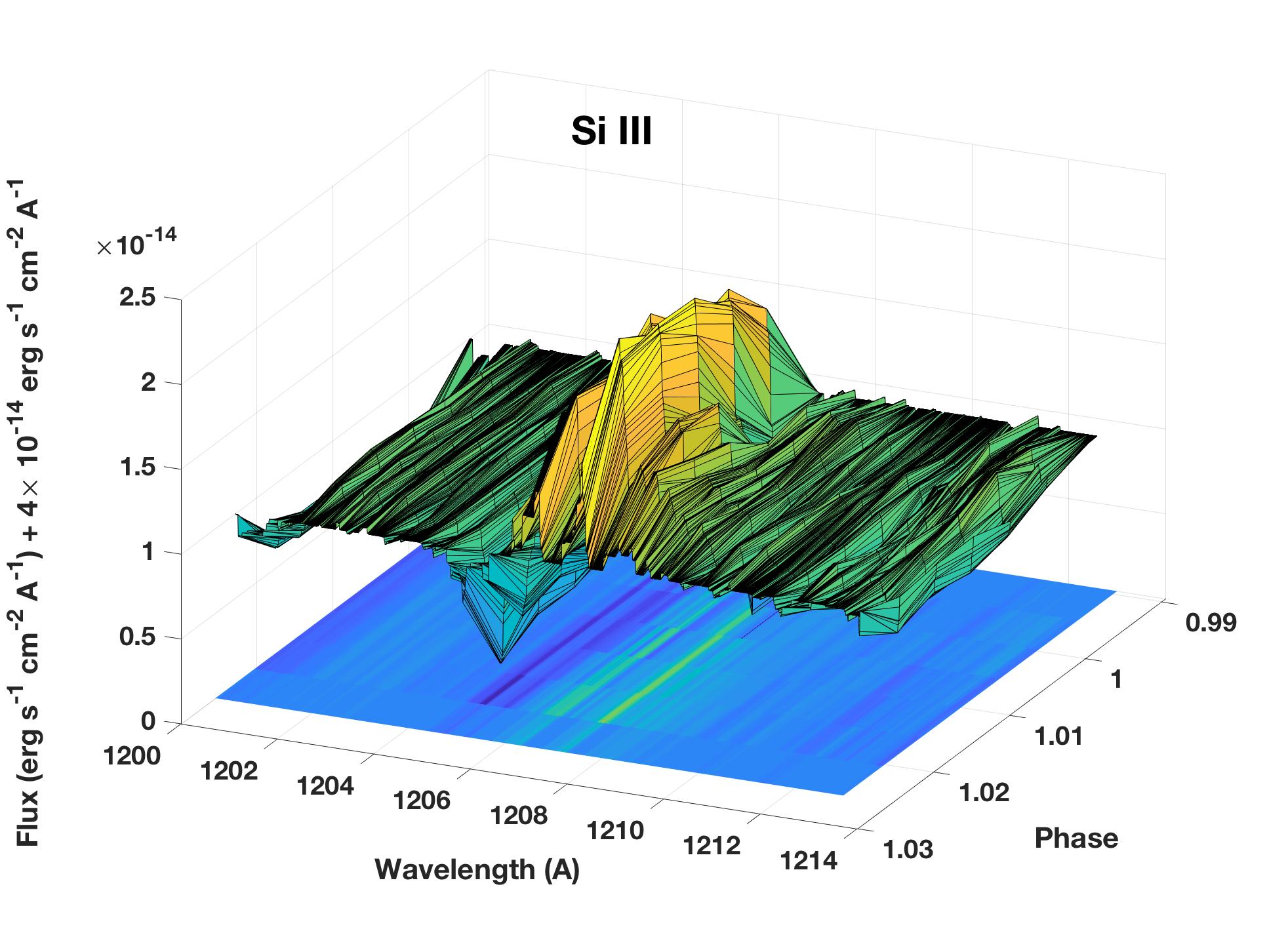}&\includegraphics[width=9cm]{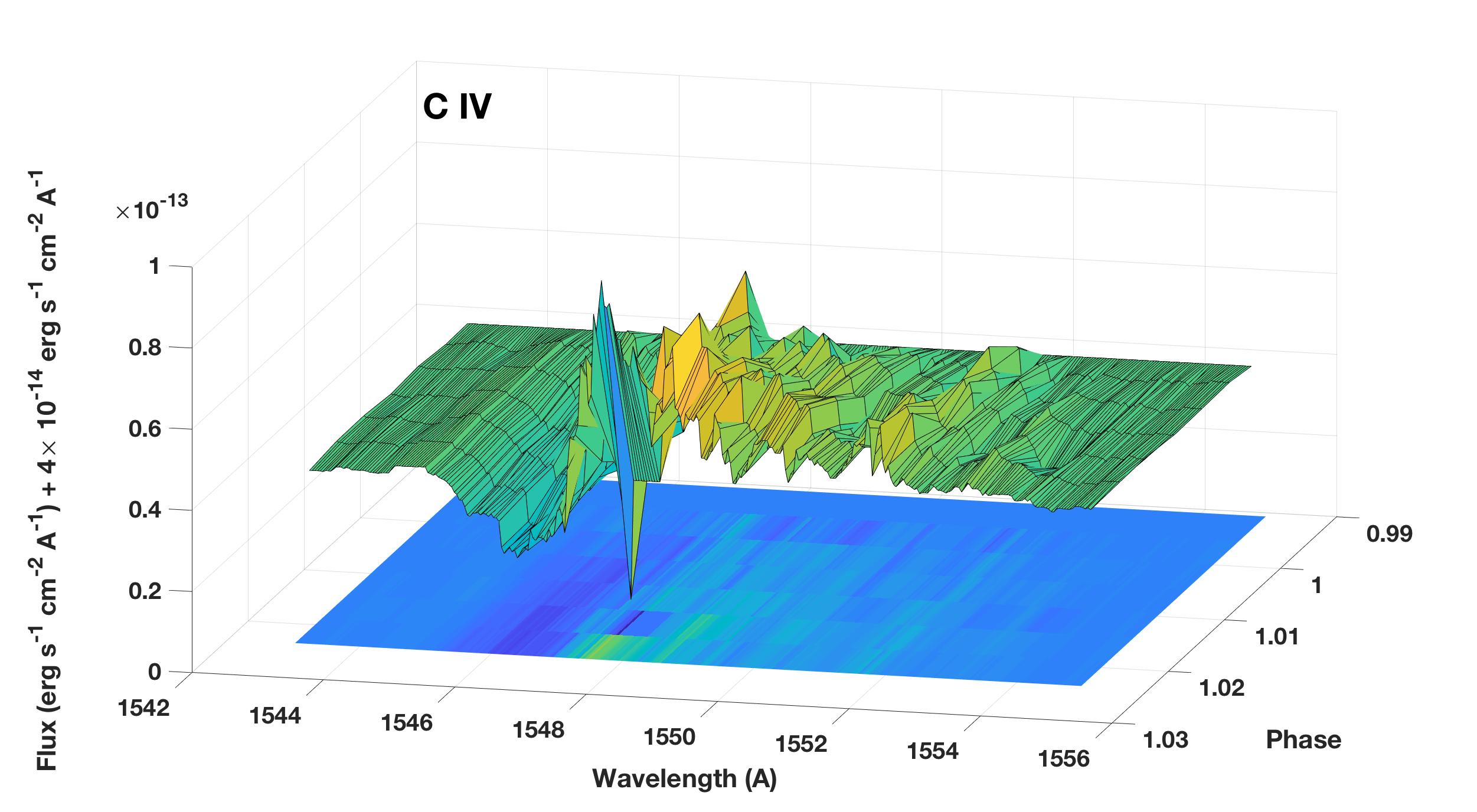}\\
\end{tabular}
\caption{Variability of the N V, Si IV, Si III and CIV profiles during COS monitoring. The evolution of the profile during periastron passage is shown by subtracting the profile at phase 0.9920 from the rest and making a 3-D plot. Note that just after periastron, the flux is enhanced. Also a weak absorption develops at high blue-wards shifted velocities (250 km/s). From top to bottom and from left to right: N V, Si IV, Si III and C IV.}
\end{center}
\label{fig:COS_si4}
\end{figure}

\newpage

\begin{figure}[h]
\centering
\includegraphics[width=20cm]{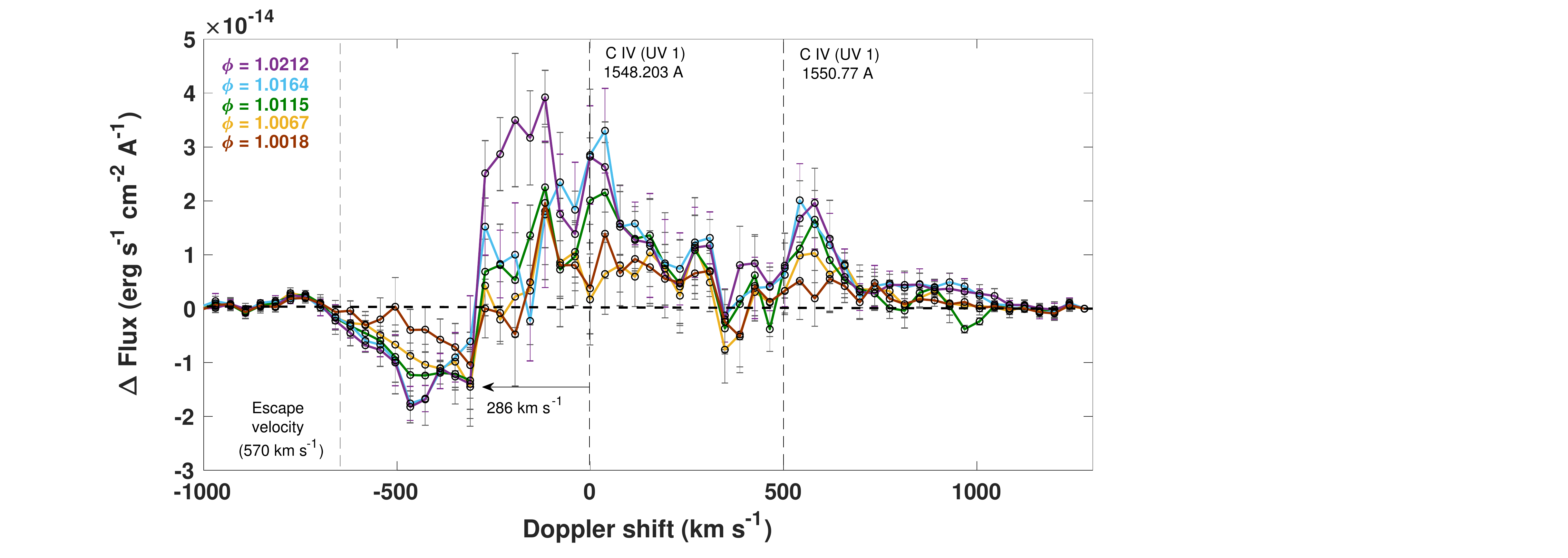} \\
\caption{Variation of the CIV profile during cycle 3. The excess CIV emission with respect to the first two observations in cycle 3 is plotted, with 1$\sigma$ error bars. The rest
wavelengths of the doublet lines, at the radial velocity of the AK Sco binary system are marked. The radial component of the orbital velocity of each component
is $\pm 90$ km~s$^{-1}$ with respect to the systemic velocity. The peak of the CIV, 1550.77~\AA\ line, is redshifted by this amount, similarly to what observed
in the He II line (see Figure 14).}
\end{figure}

\newpage

\begin{figure}[h]
\centering
\includegraphics[width=14cm]{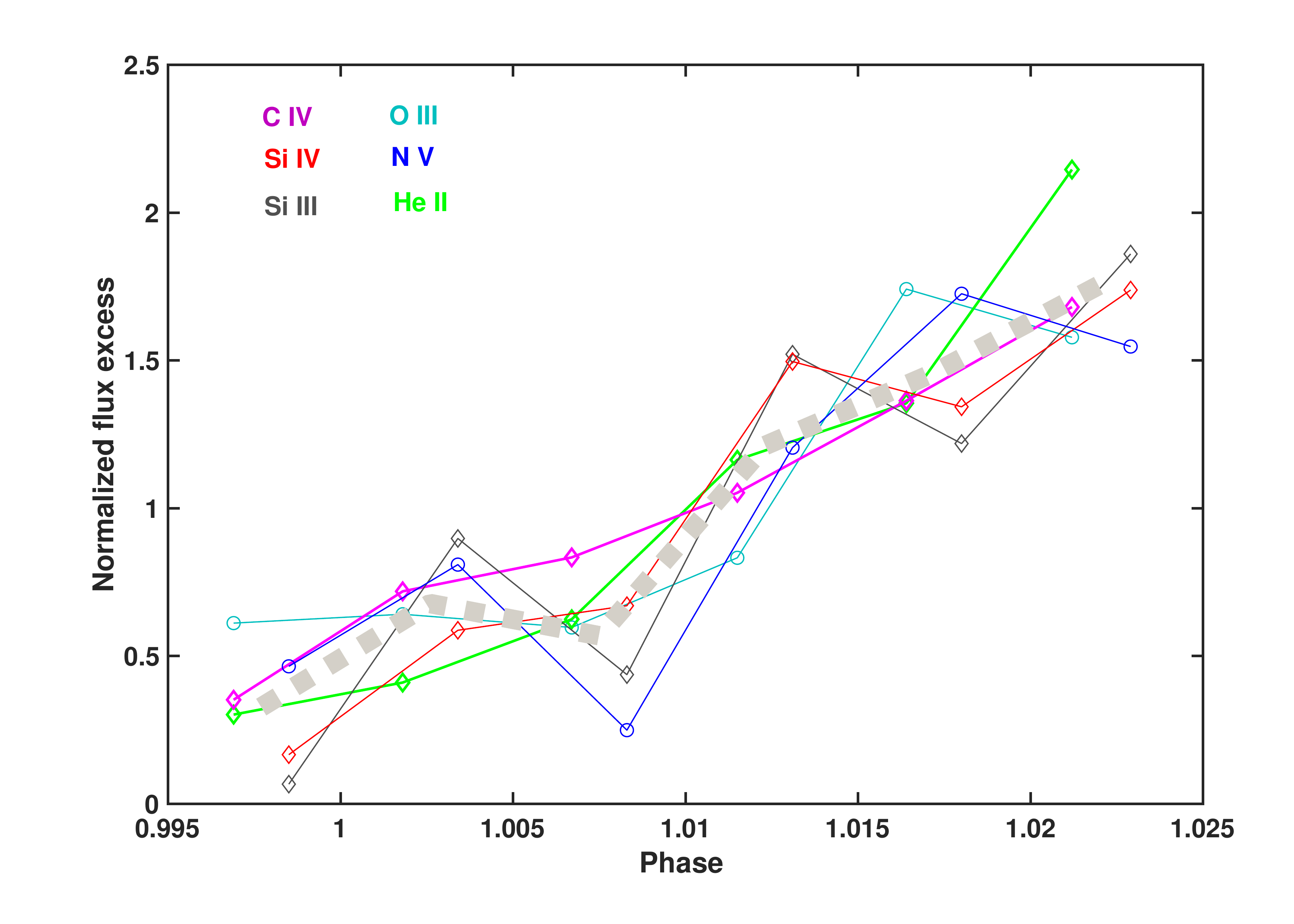} \\
\caption{Variation of the flux excess during periastron passage for the main spectral tracers.
The curves have been normalized to the average excess with values of  $6.49\times 10^{-13}$ erg~s$^{-1}$~cm$^{-2 }$, 
$1.079\times 10^{-12}$ erg~s$^{-1}$~cm$^{-2 }$, $1.311\times 10^{-12}$ erg~s$^{-1}$~cm$^{-2 }$, $1.03\times 10^{-13}$ erg~s$^{-1}$~cm$^{-2 }$,
$2.32\times 10^{-13}$ erg~s$^{-1}$~cm$^{-2 }$ and $0.98 \times 10^{-13}$ erg~s$^{-1}$~cm$^{-2 }$ for  Si~III, Si~IV, C~IV, He~II, O~III and N~V
lines, respectively. The average light curve is marked with a thick dashed line.
}
\end{figure}

\newpage

\begin{figure}[h]
\centering
\includegraphics[width=20cm]{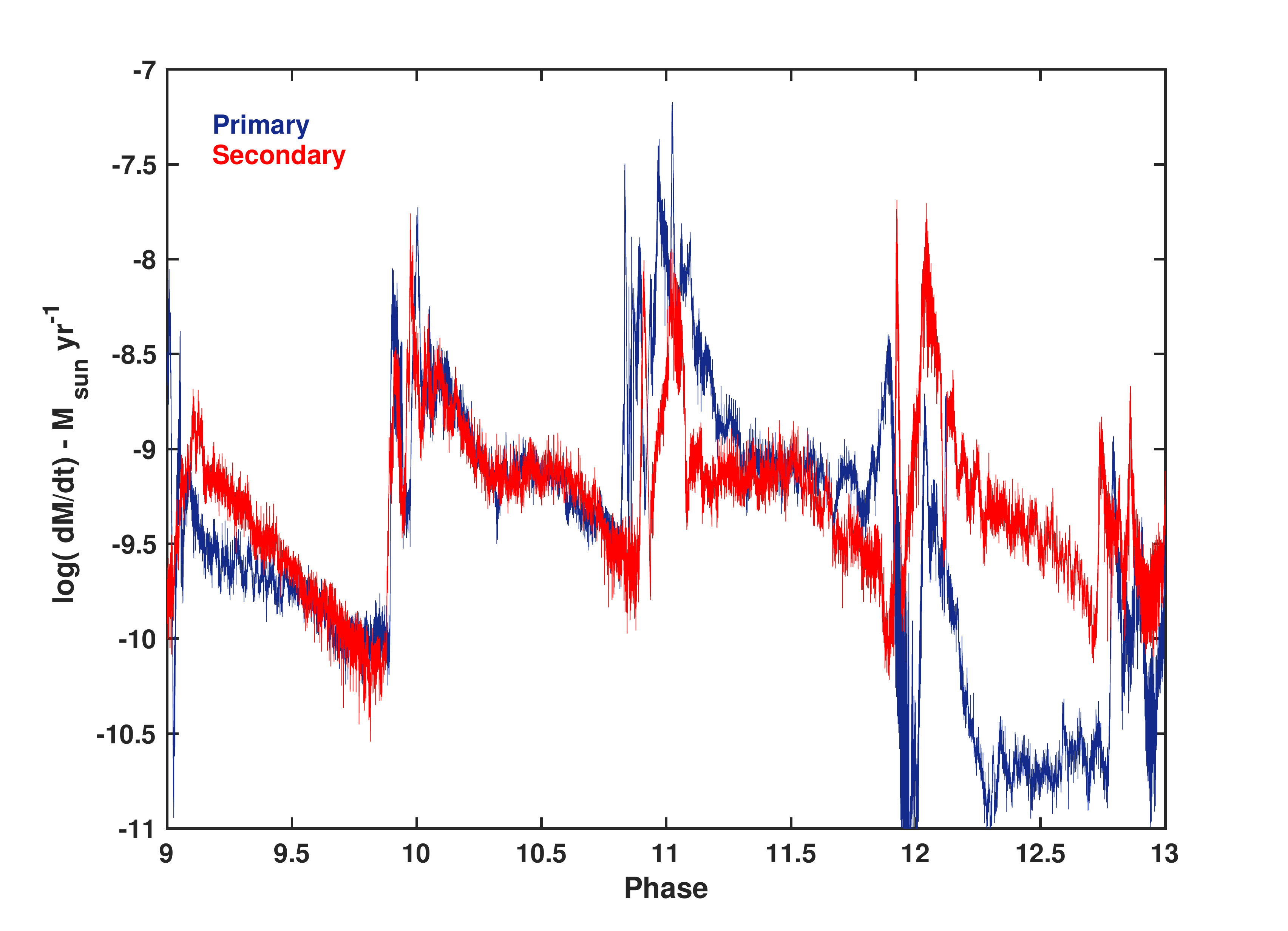} \\
\caption{Expected accretion rate onto both components of the system as inferred from numerical simulations (Paper I).}
\end{figure}

\newpage 

\begin{figure}[h]
\centering
\includegraphics[width=14cm]{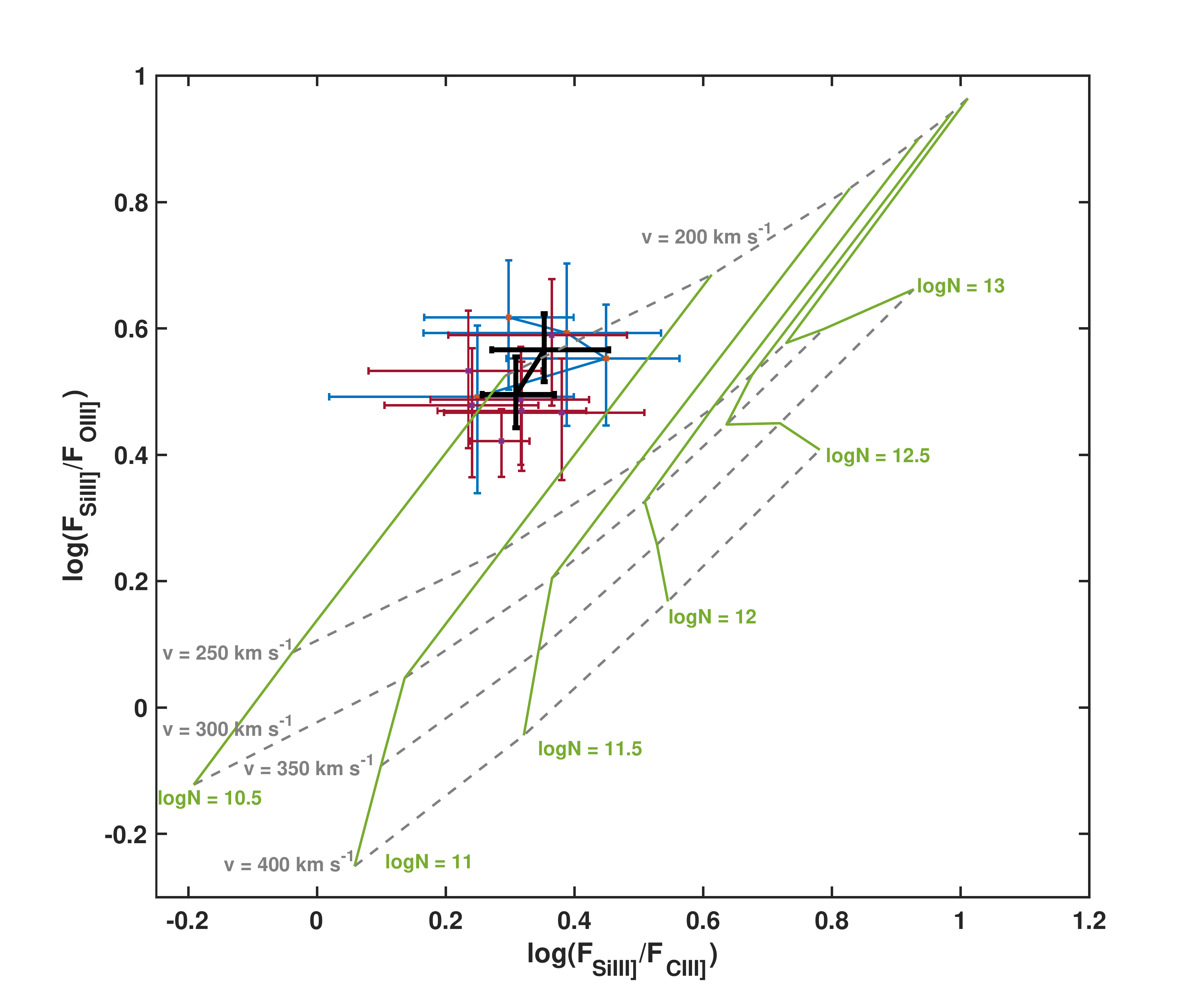} \\
\caption{Accretion shock diagnosis diagram from G\'omez de Castro \& Lamzin (1999). AK observations during cycle 1 (blue) and 2 
(red) are plotted in the diagram (1-$\sigma$ error bars). }
\label{c3shocks}
\end{figure}

\newpage

\begin{figure}
\begin{center}
\includegraphics[width=6cm]{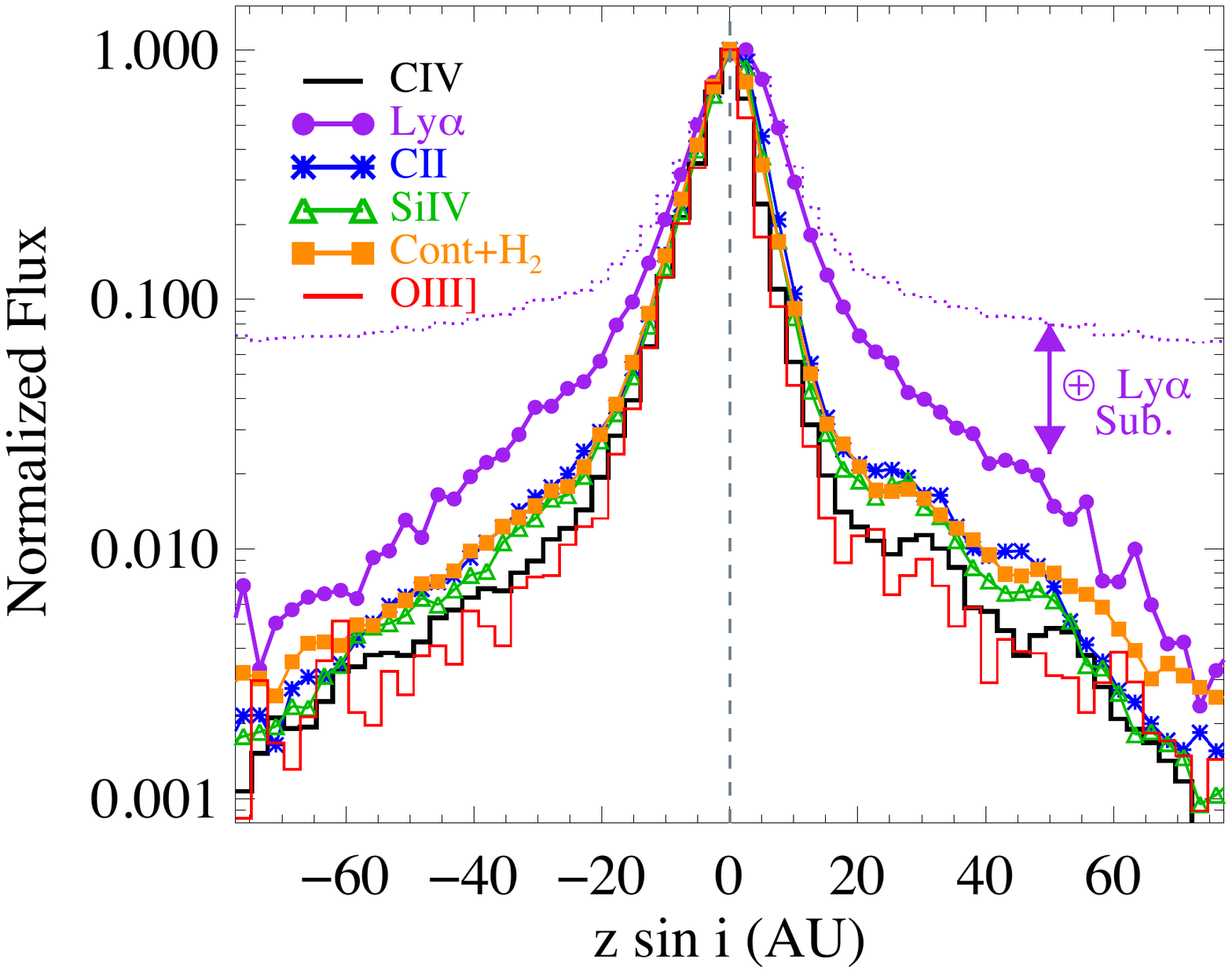}\\
\includegraphics[width=6 cm]{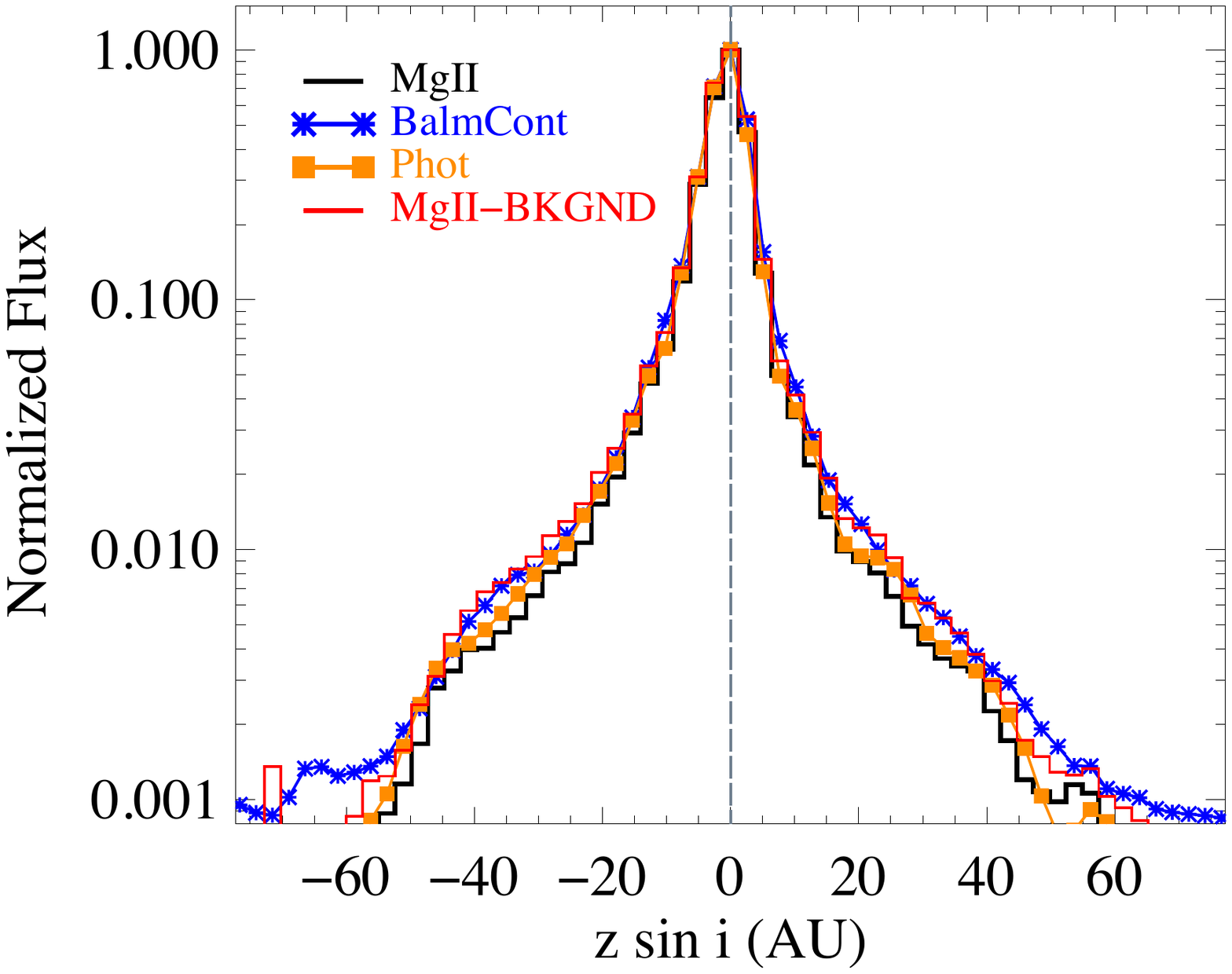}
\caption{ Spatial (cross-dispersion) profiles of the STIS G140L ($top$) and G230L ($bottom$) data in discrete bands, as labeled in the legends.  The dotted and solid Ly$\alpha$ curves represent the Ly$\alpha$ extraction before and after subtracting the geocoronal signal, respectively.  
 }
\end{center}
\label{fig:spatialprofile}
\end{figure}

\newpage
\begin{figure}
\begin{center}
\begin{tabular}{c}
\includegraphics[width = 13cm]{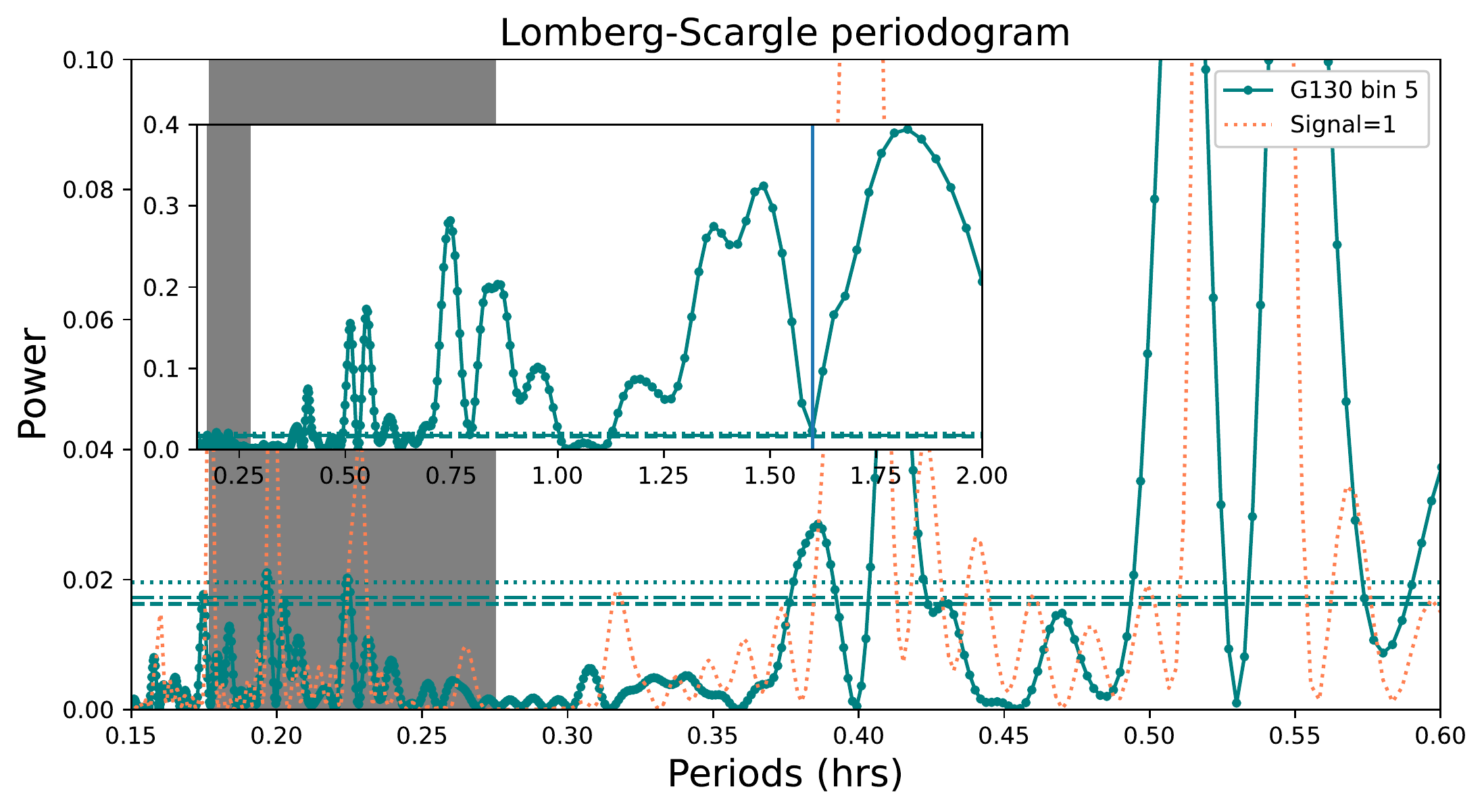} \\
\includegraphics[width = 13cm]{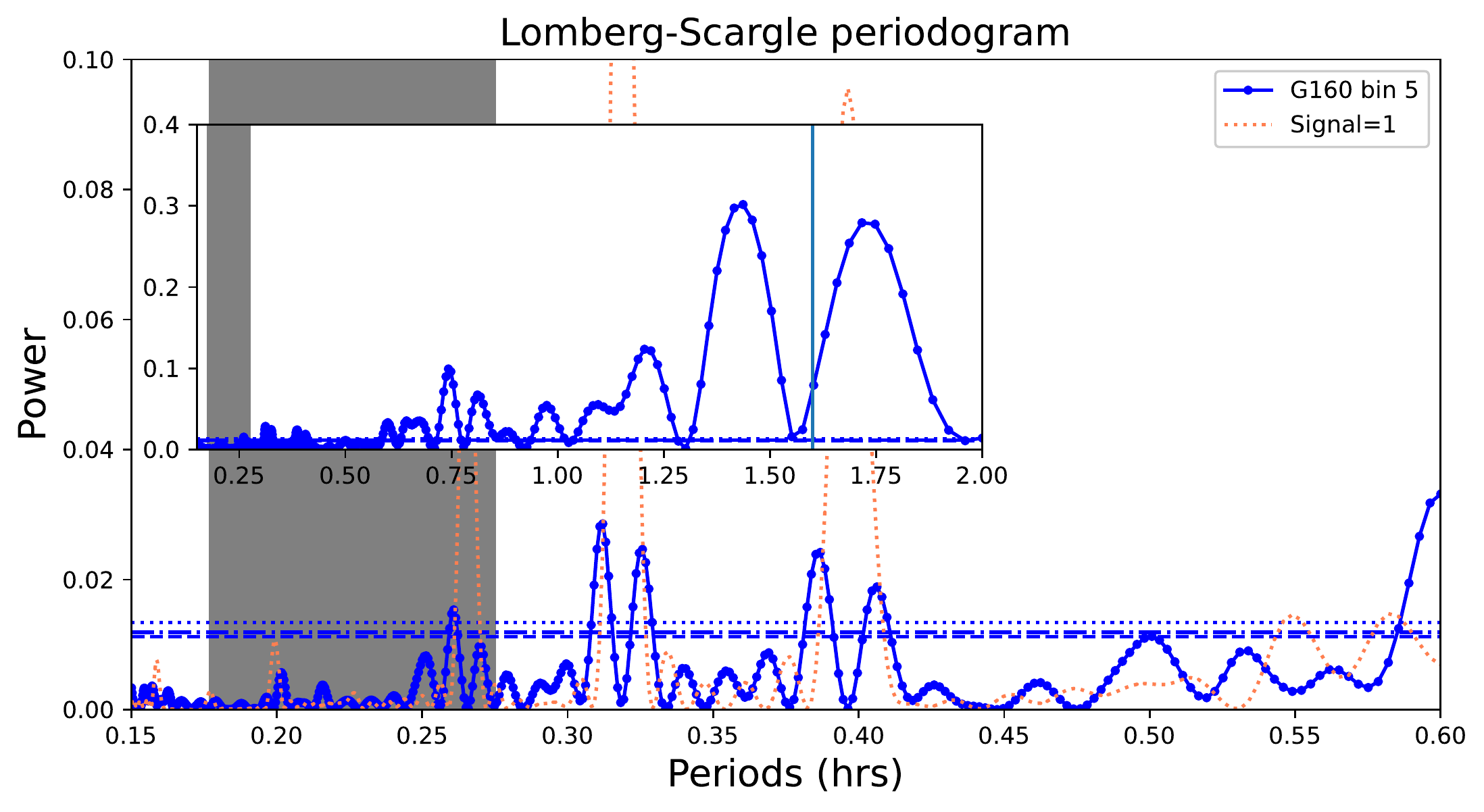}
\end{tabular}
\end{center}
\caption{Lomb-Scargle periodograms corresponding to the COS light-curves.
The upper panel corresponds to the G130M data, and the bottom panel to
the G160M data. The grey areas show 
the suspected ULF period interval. The insets show the periodograms up to a period of $2.0$hrs. The dotted pale
curves result from processing the exposure signals arbitrarily set to $1$, and follow
the same patterns that the periodograms resulting from the real signal.}
\label{fig_jcv_periodogram}
\end{figure}

\newpage
\begin{figure}
\begin{center}
\begin{tabular}{c}
\includegraphics[width = 13cm]{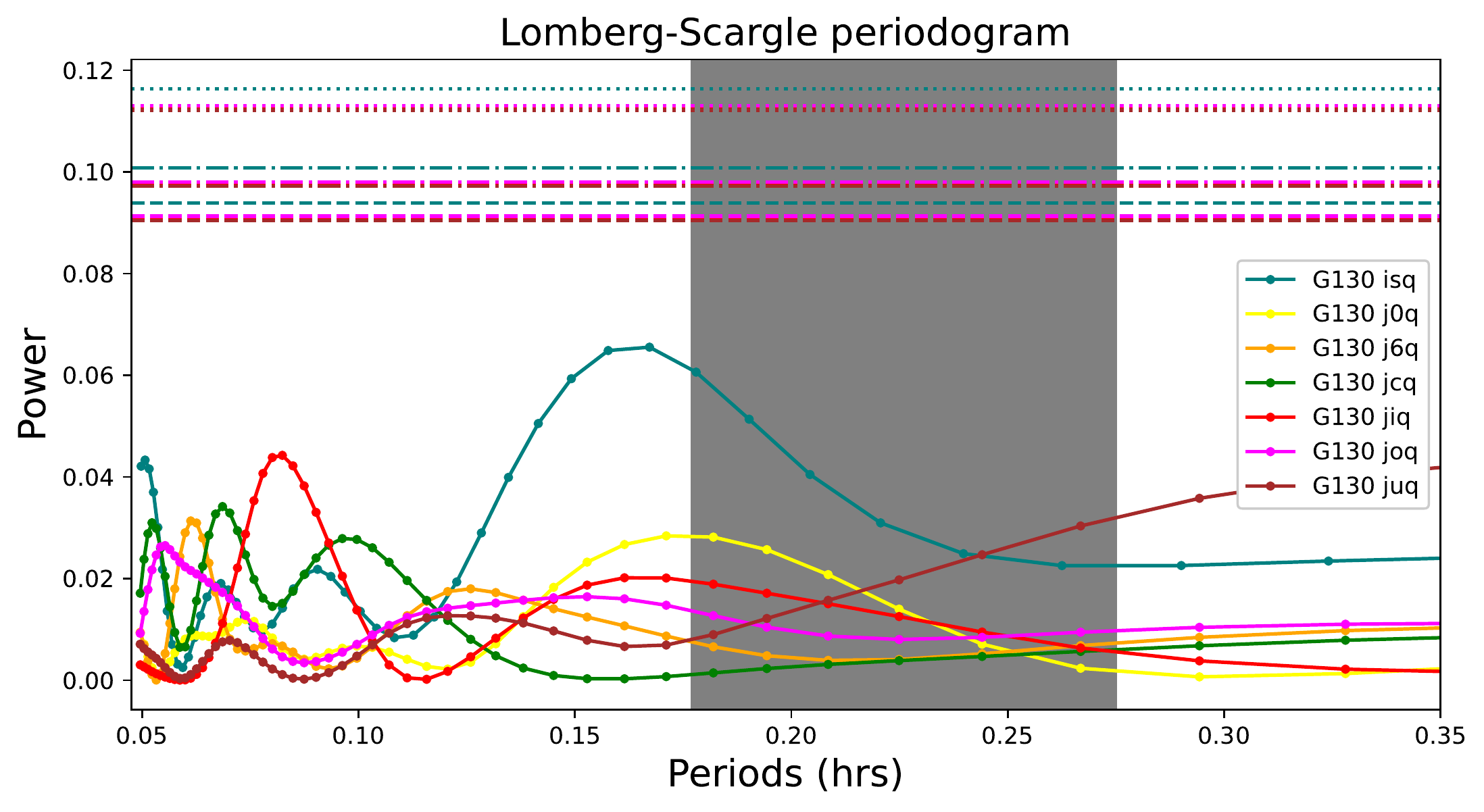} \\
\includegraphics[width = 13cm]{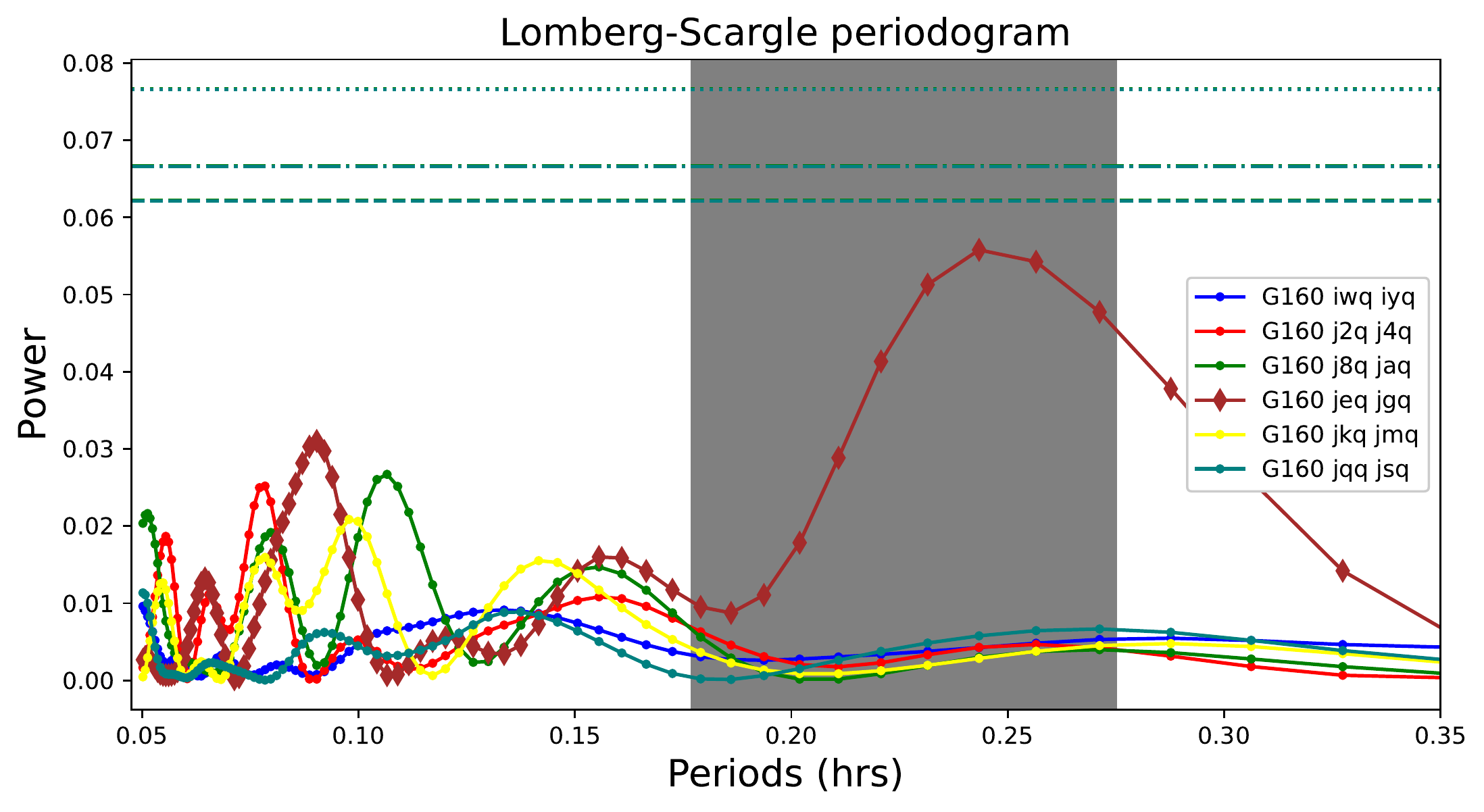} 
\end{tabular}
\end{center}
\caption{Lomb-Scargle periodograms corresponding to the G130M individual exposures
(upper panel) and to the paired exposures from G160M data (bottom panel).
The jeq-jgq pair presents a peak around $0.25$hrs, with a fap of $\%24$.}
\label{fig_jcv_periodogram_individual}
\end{figure}

\clearpage

\begin{figure}[h]
\centering
\includegraphics[width=15cm]{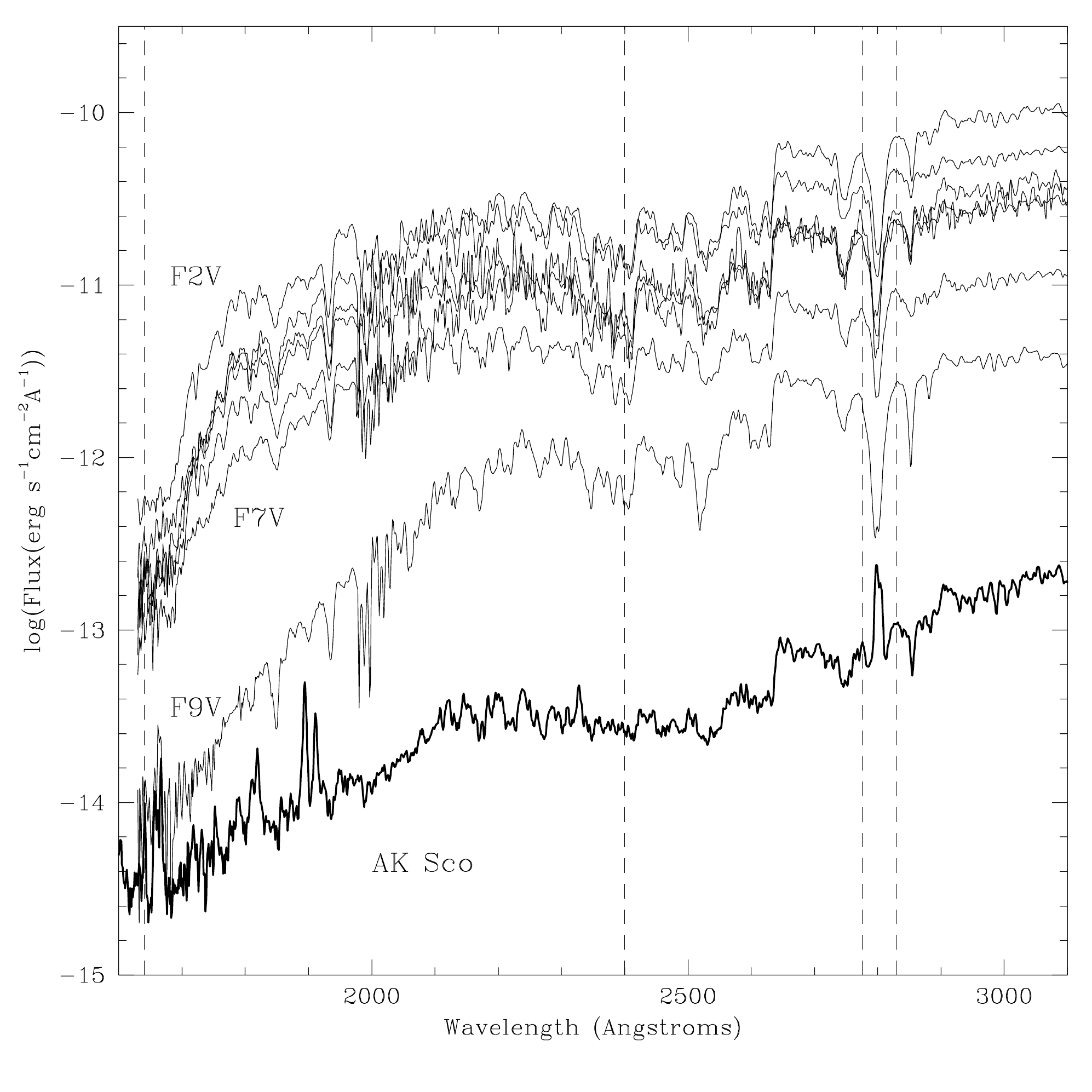}
\caption{UV spectra of main sequence F stars. AK Sco spectrum is also plotted for comparison. The borders of the windows used
to define the UV colour-colour diagrams are marked with dashed lines; note that the Mg~II feature is avoided.}
\label{fig:Fstars_SED}
\end{figure}

%%%%%TABLES
\newpage
\begin{table}
\caption{AK Sco main parameters.}
\begin{center}
\small
\begin{tabular}{lll}
\hline
Property & Value & Source \\
\hline
Projected semimajor axis & $a\sin i = 30.77 \pm 0.12$R$_{\odot}$ & Andersen et al 1989 \\
Eccentricity & e= 0.47 & Andersen et al 1989, Alencar et al. 2003 \\
Orbital period & P=13.609453$\pm$ 0.000026 d & Andersen et al 1989, Alencar et al. 2003 \\
Periastron Passage & T=2,446,654.3634 $\pm$ 0.0086 &  Alencar et al. 2003 \\
Inclination & $ i=65^o-70^o$ & Alencar et al. 2003 \\
Age & 10-30 Myrs & Alencar et al. 2003  \\
Spectral type & F5 & Alencar et al. 2003\\
Stellar Mass & $M_* = 1.35 M_{\odot} \pm 0.07 M_{\odot}$ & Alencar et al. 2003\\
Radius & $R_* = 1.59 R_{\odot} \pm 0.35 R_{\odot}$ & Alencar et al. 2003 \\
Projected rotation velocity & $v \sin i = 18.5 \pm 1.0$ km s$^{-1}$ & Alencar et al. 2003 \\
Bolometric flux & $6.33\times 10^{-9}$ erg~s$^{-1}$cm$^{-2}$ & Andersen et al 1989\\
$L_{bol}$ & 8.07 $L_{\odot}$ & \\
Extinction: A$_V$ & 0.5 mag & Manset et al. 2005 \\
Extinction: R & 4.3 & Manset et al. 2005 \\
Distance & 102.8 pc & Van Leeuwen 2007 \\
Radial velocity & -1.3 km s$^{-1}$ & Gontcharov 2006\\ 
\end{tabular}
\end{center}
\label{tab:table1}
\end{table}

\newpage
\topmargin=-40pt
\voffset=-20pt
\begin{table}
\begin{tiny}
\caption{Log of observations.\label{tab:table1}}
\begin{tabular}{llllllll}
\hline\hline
Instrument/ & Observation  &  Start Time   &Phase$^{(1)}$ & Exposure   & Dispersion & Spec. Initial & Spec. Final  \\
Grating     & ID           & (JD-2456800.0)&              & Time (sec) &          &Wavelength (\AA )& Wavelength (\AA )\\ \hline
VISIT 1:    &		   &			     &		&  &		  &		 & 		\\
STIS/G140L  &	OCA601010  &47.740671& 0.9924&	1191.185  &	1190.000  &	1140.000  &	1730.000\\
STIS/G230L  &	OCA601020  &47.760116& 0.9939 &	716.200	  &     740.000	  &	1568.000  &	3184.000\\
STIS/G140L  &	OCA601030  &47.799606& 0.9968 &	600.179	  &     1190.000  &	1140.000  &	1730.000\\
STIS/G230L  &	OCA601040  &47.812211& 0.9977 &	60.020	  &     740.000	  &	1568.000  &	3184.000\\
STIS/G140L  &	OCA601050  &47.818565& 0.9982 &	860.190	  &	1190.000  &	1140.000  &	1730.000\\
STIS/G230L  &	OCA601060  &47.834178& 0.9993 &	50.002	  &	740.000	  &	1568.000  &	3184.000\\
STIS/G140L  &	OCA601070  &47.865949& 1.0017 &	600.193	  &	1190.000  &	1140.000  &	1730.000\\
STIS/G230L  &	OCA601080  &47.878553& 1.0026 &	60.020	  &	740.000	  &	1568.000  &	3184.000\\
STIS/G140L  &	OCA601090  &47.884907& 1.0030 &	860.189	  &	1190.000  &	1140.000  &	1730.000\\
STIS/G230L  &	OCA6010A0  &47.900521& 1.0042 &	50.002	  &	740.000	  &	1568.000  &	3184.000\\ \hline
VISIT 2:    &		   &			     &		  & &		  &		 & 		\\
STIS/G140L  &	OCA602010  &61.341238& 0.9918 &	1191.198  &	1190.000  &	1140.000  &	1730.000\\
STIS/G230L  &	OCA602020  &61.360683& 0.9932 &	716.200	  &	740.000	  &	1568.000  &	3184.000\\
STIS/G140L  &	OCA602030  &61.374630& 0.9943 &	600.190	  &	1190.000  &	1140.000  &	1730.000\\
STIS/G230L  &	OCA602040  &61.414896& 0.9973 &	60.020	  &	740.000	  &	1568.000  &	3184.000\\
STIS/G140L  &	OCA602050  &61.421250& 0.9977 &	860.184	  &	1190.000  &	1140.000  &	1730.000\\
STIS/G230L  &	OCA602060  &61.436863& 0.9988 &	50.002	  &	740.000	  &	1568.000  &	3184.000\\
STIS/G140L  &	OCA602070  &61.443102& 0.9993 &	600.199	  &	1190.000  &	1140.000  &	1730.000\\
STIS/G230L  &	OCA602080  &61.474120& 1.0016 &	60.020	  &	740.000	  &	1568.000  &	3184.000\\
STIS/G140L  &	OCA602090  &61.480475& 1.0021 &	860.197	  &	1190.000  &	1140.000  &	1730.000\\
STIS/G230L  &	OCA6020A0  &61.496088& 1.0032 &	50.002	  &	740.000	  &	1568.000  &	3184.000\\
STIS/G140L  &	OCA6020B0  &61.502326& 1.0037 &	600.189	  &	1190.000  &	1140.000  &	1730.000\\
STIS/G230L  &	OCA6020C0  &61.514931& 1.0046 &	60.020	  &	740.000	  &	1568.000  &	3184.000\\
STIS/G140L  &	OCA6020D0  &61.540567& 1.0065 &	860.199	  &	1190.000  &	1140.000  &	1730.000\\
STIS/G230L  &	OCA6020E0  &61.556181& 1.0076 &	50.002	  &	740.000	  &	1568.000  &	3184.000\\ \hline
VISIT 3:    &		   &			     &		 & &		  &		 & 		\\
COS/G130M  &	LCA603010  &74.952778& 0.9920 &	1000.192  &	19000.000  &	1159.478  &	1453.067\\
COS/G160M  &	LCA603020  &74.967025& 0.9930 &	1260.160  &	19000.000  &	1401.911  &	1762.490\\
COS/G160M  &	LCA603030  &75.019109& 0.9969 &	1600.320  &	19000.000  &	1416.649  &	1777.297\\
COS/G130M  &	LCA603040  &75.041574& 0.9985 & 1037.184  &	18000.000  &	1149.968  &	1443.513\\
COS/G160M  &	LCA603050  &75.085498& 1.0018 &	1600.288  &	19000.000  &	1428.748  &	1789.398\\
COS/G130M  &	LCA603060  &75.108044& 1.0034 &	1035.200  &	18000.000  &	1136.604  &	1430.194\\
COS/G160M  &	LCA603070  &75.151852& 1.0067 &	1600.384  &	19000.000  &	1428.846  &	1789.508\\
COS/G130M  &	LCA603080  &75.174363& 1.0083 &	1035.200  &	19000.000  &	1166.476  &	1460.058\\
COS/G160M  &	LCA603090  &75.218194& 1.0115 &	1600.384  &	19000.000  &	1410.898  &	1771.532\\
COS/G130M  &	LCA6030A0  &75.240706& 1.0131 &	1035.200  &	19000.000  &	1153.969  &	1447.576\\
COS/G160M  &	LCA6030B0  &75.284560& 1.0164 &	1600.320  &	19000.000  &	1405.080  &	1765.708\\
COS/G130M  &	LCA6030C0  &75.307072& 1.0180 &	1035.200  &	18000.000  &	1149.590  &	1443.174\\
COS/G160M  &	LCA6030D0  &75.350903& 1.0212 &	1600.352  &	19000.000  &	1419.522  &	1780.184\\
COS/G130M  &	LCA6030E0  &75.373368& 1.0229 & 1035.168  &	19000.000  &	1151.647  &	1445.254\\  \hline
\end{tabular}

\begin{tabular}{ll}
(1) & According to ephemeris in Table~1.\\
\end{tabular}
\end{tiny}
\end{table}

\newpage
\begin{table}
\begin{small}
\caption{F-type stars from the I.U.E. Archive$^{(a)}$ \label{tab:tableA1}}
\begin{tabular}{llll}
\hline\hline
Star & Spectral  &  Distance & (B-V)$^{(b)}$\\ 
      &  Type &  (pc)        & mag\\ \hline
HD129502 & F2V & 18.3 & 0.37\\
HD26462 & F4V & 37.0 & 0.36 \\
HD139664 & F5V & 17.4 & 0.40\\
HD22001 & F5V & 21.7 & 0.39\\
HD30652 & F6V & 8.1 & 0.44\\
HD173667 & F6V & 19.2 & 0.46\\
HD222368 & F7V & 13.7 & 0.50\\
HD126660 & F7V & 14.5 & 0.51\\
HD11007 & F8V & 27.9 & 0.55 \\
HD114710 &F9.5V & 9.1 & 0.59 \\ 
HD70907 & F3IV/V & 286.1 & 0.46 \\
HD124850 & F7IV & 22.2 & 0.52\\
HD220657 &F8IV & 52.2 & 0.61\\
HD61110 &F3III & 51.0 & 0.40\\
HD57623 &F6II & 226.2 & 0.79 \\
HD20902 &F5Ib & 155.3 & 0.48\\
HD171635 &F7Ib &649.35  & 0.587\\
HD54605 &F8Iab: & 492.6 & 0.68\\ \hline
\end{tabular}

\begin{tabular}{ll}
(a) & Spectral types, (B-V) and distances (parallaxes) have been extracted \\
     & from the data base in the Centre de Donn\'ees Stelaires in \\
     & Strassbourg (France), using the SIMBAD interface.\\
(b) & Intrinsec (B-V) colours of main sequence F-type stars \\
     & range from 0.32 (F0), 0.35 (F2), 0.45 (F5), 0.53 (F8) \\
    & (source \url{www.stsci.edu/~inr/intrins/johnson.cols}) \\

\end{tabular}
\end{small}
\end{table}

\clearpage

\begin{equation}
\langle v_{rad} \rangle~=~\frac{ \frac{ \sum\limits_{i}\lambda_{i}F_{i}(\lambda) }{ \sum\limits_{i} F_{i}(\lambda)}  - \lambda_{o} }{ \lambda_{o}} c
\end{equation}

\end{document}